\documentclass[11pt,urlcolor=blue, linkcolor=blue]{article} 
\usepackage{cite}

\usepackage{amsmath, amsthm, amssymb,slashed}
\usepackage{ifpdf}
\ifpdf
  \usepackage[pdftex]{graphicx}
  \usepackage{epstopdf}
\else
  \usepackage[dvips]{graphicx}
\fi
\parskip=\baselineskip

\usepackage[usenames, dvipsnames]{color}

\allowdisplaybreaks[1]

\newcommand{\pin}{\textrm{Pin}}
\newcommand{\spin}{\textrm{Spin}}
\usepackage{booktabs}
\newtheorem{thm}{Theorem}

\theoremstyle{definition}

\newtheorem{claim}[thm]{Claim}
\newtheorem{rmk}[thm]{Remark}

\newcommand{\cred}[1]{\textcolor{black}{#1}}

\newcommand{\ccblue}[1]{\textcolor{blue}{#1}}
\newcommand{\ccred}[1]{\textcolor{red}{#1}}

\definecolor{mygray}{gray}{0.6}

\usepackage{upgreek}
\usepackage{bbm}

\newenvironment{myfont}[2][]{\csname#2\endcsname[#1]}{}

\usepackage{slashed}
\usepackage[makeroom]{cancel}
\usepackage[normalem]{ulem}
\usepackage{soul}
\newcommand{\stkout}[1]{\ifmmode\text{\sout{\ensuremath{#1}}}\else\sout{#1}\fi}

\usepackage[all,cmtip]{xy}
\usepackage{tikz-cd}
\usepackage{tikz}
\usetikzlibrary{matrix}

\newcommand{\bea}{\begin{eqnarray}}
\newcommand{\eea}{\end{eqnarray}}
\def\be{\begin{equation}}
\def\ee{\end{equation}}

\newcommand{\e}{\hspace{1pt}\mathrm{e}}
\newcommand{\imth}{\hspace{1pt}\mathrm{i}\hspace{1pt}}
\newcommand{\im}{\hspace{1pt}\mathrm{i}\hspace{1pt}}

\def\RP{{\mathbb{RP}}}
\def\CP{{\mathbb{CP}}}

\usepackage{color}
\usepackage[colorlinks,citecolor=blue]{hyperref}
\definecolor{red}{rgb}{1,0,0}
\definecolor{blue}{rgb}{0,0,1}
\definecolor{dblue}{rgb}{0,0,0.4}
\definecolor{green}{rgb}{0,1,0}
\definecolor{black}{rgb}{0,0,0}
\definecolor{white}{rgb}{1,1,1}

\definecolor{brn}{rgb}{.8,.4,.0}
\definecolor{redo}{rgb}{1,.5,.0}
\definecolor{ddgrn}{rgb}{0,0.4,0}
\definecolor{dgrn}{rgb}{0,0.55,0}
\definecolor{dbl}{rgb}{0,0,0.5}

\usepackage[bbgreekl]{mathbbol}
\usepackage{amscd}

\newcommand{\Z}{\mathbb{Z}}
\newcommand{\C}{\mathbb{C}}
\newcommand{\R}{\mathbb{R}}

\newcommand{\eqn}[1]{eqn.~(\ref{#1})}

\newcommand{\up}{\uparrow} 
\newcommand{\down}{\downarrow}

\newcommand{\bpm}{\begin{pmatrix}}
\newcommand{\epm}{\end{pmatrix}}
\newcommand{\bmm}{\begin{matrix}}
\newcommand{\emm}{\end{matrix}}

\newcommand{\al}{\alpha} 
\newcommand{\bt}{\beta}

\newcommand{\eps}{\epsilon}

\renewcommand{\th}{\theta} 
 
\newcommand{\si}{\sigma}

\def\Z{{\mathbb{Z}}}

\def\R{{\mathbb{R}}}
\def\C{{\mathbb{C}}}

\begin{document}
\begin{titlepage}
\begin{flushright}
\end{flushright}
\vskip 1.25in
\begin{center}



{\bf\LARGE{ 
Time Reversal, 
$SU(N)$ Yang-Mills  
and 
Cobordisms:\\[2.75mm] 
Interacting Topological Superconductors/Insulators \\[2.75mm] 
and Quantum Spin Liquids in 3+1D\\[3.75mm]
}}

\vskip0.5cm 
\Large{Meng Guo$^1$,  Pavel Putrov$^{2}$ and Juven Wang$^{2,3}$\\[2.75mm] 
} 
\vskip.5cm
{\small{\textit{$^1$Department of Mathematics, Harvard University, Cambridge, MA 02138, USA \\}}
}
 \vskip.2cm
 {\small{\textit{$^2$School of Natural Sciences, Institute for Advanced Study, Princeton, NJ 08540, USA}\\}}
 \vskip.2cm
 {\small{\textit{$^3${Center of Mathematical Sciences and Applications, Harvard University,  Cambridge, MA, USA} \\}}
}

\end{center}
\vskip.5cm
\baselineskip 16pt
\begin{abstract}


We introduce a web of strongly correlated interacting 3+1D topological superconductors/ insulators  
of 10 particular global symmetry groups of Cartan classes, 
realizable in electronic condensed matter systems, and 
their new $SU(N)$ generalizations. The symmetries include $SU(N)$, $SU(2)$, $U(1)$, fermion parity, time reversal and
relate to each other through symmetry embeddings.
We overview the lattice Hamiltonian formalism. 
We complete the list of field theories of bulk symmetry-protected topological invariants 
(SPT invariants/partition functions that exhibit boundary 't Hooft 
anomalies) via 
cobordism calculations, matching their full classification. 
We also present explicit 4-manifolds that detect these SPTs.
On the other hand, once we dynamically gauge part of their global symmetries, 
we arrive in various new phases of $SU(N)$ Yang-Mills (YM) gauge theories,
analogous to quantum spin liquids with emergent gauge fields. 
We discuss how coupling YM theories to time reversal-SPTs  affects the strongly coupled theories at low energy. 
For example, we point out 
a possibility of having two deconfined gapless time-reversal symmetric
$SU(2)$ YM theories at $\theta=\pi$ as two distinct conformal field theories, 
which although are
\cred{secretly indistinguishable by correlators of local operators on orientable spacetimes
 nor by gapped SPT states}, 
 can be distinguished on non-orientable spacetimes or potentially by correlators of extended operators.

\end{abstract}
\end{titlepage}



\tableofcontents   

\newpage

\section{Introduction and Summary}

\label{sec:intro}
Global symmetry plays a crucial role in constraining the fate of macroscopic states\footnote{
We would use \emph{phases} of matter or \emph{states} of matter interchangeably, 
and the quantum \emph{vacua} or the \emph{ground states} of system interchangeably.} 
of physical systems --- 
its constraint is applicable including but not limited to
quantum many body condensed matter (normally defined with an ultraviolet (UV) high-energy/short-distance lattice regularization cutoff), 
and quantum field theories (preferably studied at the infrared (IR) low-energy/long-distance universal behavior at fixed points), including 
gauge theories \cite{H8035}.

In condensed matter physics, one digs into how global symmetry acts
on the operators and the states in the local Hilbert space in the deep UV. For example, 
the electrons that can hop between atoms/orbitals, where the distances between atoms are essentially the UV scale lattice constant.
While quantum field theory (QFT) controls the global and large scale IR properties of systems, the condensed matter way of thinking supplements it by zooming in and examining  the ultraviolet (UV) lattice cut off --- in which we essentially see
the fairy dancing patterns of electrons within the effective ``Planck scale'' UV quantum effects. The
two principles could actually complement, like Yin and Yang, 
with each other.
In this work, we would like to implement global symmetries at both UV (deep on the lattice) and IR, 
and discuss their consequences on various systems.
%


First we consider the 10 particular global symmetries (see Table \ref{10-sym-groups} and \ref{table:sym-web}) that are mostly relevant to the fermionic electrons of condensed matter system
in 3+1 dimensional spacetime (3+1D), 
involving $SU(2)$, $U(1)$, $\Z_2^F$, \cred{or time reversal symmetries}.
If one limits these 10 global symmetries
to the quadratic Hamiltonian systems, they correspond to the 10 Cartan symmetry classes,
studied since Wigner-Dyson \cite{wigner1951,Dyson1962,AZ9742}. 
They are studied later in the context of so-called \emph{free fermion} 
topological insulators/superconductors \cite{RyuSPT,Kitaevperiod,Wen1111.6341} (See an overview 
\cite{2010RMP_HasanKane, 2011_RMP_Qi_Zhang}).\footnote{We should remind the readers
that the Cartan symmetry class combines many distinct symmetry groups in a single class. In the free fermion system,
 these distinct symmetry groups in the same Cartan symmetry class have the same topological classifications. 
However, in the interacting system, one should specify a particular symmetry group even within a Cartan symmetry class.
Distinct symmetry groups in the same Cartan symmetry class may have different classifications.
Given the same symmetry Cartan class, there could be symmetry $G_1$ and $G_2$ within this class.
 Some free fermion SPTs in $G_1$ may collapse into the same state (e.g. trivial insulator/vacuum) within the $G_1$-symmetry-preserving interactions.
However, the corresponding free fermion SPTs in $G_2$ may not collapse into the same state,
 because of the limitation of $G_2$-symmetry-preserving interactions.
In our work, 
the ``abused'' Cartan symmetry class notation for interacting SPTs  means a particular global symmetry that we will specify.
} 
The $SU(2)$ plays the role of the flavor symmetry or the spin-$\frac{1}{2}$'s $SU(2)$ rotational symmetry. 
The $U(1)$  plays the role of electromagnetic charge symmetry or the spin rotational symmetry.
The $\Z_2^F$ is the fermionic parity symmetry that flips the sign of fermionic operator $\psi \to -\psi$.
\cred{The $\Z_2^T$ or $\Z_4^T$ is the time reversal symmetry with the symmetry transformation $T$ acting on fermions respectively
as 
$T^2=+1$ or $T^2=(-1)^F$ (as the fermion parity of $\Z_2^F$).} 
However, including interactions can change the classifications and characterizations of these Symmetry-Protected Topological states (SPTs).
One of the most powerful tools can be used is finding their symmetry-protected topological invariants (SPT invariants).
By  \emph{topological invariants}, we mean the partition functions (or path integrals) in field theoretic form at IR
that capture the bulk SPTs (by coupling to \emph{background non-dynamical probed fields})
and also constrain the boundary anomalies. 
The field theoretic partition functions studied in \cite{QiHughesZhang, Wang:2014pma, Kapustin:2014dxa, Wen1410.8477, Metlitski:2015yqa, Witten:2015aba, Witten:2016cio} are good examples.  
Recently the 3+1D SPT classifications have been more-or-less completed by pioneer works 
(the bosonic cases in \cite{XieSPT4, XieSPT5, SenthilBF}, %
physical intuitive studies of 
interacting fermionic topological insulators/superconductors (TI/TSC) \cite{1306.3238WangPS, 1401.1142WS, 1406.3032Metlitski}, %
with the later corrections and refinements from cobordisms \cite{Kapustin1403.1467, Kapustin:2014dxa, 1604.06527FH} or generalized group cohomology \cite{Wen1410.8477, WangGu1703.10937}, see more References therein).
Some of their {topological invariants} are uncovered \cite{QiHughesZhang, Witten:2015aba, Witten:2016cio}, 
however other {topological invariants} (especially those involving the $SU(2)$-symmetries) are still not fully transparent.
One of the goals of our work is to  fill this gap, by providing the \emph{complete list of symmetry-protected topological invariants} 
that characterize the 10 particular global symmetries of Cartan symmetry classes.
These bulk SPT invariants also specify certain 't Hooft anomalies \cite{H8035} living on the boundaries of bulk spacetime.
We stress that ``the anomalies'' here for interacting TI/TSC are in some sense unfamiliar: 
they are non-perturbative global gauge anomalies that may mix between time-reversal and gauge anomalies. 
Since the precise definition of global symmetry is significant, along the way
we correct and refine the misused or confusing 
notations in condensed matter literature.

\cred{In addition, we also examine global symmetries and topological invariants that are pertinent to 
the $SU(N)$-generalizations of time-reversal invariant SPTs ($SU(N)$-generalizations of  TI/TSC), potentially applicable to 
non-Abelian Yang-Mills theories \cite{PhysRev.96.191-YM},}
quantum chromodynamics (QCD$_4$) or the cold atom systems with larger flavor/or spin rotational global symmetries. 
\cred{There we consider} $SU(2) \times SU(2)$ color-flavor symmetry,
$SU(3)$ symmetry, and $SU(4)$ symmetry with \cred{time-reversal}. 
Or more generally, for a larger flavor, we can characterize topological invariants of $SU(N)$ symmetry with time-reversal, when $N$ is an odd positive integer.
We achieve these goals by explicitly constructing field theoretic partition functions, and by computing the appropriate cobordism groups.
Ref.~\cite{1604.06527FH} provides especially the important guidance.

Later we can also dynamically (fully or partially) gauge the global symmetries of system, 
with the caution in mind that its Hilbert space is dramatically changed under the gauging process.
By gauging, we mean that microscopically at the deep UV, 
we insert the gauge variable on the link between the site variable associated to the global symmetry,
where the global symmetry acts on-site.
By implementing a machinery of gauging, 
we \emph{input} various quantum vacua/ground states protected by global symmetries, and \emph{output} dynamically gauge theories.

For example, gauging the $SU(N_c)$ subgroup out of the original SPTs protected by $SU(N_c) \times SU(N_f)$ global symmetry group with a 
time-reversal,\footnote{More precisely, we gauge $SU(N_c)$ out of the symmetry group
$\frac{SU(N_c) \times SU(N_f) \times \Z_4^T}{\Big(\frac{\Z_{\gcd(N_c,N_f)} \times \Z_{\gcd(N_c,2)} \times \Z_{\gcd(N_f,2)}   }{\Z_{\gcd(N_c,N_f, 2)}} \Big)}$.}
we obtain various new quantum vacua of time-reversal invariant $SU(N_c)$-Yang-Mills theories.
Our results resonate and shed new lights to the recent attempts to study time-reversal-invariant Yang-Mills or QCD$_4$, in the beautiful work of \cite{Gaiotto:2017yup, Gaiotto:2017tne}. 
For other examples, gauging the $U(1)$ subgroup of topological insulators/superconductors (SPTs with $U(1)$ and \cred{time-reversal} symmetries), we obtain the
time-reversal invariant dynamical Maxwell's $U(1)$ gauge theories.
These long-range entangled (gapless or gapped) quantum phases 
are also known as 
$SU(N)$-quantum spin liquids\footnote{For $SU(N)$-quantum spin liquids, we mean that the $SU(2)$-spin degrees of freedom on the lattice may need
to be generalized to $SU(N)$-spin degrees of freedom, in condensed matter or quantum many body systems.}
 \cite{0107071QSLWen} 
 and $U(1)$-quantum spin liquids \cite{0107071QSLWen, 1505.03520WS, 1710.00743ZWSenthil} in condensed matter.
One can also study gauging only the discrete subgroup ($\Z_N$) out of the full symmetry group (time reversal, $U(1)$, $SU(N)$) \cite{Ye:2013upa}.

To be more explicit and schematic, 
we will first study the SPT states, in Sections \ref{10SPT}, \ref{sec:SUN-Sym}, \ref{sec:sym-reduction} and \ref{sec:sym-reduction-SU(N)}, as partition functions of
probed background fields (integrating out the matter fields, in order 
to obtain the dependence on classical background fields $a$ or other characteristic classes $w_i$, etc.). 
Their partition functions $Z_{{\text{SPT}}}[a, w_i, \dots]$ are classical, depending on the \emph{non-dynamical} fields:
\bea
Z_{{\text{SPT}}}[a, w_i, \dots]=\exp\big(-S_{\text{SPT}}[a, w_i, \dots]\big).
\eea
After dynamically gauging the global symmetry (sub-)group (e.g. gauging the $SU(N)$ part but leaving the time reversal symmetry intact) of SPT states, 
we will study the dynamical gauge systems obtained from Yang-Mills theory (with an action $S_{\text{YM}}$) coupling to SPTs,  in Sec.~\ref{sec:SUN-gauge}, 
formally defined as a partition function $Z\equiv {Z_{\text{YM+SPT}}}$,
schematically as 
\bea
Z=\int [{\cal D} {a}] \exp\big(-S_{\text{YM}}[a, \dots]- S_{{\text{SPT}}}[a, \dots] \big)=
\int [{\cal D} {a}] \exp\big(-\int\limits_{M_4} (\frac{1}{4g^2}\text{Tr}\,F_a\wedge \star F_a) \big)  Z_{{\text{SPT}}}[a,\dots],
\eea
where all the allowed gauge field configurations are integrated over in the path integral measures $\int [{\cal D} {a}]$.
See further detailed discussions in Sec.~\ref{sec:SUN-gauge}. 

Here let us pause for a moment to have a deeper overview of the possible states of quantum matter.
Two key concepts can be used, Short/Long-Range Order (SRO/LRO) and Short/Long Range Entanglement (SRE/LRE):
\begin{itemize}
\item
Short-Range Order (SRO): Gapped system, where two-point correlation
functions show exponential decays.

------ Examples of SRO include trivial gapped vacuum (pure Yang-Mills theory
at $\theta=0$), SPTs (Topological Insulators), SETs (Symmetry-Enriched Topologically ordered states) 
and intrinsic Topological Order (TO), etc.

\item Long-Range Order (LRO): Gapless system, where two-point correlation
functions show power-law decays [e.g. Conformal Field Theory (CFT)] or stay constant, even
at large distance, such as spontaneous symmetry breaking (SSB) of continuous group results in gapless Goldstone modes [e.g. superfluid].

------ Examples of LRO include superfluid, gapless quantum spin liquids, metals, CFT, etc.

\item Short-Range Entangled (SRE): Can be deformed to a trivial vacuum
[product state on-site] by local unitary transformations  (LUT) through finite-depth local
quantum circuits, where each set of LUT is implemented within a finite width.

------ Examples of SRE include trivial gapped vacuum (pure Yang-Mills theory at $\theta=0$), SPT and superfluid, etc.

\item Long-Range Entangled (LRE): Cannot be deformed to a trivial vacuum
[product state on-site] by local unitary transformations (LUT) through finite-depth local
quantum circuits, where each set of LUT is implemented within a finite width.

------ Examples of LRE include SETs, Topological Order, Quantum Hall states, gapless and gapped quantum spin liquids, CFTs, and metals with Fermi surfaces, etc.

\end{itemize}
The concepts of SRO/LRO and SRE/LRE are different, so in general there are roughly four scenarios for these characterizations.

Now let us discuss
 the possibilities of \emph{gauging} SPT states. 
Note that SPTs are gapped SRO and SRE states, but what are the outcome states of gauging?

\noindent
(1) {\bf SPTs $\otimes$ gauge theory}: Here SPTs is gapped and decoupled from the dynamical gauged sector.
The gauge theory can be gapless or gapped, such as CFTs, or TO (TQFT), etc. Since two sectors are decoupled 
(thus a tensor product $\otimes$ structure), these are less-interesting. 

\noindent
(2) {\bf SSB Landau-Ginzburg order $\otimes$ \dots (gapless or gapped, with Goldstone modes or SPTs, etc)}:
SSB stands for spontaneous symmetry breaking (partially or fully), which thus characterizes Landau-Ginzburg order.
But the continuous symmetry breaking results in gapless Goldstone modes. 
Another possibly is that the remained unbroken symmetry hosts gapped SPTs.
Thus the whole states can be either gapless or gapped.

\noindent
(3) {\bf SETs} \cite{1212.0835Ran, 1212.1827Hung} (gapped, not necessarily SPTs $\otimes$ TO):
SETs are SRO but LRE. They are symmetry-protected and topologically ordered, but in general are
symmetry-enriched \emph{richer} than a tensor product SPTs $\otimes$ TO.

\noindent
(4) {\bf  SP-gapless} or {\bf  SP-CFT} (Symmetry-Protected gapless states, or Symmetry-Protected CFT):
The CFT near the fixed point somehow is nontrivially protected by global symmetries, say a 
SP-CFT, but which is \emph{not} a normal SPTs $\otimes$ CFT. 
Breaking symmetries may  result in different states than a SP-CFT.
More exotically, SP-CFTs protected by time-reversal symmetries, 
we will discuss a scenario that two CFTs with exactly same global symmetries but
\emph{cannot} be distinguished on orientable manifolds, but only distinguished on non-orientable manifolds.

\noindent
(5) {\bf  SET-gapless} or {\bf SET-CFT} (Symmetry-Enriched Topological gapless states, or Symmetry-Enriched Topological CFT):
There could be a topologically-ordered CFT, say TO-CFT, that is \emph{not} TO $\otimes$ CFT. 
Different TO-CFTs may not be distinguished by local operator product expansions (OPE), but \emph{only}
by correlation functions of \emph{extended operators}, like lines and surfaces.
Furthermore, such states could  also be \emph{symmetry-enriched} 
(e.g. enriched by unitary symmetries or anti-unitary time-reversal symmetry), just like TO and TQFT can be \emph{symmetry-enriched}.

What are the justifications and the sharp distinctions of states that we outlined as (4) SP-gapless/SP-CFT and (5) SET-gapless/SET-CFT?
How do these (1)-(5) states emerge from our studies of $SU(N)$-SPTs or $SU(N)$ Yang-Mills?
We leave these questions resonating as our \emph{prelude} into the \emph{overture}. 
In the \emph{finale}, Conclusion (Sec.~\ref{sec:Conclusion}), we will come back to discuss the quantum phases of $SU(N)$ Yang-Mills with topological terms obtained 
from gauging SPTs in terms of the framework outlined above.

The convention for our notations is listed in Appendix~\ref{sec:plan-convention}.

Our article is organized as follows. 

In this work, we start with a {self-contained overview of global symmetries on the UV lattice and IR field theories}
in Sec.~\ref{sec:global}.
We  explore the symmetry-protected quantum vacua (SPTs) protected by $SU(N), SU(2), U(1)$, 
fermion parity and time reversal symmetries, focusing on their topological terms and their complete classifications 
(some of them are known, while others, primarily involving $SU(N)$ with time reversal, are new to the literature), in 
Sec.~\ref{10SPT} and Sec.~\ref{sec:SUN-Sym}.
Then we address how these global symmetry groups and how their topological terms
can embed/break into each other, in Sec.~\ref{sec:sym-reduction} and in
Sec.~\ref{sec:sym-reduction-SU(N)}.
By examining the consequences of dynamical gauging of $SU(N)$ symmetry of the above SPTs, 
we then switch gears to study 
time-reversal symmetric $SU(N)$ Yang-Mills 
in Sec.~\ref{sec:SUN-gauge}.
We conclude in Sec.~\ref{sec:Conclusion}.


\section{Overview of Global Symmetries of the UV Lattice and IR Field Theories}
\label{sec:global}

\subsection{Global Symmetries of the UV Lattice, Hamiltonian and Hilbert space}

\label{sec:global-UV}

We start by making precise the global symmetries and how they act on the UV lattice of fermionic systems with an intrinsic $\Z_2^F$ fermion parity. 
For example, we can consider electron systems that could be non-relativistic on the lattice at UV.
The electron is a fermion in the spin-statistics relation, with an electromagnetic charge under $U(1)$, 
and is an $SU(2)$-spin doublet (spin-$\frac{1}{2}$) under $SU(2)$-fundamental representation.
Our attempt is to first make connections to global symmetries discussed in the condensed matter literature. 
We would like to emphasize the \emph{total symmetry group} $G_{\text{Tot}}$ 
which obeys the short exact sequence
$1 \to \Z_2^F \to G_{\text{Tot}} \to G \to 1 $ with $\Z_2^F$ is the fermion parity.
The $G\equiv {G_{\text{Tot}}}/ \Z_2^F$ is closely related to the symmetry group in the condensed matter notation, although
the condensed matter notation does not quite precisely match with $G$. 
Our spirit on analyzing symmetries is coherent to a rigorous insightful setup in \cite{Wen1111.6341}, 
which also emphasizes the total symmetry group that contains $\Z_2^F$.
 Instead of focusing on free quadratic Hamiltonians as in \cite{Wen1111.6341}, 
we like to extend the analysis to the interacting systems.

\emph{Definition}: Given the Hermitian operator Hamiltonian $\hat H$ and a state-vector $|\Psi \rangle$ living in the Hilbert space,\footnote{
The state-vector $|\Psi \rangle$ satisfies the time-dependent Schr\"odinger equation
$\hat H |\Psi \rangle = \im \partial_t  |\Psi \rangle$, although in the present context we focus on the 
eigenstate $\hat H |\Psi \rangle = E  |\Psi \rangle$, especially the ground state with $E$ being the minimum energy eigenvalue.
Here we use the hat symbol \, $\hat {}$ \, on $\hat M_g$ to denote a matrix operator of $M_g$ in the quantum mechanical sense.}
we define the operation of the global symmetry group $G$ as a matrix representation operator $\hat M_g$ for any group element $g \in G$, such that
their 
algebra obeys
\bea
&&\hat M_g \hat H \hat M^{-1}_g=\hat H,\\
&& \hat M_g |\Psi \rangle = |\Psi \rangle.
\eea
By satisfying the above criteria, we say the Hamiltonian system $\hat H$ and the state $ |\Psi \rangle$ respect the global symmetry $G$.
If $g \in G$ is unitary, we have $\hat M_g \hat M^{\dagger}_g=1$ and $\hat M_g \im \hat M^{-1}_g=\im$ for an imaginary number $\im$.  
If $g \in G$ is anti-unitary, we have $\hat M_g \im \hat M^{-1}_g=-\im$.  

Below we overview and define some crucial symmetries later implemented in Sec.~\ref{10SPT}. 
Since we consider the interacting systems, $\hat H$ is composed of many-body interactions between local fermionic 
annihilation/creation operators
$\hat{c}_j$ and $\hat{c}^\dagger_j$ (of site $j$) that satisfy the anti-commutation relations $\{\hat{c}_j, \hat{c}_l^\dagger \}=\im \delta_{il}$.\footnote{
Here $\hat{c}^\dagger_j |0_j\rangle =|1_j\rangle$, 
$\hat{c}_j |1_j\rangle=|0_j\rangle$,  and
$\hat{c}^\dagger_j |1_j\rangle =\hat{c}_j |0_j\rangle=0$, 
where $|0\rangle $ and $ |1\rangle$ are empty and filled fermionic state.}
For example, 
$$
\hat H= 
h_{i,j}^{(1)} \hat{c}_i^\dagger \hat{c}_j 
+h_{i,j}^{(2)}  \hat{c}_i \hat{c}_j + 
h_{i,j,k,l}^{(3)} \hat{c}_i^\dagger \hat{c}_j^\dagger \hat{c}_k \hat{c}_l 
+ h_{i,j,k,l}^{(4)}\hat{c}_i \hat{c}_j \hat{c}_k \hat{c}_l +  
h_{i,j,k,l,m,n}^{(5)} \hat{c}_i^\dagger \hat{c}_j^\dagger \hat{c}_k^\dagger \hat{c}_l \hat{c}_m \hat{c}_n +\dots,
$$
where the $\dots$ terms contain other terms, overall we have $\hat H =\hat H^\dagger$ Hermitian.
The first two terms (e.g. $h^{(1)}, h^{(2)}$) are so-called free quadratic, 
the next terms (e.g. $h^{(3)}, h^{(4)}, h^{(5)}$) are fully-interacting.

In the lattice Hamiltonian formalism here, all global symmetries studied below are made to be local \emph{onsite} symmetries,
which means that the symmetry operator $\hat M_g= \bigotimes_j \hat M_{g,j}$ can be written as a tensor product of operators at each site $j$.

All fermionic systems must  obey fermionic parity $\Z_2^F$ symmetry, 
that flips the sign of fermionic operator $\hat{c}_j$ and $\hat{c}^\dagger_j$ for each site $j$: 
\bea
\hat{c}_j \to -\hat{c}_j, \;\;\; \hat{c}^\dagger_j \to -\hat{c}^\dagger_j.
\eea
Other global symmetries are optional. Since
the electron is an $SU(2)$-spin doublet, we write the corresponding operator as $\hat{c}_{\alpha, j}$ where $\alpha$ is the $SU(2)$ doublet index, 
spin up or down $\up, \down$ (say $1$ and $2$ respectively). 

In general,  time reversal symmetry implies that under Schr\"odinger equation
$\im \frac{\partial}{\partial t} \Psi (t) =\hat{H} \Psi (t)$, the $\Psi^*(-t)$ is also a solution, and $\hat{H} =\hat{H}^*$.
Time reversal operator $\hat T =\hat U_T K$ is an anti-unitary operator where 
$\hat U_T$ is unitary and $K$ is a complex conjugate operator.
For a spin-$\frac{1}{2}$ fermion, we require that $\hat T^2=-1$ on a fermion,
thus $(\hat U_T K) (\hat U_T K)= \hat U_T \hat U_T^*=-1$.
We define the time reversal symmetry $\hat T$ on the lattice as:
\bea
\hat T_{} \hat c_{\al,j} \hat T_{}^{-1}= \eps_{\al\bt} \hat c_{\bt,j}= \imth \sigma^y_{\al\bt} \hat c_{\bt,j},\ \ \
\hat T_{} \hat c^\dag_{\al,j} \hat T_{}^{-1}= \eps_{\al\bt} \hat c^\dagger_{\bt,j}
= \imth \sigma^y_{\al\bt} \hat c^\dag_{\bt,j},
\eea
where $\sigma^y$ follows the convention of Pauli matrices. The $\hat T$ acts on the spin-$\frac{1}{2}$ doublet
as $\imth \sigma^y K$ where $K$
is complex conjugate. It is easy to see that indeed
 \bea
 \hat T^2=(-1)^{\hat{N}}\equiv (-1)^F.
\eea
 Acting by time-reversal twice on any operator multiplies it by $+1/-1$ depending on the fermionic number operator
 $\hat{N}= \sum_j {\hat{N}_j}=\sum_j \hat c^\dag_{j} \hat c_{j}$, thus it is the fermionic parity $\Z_2^F$ operator $(-1)^F$. Therefore, 
$ \hat T^4=+1$, the true time reversal symmetry is actually $\Z_4^T$ such that $\Z_2^F$ is its normal subgroup,
$\Z_4^T \supset \Z_2^F$. What condensed matter community usually denotes $\Z_2^T$ in a fermionic system, actually refers to a partial symmetry
$\Z_4^T/ \Z_2^F= {\Z_2^T}'$ instead of the full time reversal $\Z_4^T$. 
\cred{In terms of a short exact sequence, we have $1 \to \Z_2^F  \to \Z_4^T \to {\Z_2^T}' \to 1$.}

 The operator of $SU(2)$-spin rotation symmetry operator around the unit vector $\hat n$-direction 
 by a $\theta$-angle acts on a site $j$ as follows:
\bea \label{eq:SU(2)-rot}
\e^{\imth {\theta} \hat n \cdot \hat S_j}
=\e^{\imth {\theta}  (n_x  \hat S_j^x+n_y  \hat S_j^y+ n_z  \hat S_j^z)}
=\e^{\imth \frac{\theta}{2} \hat n \cdot \hat c^\dag_j \hat \si \hat c_j}
=\e^{\imth \frac{\theta}{2}  ( n_x \hat c^\dag_{\al,j} \hat \si^x_{\al\bt} \hat c_{\bt,j}+
n_y \hat c^\dag_{\al,j} \hat \si^y_{\al\bt} \hat c_{\bt,j} 
+n_z \hat c^\dag_{\al,j} \hat \si^z_{\al\bt} \hat c_{\bt,j})}.
\eea

The $U(1)$-charge symmetry associated to the fermion number $\hat{N}_j$ acts on a site $j$ as:
\bea
\e^{\imth {\theta}  \hat N_j}
=\e^{\imth {\theta}  \hat c^\dag_j \hat \si^0 \hat c_j}
=\e^{\imth {\theta}  (\hat c^\dag_{\up,j} \hat c_{\up,j}+\hat c^\dag_{\down,j} \hat c_{\down,j})}.
\eea
We remark that $SU(2)$ (here the spin rotation) contains the $\Z_2$-center 
and $U(1)$-charge also contains a $\Z_2$-subgroup. Both are precisely the fermionic parity $\Z_2^F$ symmetry,
so $SU(2) \supset \Z_2^F$ and $U(1) \supset \Z_2^F$.

Also the unitary charge conjugation symmetry $\hat C_{}$ on the lattice acts as:
\bea \label{eq:C-charge-cj}
\hat C_{} \hat c_{\al,j} \hat C_{}^{-1}= \eps_{\al\bt} \hat c_{\bt,j}^\dag= \imth \sigma^y_{\al\bt} \hat c_{\bt,j}^\dag,\ \ \
\hat C_{} \hat c^\dag_{\al,j} \hat C_{}^{-1}= \eps_{\al\bt} \hat c_{\bt,j}
= \imth \sigma^y_{\al\bt} \hat c_{\bt,j}.
\eea
Similarly $\hat C^2=(-1)^{\hat{N}}\equiv (-1)^F$ and $ \hat T \hat C = \hat C \hat T$, 
so
$ \hat C^4=+1$. The true charge conjugation symmetry is indeed $\Z_4^C \supset \Z_2^F$ such that $\Z_2^F$ is its normal subgroup.

Other than the above symmetry realization on the fermionic electron systems, 
we will also comment about \emph{other more general ways} to realize and ``regularize'' symmetry on the UV lattice later,
especially in Sec.~\ref{sec:SUN-Sym} for $SU(N)$ symmetry.

\subsubsection{Symmetries of charge/spin, fermion-pairing, and spin-orbital coupling orders}

\label{sec:global-UV-H}

Below we analyze and summarize the global symmetry systematically on several 
possible Hamiltonian terms, partially inspired by \cite{Wen1111.6341},
including charge/spin-orderings, fermion-pairings, and spin-orbital interactions:
\begin{enumerate}
\item $\hat N_i=\hat c_i^\dag    \hat \si^0 \hat c_i$ as a charge order,  invariant under time-reversal and spin rotations:
\bea
&&\hat T_{} \hat N_i \hat T_{}^{-1}
=\e^{\im {\theta} \hat n \cdot \hat S} \hat N_i \e^{-\im {\theta} \hat n \cdot \hat S} 
=+\hat N_i.
\eea

\item $(\hat c_{\up i}\hat c_{\down j}- \hat c_{\down i}\hat c_{\up j})$  $=\hat c_{i} \im \hat \sigma^y\hat c_{j}$, a spin-singlet real superconductor pairing
(expectation value $\langle \hat S \rangle=0$, $\langle \hat S^z \rangle=0$), here $i$ and $j$ can be on different sites, invariant under time-reversal and spin rotations:
\bea
&&\hat T_{} (\hat c_{\up i}\hat c_{\down j}- \hat c_{\down i}\hat c_{\up j}) \hat T_{}^{-1}
=\e^{\im {\theta} \hat n \cdot \hat S} (\hat c_{\up i}\hat c_{\down j}- \hat c_{\down i}\hat c_{\up j}) \e^{-\im {\theta} \hat n \cdot \hat S} 
=+(\hat c_{\up i}\hat c_{\down j}- \hat c_{\down i}\hat c_{\up j}).
\eea

\item $(\hat c_{\up i}\hat c_{\down j}+ \hat c_{\down i}\hat c_{\up j})$
 $=\hat c_{i}  \sigma^x\hat c_{j}$, a spin-triplet real superconductor pairing ($\langle \hat S \rangle=1$, $\langle \hat S^z \rangle=0$):
\bea
\hat T_{} (\hat c_{\up i}\hat c_{\down j}+ \hat c_{\down i}\hat c_{\up j}) \hat T_{}^{-1}
&=&-(\hat c_{\up i}\hat c_{\down j}+ \hat c_{\down i}\hat c_{\up j}), \nonumber\\
\e^{\im \pi \hat S^x} (\hat c_{\up i}\hat c_{\down j}+ \hat c_{\down i}\hat c_{\up j}) \e^{-\im \pi \hat
S^x}=\e^{\im \pi \hat
S^y} (\hat c_{\up i}\hat c_{\down j}+ \hat c_{\down i}\hat c_{\up j}) \e^{-\im \pi \hat
S^y} 
&=&-(\hat c_{\up i}\hat c_{\down j}+ \hat c_{\down i}\hat c_{\up j}), \nonumber\\
\e^{\im \theta \hat S^z} 
(\hat c_{\up i}\hat c_{\down j}+ \hat c_{\down i}\hat c_{\up j}) 
\e^{-\im \theta  \hat S^z} 
&=& {+}(\hat c_{\up i}\hat c_{\down j}+ \hat c_{\down i}\hat c_{\up j}).
\eea

\item $\hat S^a_i =( \frac{1}{2} \hat c_i^\dag  \hat \si^a \hat c_i)$, as a spin order, $a=x,y,z$.
\bea
&&
\hat T_{} \hat S^a_i \hat T_{}^{-1}= -  \hat S^a_i.
\eea
We have a \emph{coplanar} (the spin along the $z$-$x$ plane) or a \emph{collinear} (the spin along the $y$) order. 
We discuss these details in the following three cases respect to spin-flip symmetries. 

\item $(\hat c_i^\dag  n_a \cdot  \hat \si^a \hat c_i)$, say a \emph{coplanar} spin order along $a=x$ or $z$, and a \emph{collinear} along $a=y$ 
(the reason for this convention is that later we consider the spin-flip under $\e^{\im \pi \hat S^y}$): 
\bea
\hat T_{} (\hat c_i^\dag  n_a \cdot  \hat \si^a \hat c_i) \hat T_{}^{-1}
&=&-(\hat c_i^\dag  n_a \cdot  \hat \si^a \hat c_i), \nonumber\\
\e^{\im \pi \hat S^b} (\hat c_i^\dag  n_a \cdot  \hat \si^a \hat c_i) \e^{-\im \pi \hat
S^b}
&=&(-1)^{(1-\delta_{ab})}(\hat c_i^\dag  n_a \cdot  \hat \si^a \hat c_i). 
\eea
In general, if the spin order along $n_a$ is parallel to spin rotational operator $\e^{\im \pi \hat S^a}$, we have a symmetry invariant;
otherwise, for other spin rotational operators $\e^{\im \pi \hat S^b}$ with $a \neq b$, we get an odd $(-1)$ factor under operations.

\item $(c_i^\dag  n_x \cdot  \hat \si^x \hat c_j+c_i^\dag  n_y \cdot  \hat \si^y \hat c_j + c_i^\dag  n_z \cdot  \hat \si^z \hat c_j) \equiv 
\sum_a c_i^\dag  n_a \cdot  \hat \si^a \hat c_j$, 
a spin-orbital coupling term has symmetry operations as:
\bea
\hat T_{} ( c_i^\dag  n_a \cdot  \hat \si^a \hat c_j) \hat T_{}^{-1}
&=&-( c_i^\dag  n_a \cdot  \hat \si^a \hat c_j), \nonumber\\
\e^{\im \pi \hat S^b} ( c_i^\dag  n_a \cdot  \hat \si^a \hat c_j) \e^{-\im \pi \hat
S^b}
&=&(-1)^{(1-\delta_{ab})}( c_i^\dag  n_a \cdot  \hat \si^a \hat c_j). 
\eea
 
\end{enumerate}

We will come back to use these orders, couplings and interaction terms to suggest
what topological superconductors/insulators respect the same global symmetries of these orders in Sec.~\ref{10SPT}.

\subsection{Global Symmetries of the IR Field Theories, Path Integral and Cobordisms}
\label{sec:global-IR}

Now we switch gears to explore global symmetries suitable for description in terms of the IR field theories, path integrals and cobordism formalism. Note that IR here means continuum field theory that describes lattice theory at long distances. It should not be confused with ``deeper'' IR theory discussed later in the paper which arises after gauging (part of) the global symmetry at strong gauge coupling. From the point of view of the latter theory, the continuum weakly gauged field theories below can be considered as UV theories. The partition function of an SPT then can be viewed as the exponential of a classical weakly gauged action, which depends on topology\footnote{In general, it can also depend on the choice of smooth structure.} and some background gauge fields (which mathematically has meaning of some additional structure on the spacetime manifold, such as choice of a principle bundle for the symmetry group). 

Below we provide general comments before proceeding to special cases in Sec.~\ref{10SPT}. Suppose we have a global symmetry with Lie group\footnote{We use tilde to distinguish it from $G_\text{Tot}$ or $G$ which contain time-reversal symmetry.} $\tilde{G}$ and so that our system contains fermions in a faithful representation\footnote{For example, it is $N$-dimensional fundamental representation if $\tilde{G}=SU(N)$.} of $\tilde{G}$. There are two qualitatively distinct cases: $\tilde{G}$ contains $\Z_2$ center subgroup or not. In the first case we also assume that all bosonic fields, if present, transform under $\tilde{G}/\Z_2$ so that we can identify the $\Z_2$ center subgroup with the fermionic parity $\Z_2^F$. The fermion fields in $d$ dimensions therefore form a faithful representation of $\text{Pin}^{\pm}(d)\times_{\Z_2^F} \tilde{G}$ or $\text{Spin}(d)\times_{\Z_2^F} \tilde{G}$ group depending on whether time-reversal symmetry is present or not. To be specific $A \times_{\Z_2^F} B\equiv (A\times B)/\Z_2^F$ means the quotient w.r.t. the diagonal center $\Z_2^F$ subgroup, which physically means identifying fermionic parity contained both in $\text{Pin}^{\pm}(d)$ (or $\text{Spin}(d)$) and $\tilde{G}$ groups. 

We will be mostly interested in the case when time-reversal symmetry is present. In order to define such fermionic fields on a $d$-manifold $M_d$, one should equip it with $\text{Pin}^{\pm}(d)\times_{\Z_2^F} \tilde{G}$ principal bundle, or, equivalently, with $\text{Pin}^{\pm}\times_{\Z_2^F} \tilde{G}$ structure. The principle bundle is such that the $O(d)$ principle bundle obtained by projection $\text{Pin}^{\pm}(d)\times_{\Z_2^F} \tilde{G}\rightarrow \text{Pin}^{\pm}(d)/{\Z_2^F}\equiv O(d)$ is the structure bundle of the tangent bundle $TM_d$. Note that it is possible that manifold does not have a $\text{Pin}^\pm$ (or $\text{Spin}$) structure, but does have $\text{Pin}^{\pm}\times_{\Z_2^F} \tilde{G}$ (or $\text{Spin}\times_{\Z_2^F} \tilde{G}$) structure. In other words, the fermion fields locally are sections of $S\otimes V$ bundle where $S$ is the spinor bundle and $V$ is the representation bundle of $\tilde{G}$. It may happen that although both $S$ and $V$ globally do not exist but their product does. Physically this means that one can only consider fermions coupled to gauge fields on such a manifold, but not ``uncharged'' ones. The SPTs protected by fermionic parity, time-reversal and $\tilde{G}$ symmetry are then classified by the (Poincare dual to the torsion part of) bordism group of manifolds with corresponding structure: $\Omega_d^{\text{Pin}^{\pm}\times_{\Z_2^F} \tilde{G}}$.

In the case when $\tilde{G}$ does not contain a $\Z_2$ center (for example when $\tilde{G}=SU(3)$), the fermionic parity is contained only in $\text{Pin}^\pm(d)$ (or $\text{Spin}(d)$) group. The corresponding structure is then just $\text{Pin}^\pm \times \tilde{G}$ ($\text{Spin} \times \tilde{G}$). Equivalently, this means that we should consider manifolds equipped simultaneously with $\text{Pin}^\pm$ ($\text{Spin}$) structure and a principal $\tilde{G}$-bundle. The corresponding bordism group then can be also written as $\Omega_d^{\text{Pin}^{\pm}\times\tilde{G}}=\Omega_d^{\text{Pin}^\pm}(B\tilde{G})$. In the general discussion below we will assume that $\tilde{G}$ does contain $\Z_2^F$. The other scenario is much simpler. In what follows we will also assume by default that the spacetime dimension is $d=4$.

Possible topological terms that can appear in SPT action are invariants of bordism of manifolds equipped with additional structure\cite{Kapustin1403.1467, Kapustin:2014dxa, 1604.06527FH}. As we confirm by direct calculation of the bordism groups, all topological terms that can apear are either purely ``bosonic'', and can be realized as Stiefel-Whitney numbers of bundles over the manifold or can be obtained by integrating out massive fermions coupled to background gauge fields, as was described in \cite{Witten:2015aba,Witten:2016cio,Metlitski:2015yqa} for example.

To make a connection to the cobordism formalism, we will be considering Euclidean spacetime. The systematic way to do a Wick rotation of 3+1D fermions from Minkowski to Euclidean spacetime can be found in \cite{van1996euclidean} for example.

On the flat Euclidean spacetime, the path integral of a massive Dirac fermion reads
\begin{equation}
\int [{\cal D} {\psi}] [{\cal D}\bar{\psi}] \exp\left\{-S_E \right\}=\int [{\cal D} {\psi}] [{\cal D}\bar{\psi}] \exp\left\{
-\int dtd^3x\,\bar{\psi}(\gamma^\mu\partial_\mu+m)\psi
\right\},
\end{equation} 
with $dtd^3x \equiv d^4 x_E$.  More generally, for a Dirac spinor coupled to a background (probe) gauge field $a$ 
in the curved Euclidean spacetime of a metric $g_{\mu \nu}$, the path integral becomes
\begin{equation}
\int [{\cal D} {\psi}] [{\cal D}\bar{\psi}] \exp\left\{-S_E \right\}=\int [{\cal D} {\psi}] [{\cal D}\bar{\psi}] \exp\left\{
-\int d^4x_E\, \sqrt{ \det g}
\big( \bar\psi (\slashed{D}_a+ m)\psi \big)
\right\},
\end{equation} 
where \textit{locally} $\slashed{D}_a \equiv e^\mu_{\mu'} \gamma^{\mu'} (\partial_\mu+\im \omega_\mu-\im a_\mu)$, $e^\mu_{\mu'}$ is a vielbein, 
$\omega_\mu$ and $a_\mu$ are spin and gauge connections respectively. More explicitly, in components, one has $\omega_\mu=\im \omega_\mu^{\lambda\nu}[\gamma^\lambda,\gamma^\nu]/8$ and $a_\mu=a^r_\mu T_r$ where $T_r$ are generators of the Lie algebra $\text{Lie}(\tilde{G})$. In order to \textit{globally} define the Dirac operator $\slashed{D}_a$ on $M_d$ one needs to specify the transition functions that relate fields\footnote{Note that in the path integral formalism one considers $\bar{\psi}\equiv \psi^\dagger\gamma^5$ and $\psi$ as independent Grassmann fields.} $\psi,\bar{\psi}$ and $a$ on different patches. The transition function should be also such that they leave the local expression for the action above invariant.

The transition functions between the charts that preserve orientation are standard. The definition of transition functions that change orientation is more subtle. Note that the orientation-reversing transition function can be always realized as a composition of an orientation-preserving transition function and transition function that relates local coordinates as ${x^0}'\equiv t' =-t \equiv -x^0$, ${x^i}'=x^i$. The corresponding transition function for fields of the theory is then realized by time-reversal transformation. Note that when a theory is considered on a flat spacetime, in principle the notion of the time-reversal symmetry is ambiguous. A time-reversal symmetry is \textit{any} symmetry $T'$ that flips the sign of the ``time'' coordinate $t\rightarrow -t$, acts on 1-form gauge fields as $a_0(x,t)\rightarrow -a_0(x,-t), a_{i}(x,t)\rightarrow a_{i}(x,-t)$ and satisfies either $(T')^2=1$ or $(T')^2=(-1)^F$ condition (as explained in \cite{Kapustin:2014dxa}, these conditions are swapped when one does Wick rotation from Minkowski to Euclidean spacetime). If one uses $T'$ such that $(T')^2=1$ to define the theory on a unoriented manifold, this corresponds to a choice of $\text{Pin}^+\times_{\Z_2} \tilde{G}$ structure. The condition $(T')^2=(-1)^F$ corresponds to a choice of $\text{Pin}^-\times_{\Z_2} \tilde{G}$ structure.  The time reversal symmetry in principle can also be combined with an order 2 automorphism of the symmetry algebra $\text{Lie}(\tilde{G})$ (such as, for example, charge conjugation $a_\mu\rightarrow -a_\mu$ for $\tilde{G}=U(1)$ case, see details below). In general, if $T'$ is a time-reversal symmetry and $\Gamma$ is the generator of any $\Z_2$ global symmetry, then $T'\Gamma$ is also a time-reversal symmetry. However, when put on an unoriented manifold, one has to make a particular choice of $T'$ which will be used in orientation-reversing transition functions for the fields of the theory. Different choices correspond to different ways of defining the theory on unoriented space-times. Morally speaking, putting the theory on an unoriented manifold corresponds to turning on a background gauge field for $T'$ symmetry. The ambiguities/obstructions of doing that on a quantum level correspond to anomalies of $T'$ and, as usual, can be cured by coupling the original theory to a bulk SPTs in one higher dimension.

Given the global definition of the Dirac operator acting on fermionic fields (that is sections of the twisted spinor bundle associated to $\text{Pin}^+\times_{\Z_2} \tilde{G}$ structure), one can use the general expression for the Gaussian Grassmann path integral:
\begin{equation}
\int [{\cal D} {\psi}] [{\cal D}\bar{\psi}] \exp\left\{
-\int d^4x_E\, \sqrt{ \det g}
\big( \bar\psi (\slashed{D}_a+ m)\psi \big)
\right\}=\det(\slashed{D}_a+ m).
\end{equation} 
A non-trivial SPT arises in the IR when the mass $m$ of the fermion is negative. 
In order to capture only the topological degrees of freedom of this SPT, one needs to normalize the partition function above by the partition function of the Dirac fermion with positive mass (that gives a trivial gapped theory). The value of the partition function of the SPT in the presence of a background gauge field is then given by the following ratio of the determinants (in the limit of large mass), cf. :
\begin{equation}
 Z_\text{SPT}[a]= \lim_{|m|\rightarrow \infty}\frac{\det(\slashed{D}_a-|m|)}{\det(\slashed{D}_a+|m|)}\equiv
\lim_{|m|\rightarrow \infty}\prod_\lambda \frac{i\lambda -|m|}{i\lambda +|m|}
\end{equation} 
where $\lambda$ runs over eigenvalues of $-i\slashed{D}_a$ (note that $\slashed{D}_a$ is anti-Hermitian, so $\lambda$ are real). 

\subsubsection{$C, T$, and $CT$ for Dirac fermion coupled to $U(1)$ gauge field}
\label{sec:U1-Dirac-op}

Let us first review a simple case when the weakly gauged symmetry group is $\tilde{G}=U(1)$. This case has been explored in detail in a remarkable work \cite{Metlitski:2015yqa}. The action of a massive Dirac fermion coupled to $U(1)$ gauge field $a$ in Euclidean flat spacetime reads:
\begin{equation}
 S_E=\int dtd^3x\,\bar{\psi}(\gamma^\mu(\partial_\mu-ia_\mu)+m)\psi.
\end{equation} 
It has the following discrete symmetries:
\begin{equation}
 \begin{array}{c}
  C:\\
(C^2=+1)
 \end{array}
\begin{array}{c}
 \psi(x) \to C_D \bar{\psi}^T(x),\qquad
 \bar{\psi}(x) \to - \psi^T(x) C_D^{\dagger}, \qquad a_\mu (x)\rightarrow -a_\mu(x).     
    \end{array}
\end{equation} 
\begin{equation}
 \begin{array}{c}
  CT:\\
((CT)^2=+1)
 \end{array}
 \begin{array}{c}
 \psi(\vec x,t)\rightarrow \im (\gamma^0\gamma^5){\psi}(\vec x,-t),\qquad
\bar{\psi}(\vec x,t)\rightarrow -\im \bar{\psi}(\vec x,-t)(\gamma^5\gamma^0),\\ 
a_0(\vec x,t)\rightarrow -a_0(\vec x,-t),
\qquad a_{i}^a(\vec x,t)\rightarrow a_{i}^a(\vec x,-t).    \quad\quad\quad ({\text{AIII TSC}})     
    \end{array}
\end{equation} 
\begin{equation}
 \begin{array}{c}
  T:\\
(T^2=+1)
 \end{array} \qquad 
\begin{array}{c}
\psi(\vec{x}, t) \to \im \gamma^0 \gamma^5 C_D \bar{\psi}^T(\vec{x}, -t),\qquad
\bar{\psi}(\vec{x}, t) \to \im \psi^T(\vec{x}, -t) C_D^{\dagger} \gamma^5 \gamma^0,\qquad\\ 
a_0(\vec x,t)\rightarrow + a_0(\vec x,-t),
\qquad a_{i}(\vec x,t)\rightarrow - a_{i}(\vec x,-t).     \quad\quad\quad ({\text{AII TI}})  
  \end{array}
\end{equation} 
Where $C_D$ is the unitary ``charge conjugation'' matrix acting on Dirac spinors that satisfies $C_D(\gamma^\mu)^TC_D^\dagger=-\gamma^\mu,\;C_D(\gamma_5)^TC_D^\dagger=\gamma^5, C_DC_D^*=-1$.  As was already mentioned, in Euclidean path integral formalism, one treats $\psi$ and $\bar{\psi}$ as independent fields, so that $C$, $CT$, and $T$ above are unitary symmetries. Moreover, they commute with each other.\footnote{
Since $\psi$ and $\bar{\psi}$ are independent the transformations above can be written as the following linear transforms acting on the pair of vectors $\psi$ and $\bar{\psi}^T$:
\begin{equation}
 C:\,\left(\begin{array}{c}
      \psi \\
\bar{\psi}^T
     \end{array}\right)
\;
\longrightarrow
\;
\left(\begin{array}{cc}
      0 & C_D \\
 -C^*_D & 0
     \end{array}\right)
\left(\begin{array}{c}
      \psi \\
\bar{\psi}^T
     \end{array}\right),
\end{equation} 
\begin{equation}
 CT:\,\left(\begin{array}{c}
      \psi \\
\bar{\psi}^T
     \end{array}\right)
\;
\longrightarrow
\;
\left(\begin{array}{cc}
      i\gamma^0\gamma^5 & 0 \\
 0 & -i{\gamma^0}^T{\gamma^5}^T
     \end{array}\right)
\left(\begin{array}{c}
      \psi \\
\bar{\psi}^T
     \end{array}\right)
\end{equation} 
The fact that $C$ and $CT$ are unitary and commute can then be easily seen using the properties of $C_D$ and gamma-matrices.} Both $T$ and $CT$ involve $t\rightarrow -t$ so they both can be interpreted as time-reversal transforms. However, one can see that $CT$ is actually a more natural choice, since the one-form field $a_\mu$ transforms naturally (so that $a_\mu dx^\mu$ is invariant), while for $T$ this transformation is complemented by $a_\mu\rightarrow -a_\mu$ which can be interpreted as $\Z_2$ automorphism of $U(1)$ (i.e. charge conjugation). In particular, $CT$ commutes with the $U(1)$ action so that if one uses $T'=CT$ to define a theory on a unoriented 4-manifold, this will correspond to $\text{Pin}^+\times_{\Z_2^F} U(1)\equiv \text{Pin}^c$ structure. Alternatively, one can use $T'=T$ which corresponds to $\text{Pin}^+\ltimes U(1)\equiv \text{Pin}^{\tilde{c}+}$ structure. The semi-direct product structure reflects the fact that the time-reversal transformation involves charge conjugation.

\subsubsection{$C, T$, and $CT$ for Dirac fermion coupled to $SU(2)$ gauge field}
\label{sec:SU2-Dirac}

Consider now the case of weakly gauged symmetry group $\tilde{G}=SU(2)$. Let $\psi$ be a Dirac fermion transforming as a doublet of $SU(2)$ (i.e. in fundamental representation). Explicitly, the action of one such massive Dirac fermion on flat Euclidean spacetime reads
\begin{equation}
 S_E=\int dtd^3x\,\bar\psi((\gamma^\mu\otimes \mathbf{1})\partial_\mu-\im a_\mu^a(\gamma^\mu\otimes \sigma_a))\psi+m\bar\psi\psi 
\label{massive-dirac-action}
\end{equation} 
where $\sigma_a$ a Pauli matrices, the generators of $su(2)$ Lie algebra, $\psi$ belongs to the tensor product of 4 dimensional complex (i.e. Dirac) representation of $so(4)$ algebra and 2 dimensional representation of $SU(2)$ symmetry, and $a_\mu^a$ are components of the background  $SU(2)$ gauge field.

The action (\ref{massive-dirac-action}) has the following discrete symmetries (which would not be broken by Yang-Mills action for gauge field $a$):

\begin{equation}
 \begin{array}{c}
  C:\\
(C^2=(-1)^F)
 \end{array}
\begin{array}{c}
 \psi(x)\rightarrow (C_D\otimes C_{SU(2)})\bar{\psi}^T(x),\qquad
\bar{\psi}(x)\rightarrow -\psi^T(x)(C_D^\dagger\otimes C_{SU(2)}^\dagger),\\ a_\mu^a(x)\rightarrow a_\mu^a(x).     
    \end{array}
\label{CSU2}
\end{equation} 
\begin{equation}
 \begin{array}{c}
  CT:\\
((CT)^2=1)
 \end{array}
 \begin{array}{c}
 \psi(\vec x,t)\rightarrow \im (\gamma^0\gamma^5\otimes \mathbf{1}){\psi}(\vec x,-t),\qquad
\bar{\psi}(\vec x,t)\rightarrow -\im \bar{\psi}(\vec x,-t)(\gamma^5\gamma^0\otimes \mathbf{1}),\\ 
a_0^a(\vec x,t)\rightarrow -a_0^a( \vec x,-t),
\qquad \qquad a_{i}^a(\vec x,t)\rightarrow a_{i}^a(\vec x,-t).    \quad\quad \quad \quad ({\text{CI TSC}})     
    \end{array}
\end{equation} 
\begin{equation}
 \begin{array}{c}
  T:\\
(T^2=(-1)^F)
 \end{array} \qquad \qquad
\begin{array}{c}
 \psi(\vec x,t)\rightarrow  -\im (\gamma^0\gamma^5\otimes \mathbf{1})(C_D\otimes C_{SU(2)})\bar{\psi}^T(\vec x,-t),\qquad\\
\bar{\psi}(\vec x,t)\rightarrow -\im \psi^T(\vec x,-t)(C_D^\dagger\otimes C_{SU(2)}^\dagger)(\gamma^5\gamma^0\otimes \mathbf{1}),\qquad\qquad\\ 
a_0^a(\vec x,t)\rightarrow -a_0^a(\vec x,-t),
\qquad a_{i}^a(\vec x,t)\rightarrow a_{i}^a(\vec x,-t).     \quad\quad ({\text{CII TI}})  
  \end{array}
\end{equation} 
Here $C_D$ is the unitary ``charge conjugation'' matrix acting on Dirac spinors, same as in the $U(1)$ case, and $C_{SU(2)}=e^{\im\frac{\pi }{2}\sigma_2}\in SU(2)$ is the matrix that provides an isomorphism between fundamental representation of $SU(2)$ and its conjugate. In particular, it satisfies $C_{SU(2)}\sigma_a C_{SU(2)}^{-1}=-\sigma_a^T$. Similarly to the $U(1)$ case, one can see that $C$, $CT$ and $T$ are unitary symmetries and commute with each other, if one treats $\psi$ and $\bar{\psi}$ as independent Grassmann fields in the path integral.

Again, in Euclidean spacetime, $CT$, as defined above, provides the most natural choice of time-reversal symmetry acting on the Dirac fermions, because it does not involve complex conjugation. Therefore, if the Dirac fermions form a complex representation of some symmetry group, the $CT$-transformed fermions will be in the same representation, not the conjugated one. 

The choices of $T'=CT$ or $T'=T$ to define the theory on unoriented manifolds correspond to $\text{Pin}^+\times_{\Z_2^F} SU(2)$ or $\text{Pin}^-\times_{\Z_2^F} SU(2)$ structures respectively because $(CT)^2=1$ and $T^2=(-1)^F$.

Note that in principle one could define $C$ differently (this would change $T$ correspondingly if we keep $CT$ to be the same), by keeping the same action on fermions as in the $U(1)$ case. In order to make the action invariant, this then would require a non-trivial transformation of $a_\mu\rightarrow a'_\mu$ (such that $(a')^a_\mu \sigma_a^T=-a^a_\mu \sigma_a$) corresponding to $\Z_2$ (inner) automorphism of $SU(2)$.

\subsubsection{$SU(N)$ and more general groups}
\label{sec:2-SU(N)}

Since $CT$ transformation does not involve any charge conjugation matrix, it can be generalized as the symmetry of the Dirac action for massive fermion in arbitrary representation of any gauge group $\tilde{G}$:
\begin{equation}
 \begin{array}{c}
  CT:\\
((CT)^2=1)
 \end{array}
 \begin{array}{c}
 \psi(\vec x,t)\rightarrow \im (\gamma^0\gamma^5\otimes \mathbf{1}){\psi}(\vec x,-t),\qquad
\bar{\psi}(\vec x,t)\rightarrow -\im \bar{\psi}(\vec x,-t)(\gamma^5\gamma^0\otimes \mathbf{1}),\\ 
a_0^a(\vec x,t)\rightarrow -a_0^a( \vec x,-t),
\qquad \qquad a_{i}^a(\vec x,t)\rightarrow a_{i}^a(\vec x,-t).  
    \end{array}
\end{equation} 
Therefore one can always use $T'=CT$ to define Dirac operator on a manifold with $\text{Pin}^+\times_{\Z_2^F}\tilde{G}$ (Or $\text{Pin}^+\times\tilde{G}$, if $\tilde{G}$ does not contain $\Z_2^F$ center, as for example when $\tilde{G}=SU(N)$ and $N$ is odd).

Note that one cannot generalize (\ref{CSU2}) to $\tilde{G}=SU(N)$ an arbitrary group, because there is no direct analog of $C_{SU(2)}$. The case of $SU(2)$ is special because the fundamental representation and its conjugate are equivalent, which is not the case for $SU(N),\;N>2$. When $\tilde{G}=SU(N)$, in principle, one can define $C$ transform so that it acts on the fermionic fields the same way as in the $U(1)$ case, but then one needs to transform the gauge field according to a non-trivial order 2 outer automorphism of $SU(N)$ (which corresponds to $\Z_2$ symmetry of the $A_{N-1}$ Dynkin diagram and swaps fundamental representation with anti-fundamental). The corresponding $T\equiv C\cdot CT$ then can be used to define $\text{Pin}^+\ltimes_{\Z_2} SU(N)$ structure (for even $N$).

\section{{Complete List of Symmetry-Protected Topological Invariants for 10 global symmetries of 
Cartan classes} 
} \label{10SPT}

Follow Sec.~\ref{sec:global}, here we would like to discuss the 10 Cartan classes with \emph{particular global symmetries} in the interacting cases.
We will first present the UV symmetry at the lattice scale as in Sec.~\ref{sec:global-UV}, then the IR symmetry in 
Sec.~\ref{sec:global-IR} that are suitable for continuum field theory/path integrals. 
We then present the Symmetry-Protected Topological invariants (SPT invariants) as the bulk field partition functions
that match the full interacting classifications of these SPTs (interacting topological superconductors/insulators) obtained from cobordism classifications.
We summarize the global symmetries and their notations at UV lattice/IR Minkowski/IR Euclidean signatures, and their
SPT invariants in Table \ref{10-sym-groups}.

\begin{table*}[h!] 
 \centering
 \makebox[\textwidth][r]{
 \begin{tabular}{ |c| l | c|  c| }
 \hline
Cartan  & 
\begin{minipage}[c]{2.6in}
Condensed Matter Symmetry\\
``misused'' notation (not $G_{\text{Tot}}$) \\
(for fermionic electrons)
\end{minipage} & 
\begin{minipage}[c]{1.67in} Full Symmetry $G_{\text{Tot}}$: \\ 
($G_{\text{Tot}}/ \Z_2^F = G $)\\
\ccblue{Minkowski} vs. \ccred{Euclidean}\\[-2mm]
\end{minipage} & 
\begin{minipage}[c]{1.678in}
Cobordism $\Omega^4$;  \\
Classification (3+1D)
\end{minipage} \\
\hline
\hline
CII & 
\begin{minipage}[c]{2.8in}
fTI (${T^2=C^2=(-1)^{F}}$, $C \in \Z^{{C}}_{2}$): \\
$U(1)^{\text{c}} \rtimes [\Z_2^T \times \Z^{{C}}_{2}]$\\  
$[U(1)^{\text{c}} \rtimes \Z_2^C   ] \times \Z^{{CT}}_{2}$  
\end{minipage} 
& 
\begin{minipage}[c]{2.in}
\ccblue{{$\frac{[U(1)^{\text{c}}  \rtimes \Z_4^C ] }{\Z_2}  \times \Z^{{CT}}_{2} $}}, \ccblue{${SU(2)^{\text{c}} \times \Z_2^{T}}$}
\;\;\;\;\;\;\;\; vs.\,\\
\ccred{$\frac{[U(1)^{\text{c}}  \rtimes \Z_4^C  ]  \times \Z^{{CT}}_{4} }{(\Z_2)^2}$}~or~\ccred{$\frac{SU(2) \times \Z_4^{T}}{\Z_2}$}
\end{minipage} 
& 
\begin{minipage}[c]{1.678in}
${\mathrm{Pin}^{-}\times_{\Z_2^F} SU(2)}$;\\[2mm]
$(\nu_{\text{CII}}, \alpha,\beta) \in \Z_2 \times \Z_2 \times \Z_2$
\end{minipage} 
\\
\hline
C
& fTSC: $SU(2)  \supset \Z_2^F$ & $SU(2)$& ${\mathrm{Spin} \times_{\Z_2^F} SU(2)}$; No class\\
\hline
CI & 
\begin{minipage}[c]{2.85in}
fTSC ($T^2=C^2=(-1)^{F}$, $C \in \Z^\text{spin}_{2,y}$): \\
$SU(2)^{\text{spin}} \times \Z_2^T$, 
\\
$[U(1)^\text{spin}_{z} \rtimes \Z^\text{spin}_{2,y}] \times \Z_2^T$,\\ 
$U(1)^\text{spin}_{z} \rtimes [\Z^\text{spin}_{2,y} \times \Z_2^{CT}]$ 
 \end{minipage}
& \begin{minipage}[c]{1.8in}  
\ccblue{$\frac{SU(2)^{\text{spin}} \times \Z_4^T}{\Z_2}$}\,vs.\,\ccred{${SU(2) \times \Z_2^{T}}$}, \\ 
\ccblue{$\frac{{ [U(1)^\text{spin}_{z} \rtimes \Z^\text{spin}_{4,y}] \times \Z_4^{T}}}{(\Z_2)^2}$}~vs.\\
\ccred{$\frac{{ [U(1)^\text{spin}_{z} \rtimes \Z^\text{spin}_{4,y}] \times \Z_2^{T}}}{\Z_2}$},\\
\ccblue{$\frac{U(1)^\text{spin}_{z} \rtimes [\Z^\text{spin}_{4,y} \times \Z_2^{CT}]}{\Z_2}$}~vs.\\
\ccred{$\frac{U(1)^\text{spin}_{z} \rtimes [\Z^\text{spin}_{4,y} \times \Z_4^{CT}]}{(\Z_2)^2}$}
\end{minipage}
& 
\begin{minipage}[c]{1.678in}
${\mathrm{Pin}^{+}\times_{\Z_2^F} SU(2)}$;\\[.0mm] 
$(\nu_{\text{CI}}, \alpha) \in \Z_4 \times \Z_2 $
\end{minipage}\\
\hline
AI &
\begin{minipage}[l]{2.2in} 
fTSC~(${T^2=+1}$):\\
$U(1)^{\text{c}}   \rtimes \Z^{{T}}_{2}$ 
\end{minipage}
&  
\ccblue{$U(1)   \rtimes \Z^{{T}}_{2}$}
vs.
\ccred{$\frac{U(1) \rtimes \Z_4^T}{\Z_2}$}
& 
\begin{minipage}[c]{1.678in}
${\mathrm{Pin}^{-}\ltimes_{\Z_2^F} U(1)}$;\\[1mm]
$\alpha \in \Z_2$
\end{minipage} \\
\hline
BDI & fTSC ($T^2=+1$): {$\Z_2^T \times \Z_2^F$ 
} &
\ccblue{$\Z_2^T \times \Z_2^F$} vs.
\ccred{$\Z_4^T $}
& 
\begin{minipage}[c]{1.7in}
${\mathrm{Pin}^{-}}$; No class
\end{minipage} 
\\
 \hline
D & No symmetry except only $\Z_2^F$& 
{$\Z_2^F$}
& 
 \begin{minipage}[c]{1.7in}
${\mathrm{Spin}^{} }$; No class
\end{minipage} 
\\
 \hline
DIII &
fTSC ($T^2=(-1)^{F}$) 
&
\ccblue{$\Z_4^T $} vs. \ccred{$\Z_2^T \times \Z_2^F$}  
& 
\begin{minipage}[c]{1.678in}
${\mathrm{Pin}^{+} }$;\\
$\nu_{\text{DIII}} \in \Z_{16}  $\\[-3mm]
\end{minipage} \\
 \hline
AII &
\begin{minipage}[l]{2.6in} 
fTI (${T^2=(-1)^{F}}$):\\
$U(1)^{\text{c}}   \rtimes \Z^{{T}}_{2}$  
\end{minipage}
&
\ccblue{$\frac{U(1)   \rtimes \Z^{{T}}_{4}}{\Z_2}$} vs.
\ccred{$U(1)   \rtimes \Z^{{T}}_{2} $}
& 
\begin{minipage}[c]{1.678in}
${\mathrm{Pin}^{+}\ltimes_{\Z_2^F} U(1) }$;\\[1mm]
$(\nu_{\text{AII}}, \alpha,\beta) \in \Z_2 \times \Z_2 \times \Z_2$
\end{minipage} 
\\
 \hline
\hline
\hline
A & $U(1)  \supset \Z_2^F$ & $U(1)$ & 
 \begin{minipage}[c]{1.7in}
${\mathrm{Spin}^{c} }$; No class
\end{minipage} 
\\
\hline
AIII & 
 \begin{minipage}[c]{2.85in}
fTSC ($T^2=(-1)^{F}$): \\
$U(1)^{\text{spin}}_{z} \times \Z^{{T}}_{2}$ 
\end{minipage} 
&
{$\frac{U(1)^{\text{spin}}_{z} \times \Z^{{T}}_{4}}{\Z_2}$} vs.
{$U(1)^{} \times \Z^{{T}}_{2}$}
& 
\begin{minipage}[c]{1.678in}
${\mathrm{Pin}^{c} }={\mathrm{Pin}^{\pm}\times_{\Z_2^F} U(1)}$;\\
$(\nu_{\text{AIII}}, \alpha) \in \Z_8 \times \Z_2 $\\[-3mm]
\end{minipage} 
\\
 \hline
\hline
 \end{tabular}
 }\hspace*{-32mm}
\caption{{
We list down symmetry groups related to 10 Cartan symmetry classes
 that contain $U(1)$, time reversal $T$, and/or
charge/spin conjugation $C$ symmetries.
The first column shows the Cartan symmetry class notation.
The second column shows (less-precise or misused) symmetry notation in condensed matter.
fTI/fTSC means fermionic Topological Insulator/Superconductor.
The third column shows
$G_{\text{Tot}}$, the total symmetry group containing a normal subgroup fermion parity $\Z_2^F$,
which follows that $1 \to \Z_2^F \to G_{\text{Tot}} \to G \to 1 $.
The $G\equiv G_{\text{Tot}}/ \Z_2^F$ is roughly speaking the global symmetry group without $\Z_2^F$.
The $G$ is closely related to the symmetry group in the condensed matter notation, although
the condensed matter notation does not quite precisely match with $G$.
The $U(1)^{\text{c}}$ means the electromagnetic $U(1)^\text{charge}$ symmetry.
The $SU(2)^{\text{c}}$ means the approximate charge symmetry, but there is no obvious
$SU(2)$-charge symmetry from the electronic condensed matter. 
The $U(1)^{\text{spin}}$ means the spin or orbital like $U(1)$ symmetry. 
}
The $\rtimes$ and $\ltimes$ are semi-direct products.
%
The last column shows groups for cobordism calculation, their cobordism classifications,
and their indices for SPT invariants.
}
 \label{10-sym-groups}
\end{table*}

{In the sub-section titles below, we list down the full symmetry ${G_{\text{Tot}}}$ 
on the UV lattice and the IR Euclidean global symmetry that can be used to compute the cobordism groups.}

\subsection{CI class: {$\frac{SU(2) \times \Z_4^T}{\Z_2}$}
and ${\mathrm{Pin}^{+}\times_{\Z_2^F} SU(2)}$}

\label{sec:CI-sym}

The Cartan CI class corresponds to the following 
several global symmetries that can be realized in the fermionic electronic condensed matter system.
We would like to enumerate them one by one, beginning with the largest global symmetry containing
$SU(2)$-spin/flavor and time reversal symmetries.

\begin{enumerate}

\item UV lattice symmetry {${G_{\text{Tot}}}=\frac{SU(2) \times \Z_4^T}{\Z_2}$},
IR Euclidean ${\mathrm{Pin}^{+}\times_{\Z_2^F} SU(2)}$: Topological Superconductor. 

The first example, on the lattice, we have the full spin $SU(2)$ rotation under operator 
$\e^{\imth {\theta} \hat n \cdot \hat S_j}$ in eqn.~(\ref{eq:SU(2)-rot}), and time reversal.
Since $\hat T \e^{\imth {\theta} \hat n \cdot \hat S_j} \hat T^{-1}=\e^{+\imth {\theta} \hat n \cdot \hat S_j}$,
and  $\hat T^2=(-1)^F$ thus $\hat T^4=+1$,
the full symmetry group is actually:
{$\frac{SU(2)^\text{spin} \times \Z_4^T}{\Z_2}$}.\footnote{
In the electronic system, $SU(2)$ is a spin-rotational symmetry denoted $SU(2)^{\text{spin}}$.}
The ${\Z_2}$ in the denominator is exactly ${\Z_2^F}$. We can convert this Hamiltonian operator
symmetry to the symmetry of IR Euclidean field theory on the spacetime:
By flipping $T^2=(-1)^F$ in Minkowski to $T^2=+1$ in Euclidean, we get the full symmetry ${\mathrm{Pin}^{+}\times_{\Z_2^F} SU(2)}$
for the Cobordism theory.

\item UV lattice symmetry {${G_{\text{Tot}}}=\frac{{ [U(1)^\text{spin}_{z} \rtimes \Z^\text{spin}_{4,y}] \times \Z_4^{T}}}{(\Z_2)^2}$},
IR Euclidean ${\mathrm{Pin}^{+}\times_{\Z_2^F} [\frac{U(1) \rtimes \Z_{4}}{\Z_2} }]$: Topological Superconductor. 

The second example, on the electronic lattice, we can consider the spin-$U(1)$ rotation along the $z$-axis, the 
spin-flip realized as half-rotation along $y$-axis, and also the time reversal,
under three symmetry operators respectively $\e^{\im {\theta} \hat S_j^z}
=\e^{\imth \frac{\theta}{2}  (\hat c^\dag_{\al,j} \hat \si^z_{\al\bt} \hat c_{\bt,j})}$, 
$\e^{\imth {\pi}   \hat S_j^y}$, and $\hat T$.
The spin rotation $\theta$ has $4 \pi$-periodicity, and we can modify the periodicity to $2 \pi$
by considering $U(1)$-symmetry $\e^{\imth {\theta} (2\hat S_j^z)}$ as $U(1)^\text{spin}_{z} $ where $(2\hat S_j^z)$ is quantized as an integer.
The spin-flip symmetry obeys $(\e^{\imth {\pi}   \hat S_j^y})^4=1$ an generates a finite group $\Z^\text{spin}_{4,y}$.
Since $\hat T \e^{\imth {\theta} \hat n \cdot \hat S_j} \hat T^{-1}=\e^{+\imth {\theta} \hat n \cdot \hat S_j}$, 
the only two non-commutative symmetry generators follow the rule 
$(\e^{\imth {\pi}   \hat S_j^y}) \e^{\imth {\theta} (2\hat S_j^z)} (\e^{\imth {\pi}   \hat S_j^y})^{-1} = \e^{-\imth {\theta} (2\hat S_j^z)}$,
which defines a \emph{semi-direct product} $\rtimes$ structure in $U(1)^\text{spin}_{z} \rtimes \Z^\text{spin}_{4,y}$.
We derive the overall total symmetry group as
{$\frac{{ [U(1)^\text{spin}_{z} \rtimes \Z^\text{spin}_{4,y}] \times \Z_4^{T}}}{(\Z_2)^2}$}.
The ${(\Z_2)^2}$ mod-out factors are again the $({\Z_2^F})^2$ redundantly appearing in the three symmetry generators.

By flipping $T^2=(-1)^F$ in Minkowski to $T^2=+1$ in Euclidean, we get the full symmetry 
${\mathrm{Pin}^{+}\times_{\Z_2^F} [\frac{U(1) \rtimes \Z_{4}}{\Z_2} }]$
for the cobordism theory.

\end{enumerate}

Potentially we can realize CI class topological superconductor with $SU(2)$-spin rotational and real-pairing symmetry, and additional symmetry-preserving interaction terms (See Sec.~\ref{sec:global-UV-H} and \cite{Wen1111.6341}).

There are 8 different symmetry-protected vacua, forming a group structure  $(\nu_{\text{CI}}, \alpha)$ $\in$  {$\Omega^4_{\mathrm{Pin}^{+}\times_{\Z_2^F} SU(2)}$$=\Z_4 \times \Z_2$ for a complete classification. This has been firstly computed in \cite{1604.06527FH}. 
Our Appendix \ref{sec:App} provides further details and calculations. 
We explore their field theories, topological terms and physics in the next subsection.

\subsubsection{SPT vacua and topological terms}
\label{sec:SU2-terms}

Consider the case of $\nu$ massive Dirac fermions transforming under an orientation reversal map by $T'=CT$ matrix described in Section \ref{sec:SU2-Dirac}. Since $(CT)^2=1$ this requires a choice of $\mathrm{Pin}^{+}\times_{\Z_2} SU(2)$ structure on the manifold. One can consider the forgetful map $\text{Pin}^{+}\times_{\Z_2} SU(2)\rightarrow SU(2)/\Z_2\cong SO(3)$. So that any $\text{Pin}^{+}\times_{\Z_2} SU(2)$ structure on a 4-manifold $M_4$ gives an $SO(3)$ bundle $V_{SO(3)}$ on $M_4$. The $SO(3)$ bundle can be lifted to an $SU(2)$ bundle if $w_2(V_{SO(3)})=0$. This should be possible if the manifold admits $\text{Pin}^+$ structure. The obstruction to the existence of $\text{Pin}^+$ structure is also $w_2$ of the tangent bundle $TM_4$. If $w_2(TM_4)\neq 0$, then to define  a $\text{Pin}^{+}\times_{\Z_2} SU(2)$ structure one can choose an $SO(3)$ bundle with $w_2(V_{SO(3)})=w_2(TM_4)$ (which is always possible), and lift it to $\text{Pin}^{+}\times_{\Z_2} SU(2)$.

 Consider partition function of such fermions with negative mass $m$ (normalized by partition function of fermions with positive mass) in presence of background $SU(2)$ connection $a$ (cf. \cite{Metlitski:2015yqa}):

\begin{equation} \label{eq:SU(2)-CI-top-term}
 Z^{\nu}_{SU(2)}[a] = \left(\frac{\det(\slashed{D}_a-|m|)}{\det(\slashed{D}_a+|m|)}\right)^\nu
\stackrel{|m|\rightarrow \infty}{\longrightarrow} \exp(2\pi \im \nu \eta_{SU(2)}[a])
\end{equation} 
where {$\slashed{D}_a \equiv e^\mu_{\mu'} \gamma^{\mu'} (\partial_\mu+\im \omega_\mu-\im a_\mu)$} is the Dirac operator, its global definition was discussed in section \ref{sec:global-IR}. Its $\eta$-invariant is defined, as usual, by the following formula 
\begin{equation}
 \eta_{SU(2)}=\frac{1}{2}(N_0+\lim_{s\rightarrow \cred{0}+}\sum_{\lambda\neq 0} \text{sign}\,\lambda |\lambda|^{-s})
\label{eta-su2}
\end{equation} 
where $\lambda$ are eigenvalues of $\slashed{D}_a$ and $N_0$ are the number of its zero modes. Since $\exp(2\pi \im \nu \eta_{SU(2)})$ is cobordism invariant, the calculation of the bordism group tell us that $\eta_{SU(2)}\in \frac{1}{4}\mathbb{Z}$ and non-trivial fSPT classes generated by such massive Dirac fermions are effectively labelled by $\nu\in \Z_4$. In the case of $\text{Pin}^+$ manifold the fact $\eta_{SU(2)}\in \frac{1}{4}\mathbb{Z}$ follows from the the general analysis in \cite{Witten:2016cio}
\footnote{In this case the fermions are sections of $V_{SU(2)}\otimes S$ where both $V_{SU(2)}$ and $S$ (the spinor bundle associated to the $\text{Pin}^+$ structure) are globally defined. Then from the fact that $w(V_{SU(2)})=1+w_4(V_{SU(2)})$ it follows that $w(V_{SU(2)}\oplus V_{SU(2)})=1$ and  $V_{SU(2)}\oplus V_{SU(2)}$ is stably trivial (considered as a real bundle). Therefore $\exp(4\pi \im \eta_{SU(2)})=\exp(8\pi \im \eta_{\text{Pin}^+})$ where $\eta_{\text{Pin}^+}$ is the usual (untwisted) Dirac operator defined for a given $\text{Pin}^+$ structure. The statement $\eta_{SU(2)}\in \frac{1}{4}\mathbb{Z}$ follows from $\eta_{\text{Pin}^+} \in \frac{1}{8}\Z$ \cite{gilkey1985eta}. (see also \cite{Kapustin:2014dxa,Witten:2015aba}). 
}.

This can be compared to the cases with just $U(1)$ or no global symmetry (apart from time-reversal and fermionic), where the fSPTs generated by massive fermions in the bulk are classified by $\Z_8$ and $\Z_{16}$ respectively.  When $SU(2)$ is broken to a maximal torus $U(1)$ (that is $a$ can be made globally diagonal), the Dirac doublets split into pairs of $\pm 1$ charged Dirac fermions. Each of the single fermions contributes $\exp(2\pi \im \eta_{\text{Pin}^{c}})$ where $\eta_{\text{Pin}^{c}}\in \frac{1}{8}\Z$ is the $\eta$-invariant of a defined Dirac operator for a given Pin$^c$ structure  \cite{gilkey1985eta}. That is
\footnote{
To be more precise, we actually have
\begin{equation}
 e^{2\pi \im \eta_{SU(2)}}=e^{2\pi \im (\eta_{\text{Pin}^{c}}(\mathfrak{s})+\eta_{\text{Pin}^{c}}(\mathfrak{s}'))}
\end{equation} 
where $\mathfrak{s}'$ is the ``opposite'' Pin$^c$ structure, that is the one with $\det(\mathfrak s')=\det(\mathfrak s)^{-1}$, or where $\det$ is the injective map $\det: \text{Pin}^c\rightarrow \text{Pic}(M_4)\cong H_2(M_4,\Z)$ that corresponds to taking square of the $U(1)$ part in $\text{Pin}^{c}\cong \text{Pin}^+\times_{\Z_2} U(1)$. Suppose 
$e^{2\pi \im (\eta_{\text{Pin}^{c}}(\mathfrak{s})-\eta_{\text{Pin}^{c}}(\mathfrak{s}'))}$ is non-trivial. But then it should be a Pin$^{c}$ bordism invariant, and therefore of the form:
\begin{equation}
 e^{2\pi \im (\eta_{\text{Pin}^{c}}(\mathfrak{s})-\eta_{\text{Pin}^{c}}(\mathfrak{s}'))}=
e^{2\pi \im \mu \eta_{\text{Pin}^{c}}(\mathfrak{s})}(-1)^{\lambda w_2^2}
\end{equation}  
for some universal {$(\mu, \lambda)\in \Z_8\times \Z_2$}. By considering $\mathbb{RP}^4$, which has Pin$^+$ structure, and therefore one can choose $\det(\mathfrak s)=1$, we see that $\lambda=0$. Further, on an orientable manifold one can use an appropriate index theorem that tells us that 
\begin{equation}
e^{2\pi \im \eta_{\text{Pin}^{c}}(\mathfrak{s})} = (-1)^{(\sigma(M_4)-c_1^2(\det(\mathfrak{s})))/8} 
\end{equation} 
to show that $\mu=0$ as well. This implies that 
\begin{equation}
e^{2\pi \im  \eta_{\text{Pin}^{c}}(\mathfrak{s}')} =e^{2\pi \im \eta_{\text{Pin}^{c}}(\mathfrak{s})}.
\end{equation} 
\label{foot:Pinc}
}
$\eta_{SU(2)}=2\eta_{\text{Pin}^{c}}\mod 1$. Therefore such symmetry breaking corresponds to the embedding map $\Z_4\rightarrow  \Z_8$ ($\nu \mapsto 2\nu$), where $\Z_8$ is a factor of $\Omega_{\mathrm{Pin}^{c}}^4\cong\Z_8\times \Z_2$ bordism group classifying fermionic SPTs with $U(1)$ and time-reversal symmetry of AIII class. 

If the whole $SU(2)$ is broken (i.e. one can globally set $a$ to zero), one can further split each fermion into a pair of Majorana fermions. A massive Majorana fermion contributes $\exp(\pi i \eta_{\text{Pin}^+})$ to the partition function. It is the generator of $\Z_{16}\cong\Omega_{\mathrm{Pin}^{+}}^4$ bordism group classifying fermionic SPTs with just time-reversal symmetry (DIII class). Thus complete symmetry breaking corresponds to the embedding $\Z_4\rightarrow  \Z_{16}$ ($\nu \mapsto 4\nu$). 

Similarly to the case with $U(1)$ or no symmetry, we have $2\eta_{SU(2)}=w_1^4(TM_4)/2 \mod 1$ (cf. $4\eta_{\mathrm{Pin}^{+}}=w_1^4(TM_4)/2 \mod 1$ \cite{Kapustin:2014dxa}) $\nu=2$ fSPT is equivalent to a bosonic SPT. The full classification of fermionic SPTs is given by $\Omega_{\mathrm{Pin}^{+}\times_{\Z_2} SU(2)}^4\equiv \text{Hom}(\Omega^{\mathrm{Pin}^{+}\times_{\Z_2} SU(2)}_4,U(1))\cong \Z_4\times \Z_2$ where the $\Z_2$ factor corresponds to purely bosonic SPTs generated by the SPT with action $w_2^2(TM_4)$. After inclusion of such bosonic phases the fSPT partition function in general read as:
\begin{equation}
 Z^{\nu,\alpha}[a]=\exp(2\pi \im \nu \eta_{SU(2)}[a])\,(-1)^{\alpha\,w_2^2(TM_4)},\footnote{
 Integration of characteristic classes of top degree over the base manifold (that is paired with the fundamental class in homology) is implicitly assumed throughout the paper.
 Thus $w_2^2(TM_4)$ here means $\int_{M_4} w_2^2(TM_4)$ integrating over the spacetime manifold $M_4$.}
 \qquad (\nu,\alpha) \in \Z_4\times \Z_2.
\end{equation} 
Note that the multiplication by 2 maps $\Z_4\rightarrow  \Z_8\rightarrow  \Z_{16}$ in the paragraph above are given by pullbacks of homomorphisms between bordism groups naturally defined by embeddings $\Z_2^F={\pm 1}\subset U(1) \subset SU(2)$:
\begin{equation}
\begin{array}{ccccc}
 \Omega^{\mathrm{Pin}^{+}}_4 & \rightarrow & \Omega^{\mathrm{Pin}^c}_4 & \rightarrow & \Omega^{\mathrm{Pin}^{+}\times_{\Z_2} SU(2)}_4 \\
 \Z_{16} & \rightarrow & \Z_8\times \Z_2 & \rightarrow & \Z_4\times \Z_2 
\end{array}.
\end{equation} 
On the other hand, the embedding of purely bosonic SPTs $\Z_2\times \Z_2\hookrightarrow \Z_4\times \Z_2$ ($(\nu',\alpha)\mapsto (2\nu',\alpha)$) is given by the pullback of the forgetful map
\begin{equation}
\begin{array}{ccc}
 \Omega^{\mathrm{Pin}^{+}\times_{\Z_2} SU(2)}_4 & \rightarrow & \Omega^{O}_4 \\
 \Z_4\times \Z_2 & \rightarrow & \Z_2 \times \Z_2 
\end{array}.
\end{equation} 

Suppose $M_4$ is oriented and spin $w_1(TM_4)=w_2(TM_4)=0$. Then one can split $\mathrm{Pin}^{+}\times_{\Z_2} SU(2)$ structure into the product of spin-structure and an $SU(2)$ bundle $V_{SU(2)}$ (the lift of $SO(3)$ bundle $V_{SO(3)}$, possible due to $w_2(V_{SO(3)})=0$ ) with connection $a$. On oriented manifolds, the second term in (\ref{eta-su2}) vanishes and $e^{2\pi \im \eta}=(-1)^{N_0}\equiv (-1)^{N_++N_-}=(-1)^{N_+-N_-}$ where $N_\pm$ are the numbers of zero modes of $\slashed{D}_a$ with given chirality. From the index theorem we have:
\begin{equation}
 N_+-N_-=p_1(TM_4)/12-c_2(V_{SU(2)})
\end{equation} 
Therefore, on oriented manifolds we have:
\begin{equation}
 Z^{\nu,\alpha}=(-1)^{\nu\,c_2(V_{SU(2)})},\qquad (\nu,\alpha) \in \Z_4\times \Z_2.
\end{equation} 
 where we have also used the fact that $p_1(TM_4)/3=\sigma(M_4)$ is a multiple of 16 on smooth spin 4-manifolds). Therefore odd $\nu$ phases have $\theta=\pi$ topological terms for $SU(2)$ background gauge field. 

{More precisely, on oriented spin manifolds or on a flat spacetime, odd $\nu$ phases reduce to
 \begin{equation} \label{eq:SU(2)-theta-pi-probe}
 \exp\big(-S[a] \big)=
\exp\big( \int\limits_{M_4}\frac{\rm i \theta}{8 \pi^2}  \text{Tr}\,F_a\wedge F_a \big),
 \end{equation}  
 at $\theta=\pi$, 
 with $a$ is \emph{not} presented in the path integral measure (i.e. no $\int [{\cal D} {a}]$) thus only a non-dynamical background gauge field. 
 The $F_a$ is the field strength of $a$ under the generic (flat or curved) spacetime.
 In condensed matter, the non-dynamical background gauge field means a probe field, and this field is able to probe/couple to SPTs 
 thus to characterize the SPTs.
 Thus this \emph{topological term} specifies the \emph{SPT vacua}.
 In field theory language, it can also be understood as a \emph{bulk anomaly} term, or a an anomaly-cancellation \emph{counter term} for some anomalous 2+1D QFT.
 }

For an oriented non-spin manifold using the appropriate generalization of the index theorem we get:
\begin{equation}
 Z^{\nu,\alpha}=(-1)^{\nu(\sigma(M_4)+p_1(V_{SO(3)}))/4+\alpha\,w_2^2(TM_4)}.
\end{equation} 
which is well defined due to 
\begin{equation}
 p_1(V_{SO(3)})= \mathcal{P}_2(w_2(V_{SO(3)})) = \mathcal{P}_2(w_2(V_{TM_4})) = p_1(TM_4)=-\sigma(M_4) \mod 4 
\end{equation} 
where $\mathcal{P}_2$ is Pontryagin square.

In Table \ref{table2}, we compute the partition functions on various 4-manifolds
that distinguish all SPT states 
    of Cartan CI class with this particular global symmetry.

\begin{table*}[h!]
\centering
    \begin{tabular}{|c | c|}
    \hline
$(M_4,\;V_{SO(3)})$ & $Z^{\nu,\alpha}$
 \\ \hline  \hline  
$(S^4, H)$ & $(-1)^\nu$
\\ \hline    
$(\mathbb{CP}^2, L_\C+1)$ & $(-1)^\alpha$
\\ \hline    
$(\mathbb{RP}^4, 3)$ & $e^{\pi \im \nu/2}$
\\ \hline    
    \end{tabular}
    \caption{Some examples of non-trivial values of the partition function on 4-manifolds 
    $S^4, \mathbb{CP}^2$, and $\mathbb{RP}^4$ 
    that distinguish all SPT states 
    of Cartan CI class with a particular global symmetry {$\frac{SU(2) \times \Z_4^T}{\Z_2}$} for Hamiltonian, or based on
    Cobordism $\Omega_{\mathrm{Pin}^{+}\times_{\Z_2} SU(2)}^4$.}
              \label{table2}
    \end{table*}

\subsection{CII class: {${SU(2) \times \Z_2^{T}}$}
and ${\mathrm{Pin}^{-}\times_{\Z_2^F} SU(2)}$ }

\label{sec:CII-sym}

The CII class corresponds to a different kind of $SU(2)$ and time reversal symmetries.

\begin{enumerate}

\item UV lattice symmetry {${G_{\text{Tot}}}={SU(2) \times \Z_2^{T}}$},
IR Euclidean ${\mathrm{Pin}^{-}\times_{\Z_2^F} SU(2)}$. 

Ideally we hope to construct the larger CII class symmetry requiring the total group {${SU(2) \times \Z_2^{T}}$}.
Here $SU(2)$ contains $\Z_2^F$ in the center, and $T^2=+1$. Thus this symmetry cannot be realized, at least not easily without modifying the definition of $T$-symmetry, in the lattice of fermionic electrons alone.

By flipping $T^2=+1$ in Minkowski to $T^2=(-1)^F$ in Euclidean, we get the full symmetry 
${\mathrm{Pin}^{-}\times_{\Z_2^F} SU(2)}$
for the Cobordism theory.

\item UV lattice symmetry {${G_{\text{Tot}}}=\frac{U(1)^{\text{c}} \rtimes [\Z_4^T \times \Z^{{C}}_{4}]}{(\Z_2)^2}$}
or {$\frac{[U(1)^{\text{c}}  \rtimes \Z_4^C ] }{\Z_2}  \times \Z^{{CT}}_{2} $},
IR Euclidean ${\mathrm{Pin}^{-}\times_{\Z_2^F} [\frac{U(1)^{\text{c}}  \rtimes \Z_4^C}{\Z_2} ] }$: Topological Insulator. 

We can consider a different (smaller) symmetry group  realization of CII class
that is exactly the fermionic electron symmetry.
The $U(1)^{\text{c}}$-fermion number charge symmetry acting as
$\e^{\imth {\theta}  \hat N_j}$, the charge conjugation $\hat{C}$ in eqn.~(\ref{eq:C-charge-cj}) generating $\Z_4^C$, and time reversal $\Z_4^T$.
We can define a semi-direct product relation $\rtimes$ based on $\hat T \e^{\imth \theta \hat N} \hat T^{-1} = \e^{-\imth \th \hat N}$ and 
$\hat C \e^{\imth \theta \hat N} \hat C^{-1} = \e^{-\imth \th \hat N}$,
so the total symmetry group {${G_{\text{Tot}}}=\frac{U(1)^{\text{c}} \rtimes [\Z_4^T \times \Z^{{C}}_{4}]}{(\Z_2)^2}$},
again the denominator ${(\Z_2)^2}=({\Z_2^F})^2$ is the redundancy appearing in the three symmetry generators.
We can redefine $\hat C \hat T$ as a new anti-unitary symmetry generator such that $(\hat C \hat T)^2=+1$ generating
a $\Z^{{CT}}_{2}$, and
$(\hat C \hat T) \e^{\imth \theta \hat N} (\hat C \hat T)^{-1} = \e^{\imth \th \hat N}$,
so the total symmetry group can be rewritten 
{${G_{\text{Tot}}}=\frac{[U(1)^{\text{c}}  \rtimes \Z_4^C ] }{\Z_2}  \times \Z^{{CT}}_{2} $}.

By flipping $T^2=+1$ in Minkowski to $T^2=(-1)^F$ in Euclidean, we get the full symmetry 
${\mathrm{Pin}^{-}\times_{\Z_2^F} [\frac{U(1)^{\text{c}}  \rtimes \Z_4^C}{\Z_2} ] }$
for the cobordism theory. 

\end{enumerate}

{Even though this symmetry group in the option 1 above 
is not exactly the fermionic electron symmetry, for convenience, 
we consider the SPT invariants of this group for CII class\footnote{Note that the SPT classifications via cobordisms for the two symmetries of CII class in principle could be different. The symmetry group in option 2 can be embedded in the symmetry group in option 1. In general this gives a homomorphism between the corresponding cobordism groups in the opposite direction, as discussed in section \ref{sec:sym-emb-cob}. A priori it is not obvious whether it is an isomorphism in this case or not.}.

Potentially we can realize CII class SPTs as topological insulator 
with inter-sublattice hopping and inter-sublattice spin-orbit coupling terms:
$\hat H= \sum_{j,A,B}(\hat c_{j_A}^\dag  \hat c_{j_B} 
+  \im \hat c_{i_A}^\dag  n_x   \cdot \hat \si^x \hat c_{i_B} +\im \hat
c_{j_A}^\dag  n_y \cdot \hat \si^y  \hat c_{j_B} +\im \hat c_{j_A}^\dag  n_z
\cdot \hat \si^z \hat c_{j_B} + \mathrm{h.c.}) + \dots$ with Hermitian conjugate (h.c.) terms., and additional 
symmetry-preserving interaction terms 
(See Sec.~\ref{sec:global-UV-H} and \cite{Wen1111.6341}).

There are 8 different symmetry-protected vacua, forming a group structure  $(\nu_{\text{CII}}, \alpha,\beta) \in$ {$\Omega^4_{\mathrm{Pin}^{-}\times_{\Z_2^F} SU(2)}$} $=\Z_2 \times \Z_2 \times \Z_2$ for a complete classification, firstly computed in \cite{1604.06527FH}. 
Our Appendix \ref{sec:App} provides further details and calculations. 
We explore their field theories, topological terms and physics in the next subsection.

\subsubsection{SPT vacua and topological terms}

Consider again $\nu$ doublets of Dirac fermions transforming under $SU(2)$. 
But now let them transform under orientation reversal map by $T'=T$ matrix described in Section \ref{sec:SU2-Dirac}. Since $T^2=(-1)^F$ this requires a choice of $\mathrm{Pin}^{-}\times_{\Z_2} SU(2)$ structure on the manifold. One can again consider the forgetful map $\text{Pin}^{-}\times_{\Z_2} SU(2)\rightarrow SU(2)/\Z_2\cong SO(3)$.  The obstruction to the existence of $\text{Pin}^-$ structure is $w_1^2(TM_4)+w_2(TM_4)$. To define  a $\text{Pin}^{-}\times_{\Z_2} SU(2)$ structure one can choose an $SO(3)$ bundle $V_{SO(3)}$ with $w_2(V_{SO(3)})=w_1^2(TM_4)+w_2(TM_4)$ and lift it to $\text{Pin}^{+}\times_{\Z_2} SU(2)$. In this case each eigenvalue of the Dirac operator is accompanied by an opposite one (this can be shown by an argument similar to the one in \cite{Metlitski:2015yqa}, that is by presenting an operator that anti-commutes with the Dirac operator and commutes with the transition functions) and therefore the second term in (\ref{eta-su2}) of the corresponding $\eta$-invariant identically vanishes. A similar calculation gives then:
\begin{equation} \label{eq:SU(2)-CII-top-term}
 Z^{\nu}_{SU(2)}[a] = \left(\frac{\det(\slashed{D}_a-|m|)}{\det(\slashed{D}_a+|m|)}\right)^\nu
\stackrel{|m|\rightarrow \infty}{\longrightarrow} (-1)^{\nu N_0'}
\end{equation} 
where $N_0'$ is the number of the zero modes of the Dirac operator. Its value mod 2 is a spin-topological invariant known as mod 2 index. The non-trivial fSPT classes generated by such massive Dirac fermions are effectively labelled by $\nu\in \Z_2$. The full classification of fermionic SPTs is given by $\Omega_{\mathrm{Pin}^{-}\times_{\Z_2} SU(2)}^4\equiv \text{Hom}(\Omega^{\mathrm{Pin}^{-}\times_{\Z_2} SU(2)}_4,U(1))\cong (\Z_2)^3$ where the other two $\Z_2$ factors corresponds to purely bosonic SPTs generated by the SPTs with actions $w_2^2(TM_4)$ and $w_1^4(TM_4)$. After inclusion of such bosonic phases the fSPT partition function in general read as:
\begin{equation}
 Z^{\nu,\alpha,\beta}=(-1)^{\nu N_0'+\alpha\,w_2^2(TM_4)+\beta\,w_1^4(TM_4)},\qquad (\nu,\alpha,\beta) \in \Z_2\times \Z_2\times \Z_2.
\end{equation} 

The embedding of purely bosonic SPTs $(\Z_2)^2 \hookrightarrow (\Z_2)^3$ ($(\beta,\alpha)\mapsto (0,\alpha,\beta)$) is given by the pullback of the forgetful map
\begin{equation}
\begin{array}{ccc}
 \Omega^{\mathrm{Pin}^{-}\times_{\Z_2} SU(2)}_4 & \rightarrow & \Omega^{O}_4 \\
 (\Z_2)^3 & \rightarrow & (\Z_2)^2
\end{array}.
\end{equation} 

When $M_4$ is oriented we have $N_0'=N_0=N_++N_-$ and one can again use the index theorem. As in the CI case, we have 
\begin{equation}
 Z^{\nu,\alpha,\beta}=(-1)^{\nu\,c_2(V_{SU(2)})}
\end{equation} 
for spin $M_4$, where $V_{SU(2)}$ is the $SU(2)$ bundle which is the lift of $V_{SO(3)}$ (possible due to vanishing $w_2$), and, more generally
\begin{equation}
 Z^{\nu,\alpha,\beta}=(-1)^{\nu(\sigma(M_4)+p_1(V_{SO(3)}))/4+\alpha\,w_2^2(TM_4)}.
\end{equation} 
for non-spin $M_4$.  
{Again, this means that odd $\nu$ fSPTs have $\theta=\pi$ terms.
The partition function on oriented spin manifold becomes the same as $ \exp\big(-S[a] \big)$ in \eqn{eq:SU(2)-theta-pi-probe}.
}
 
 Note that the fact CII is the same as in the CI case on oriented manifolds
 is not surprising because Pin$^+$ and Pin$^-$ become equivalent on oriented manifolds.

In Table \ref{table3}, we compute the partition functions on various 4-manifolds
that distinguish all SPT states 
    of Cartan CII class with this particular global symmetry.

\begin{table*}[h!]
\centering
    \begin{tabular}{|c | c|}
    \hline
$(M_4,\;V_{SO(3)})$ & $Z^{\nu,\alpha,\beta}$
 \\ \hline  \hline  
$(S^4, H)$ & $(-1)^\nu$
\\ \hline    
$(\mathbb{CP}^2, L_\C+1)$ & $(-1)^\alpha$
\\ \hline    
$(\mathbb{RP}^4, 2L_\mathbb{R}+1)$ & $(-1)^\beta$
\\ \hline    
    \end{tabular}
    \caption{Some examples of non-trivial values of the partition function on 4-manifolds 
    $S^4, \mathbb{CP}^2$, and $\mathbb{RP}^4$ 
    that distinguish all SPT states  
    of Cartan CII class with a particular global symmetry {${SU(2) \times \Z_2^T}$} for Hamiltonian, or based on
    Cobordism $\Omega_{\mathrm{Pin}^{-}\times_{\Z_2} SU(2)}^4$.}
                      \label{table3}
    \end{table*}

\subsection{C class: $SU(2)$ and ${\mathrm{Spin} \times_{\Z_2^F} SU(2)}$}

We can take the full spin rotation ${G_{\text{Tot}}}=SU(2)$ under operator 
$\e^{\imth {\theta} \hat n \cdot \hat S_j}$ in eqn.~(\ref{eq:SU(2)-rot}) with $SU(2) \supset \Z_2^F$. Without time reversal,
we get the full symmetry 
${\mathrm{Spin} \times_{\Z_2^F} SU(2)}$
for the cobordism theory. Apart from the trivial vacuum, there are no other non-trivial symmetry-protected vacua because {$\Omega^4_{\mathrm{Spin} \times_{\Z_2} SU(2)}=0$} \cite{1604.06527FH}.

\subsection{AI class: $U(1)  \rtimes \Z^{{T}}_{2}$ and ${\mathrm{Pin}^{-}\ltimes_{\Z_2^F} U(1)}$}

The AI class corresponds to $U(1)$ and time reversal symmetries with $T^2=+1$.
We list two ways to realize the full symmetry group $U(1)  \rtimes \Z^{{T}}_{2}$ on fermionic electrons:

\begin{enumerate}

\item UV lattice symmetry {${G_{\text{Tot}}}={U(1)^{\text{c}}  \rtimes {\Z^{{T}}_{2}}'}$},
IR Euclidean ${\mathrm{Pin}^{-}\ltimes_{\Z_2^F} U(1)}$: Topological Insulator. 

We can take the $U(1)^{\text{c}}$-fermion number charge symmetry operator as
$\e^{\imth {\theta}  \hat N_j}$, and a modified time reversal symmetry
$\hat T'=  \e^{\imth \pi \hat S_y} \hat T_{}$, so that $\hat T'^2=+1$ 
and
$\hat T' \e^{\imth \theta \hat N} \hat T'^{-1} = \e^{-\imth \th \hat N}$.
The full onsite symmetry is {${G_{\text{Tot}}}={U(1)^{\text{c}}  \rtimes {\Z^{{T}}_{2}}'}$}.

By flipping $T^2=+1$ in Minkowski to $T^2=(-1)^F$ in Euclidean, we get the full symmetry 
${\mathrm{Pin}^{-}\ltimes_{\Z_2^F} U(1)}$
for the cobordism theory. 

\item UV lattice symmetry {${G_{\text{Tot}}}={U(1)^\text{spin}_{z}  \rtimes {\Z^{{T}}_{2}}'}$},
IR Euclidean ${\mathrm{Pin}^{-}\ltimes_{\Z_2^F} U(1)}$: Topological Superconductor.

We consider $U(1)$-spin symmetry $\e^{\imth {\theta} (2\hat S_j^z)}$ as $U(1)^\text{spin}_{z}$, 
and a modified time reversal symmetry
$\hat T'=  \e^{\imth \pi \hat S_y} \hat T_{}$, so that $\hat T'^2=+1$ 
and
$\hat T' \e^{\imth \theta \hat N} \hat T'^{-1} = \e^{-\imth \th \hat N}$.
The full onsite symmetry is {${G_{\text{Tot}}}={U(1))^\text{spin}_{z}  \rtimes {\Z^{{T}}_{2}}'}$}.
We have the same full symmetry ${\mathrm{Pin}^{-}\ltimes_{\Z_2^F} U(1)}$
for the Cobordism theory. 
\end{enumerate}

Potentially we can realize AI class SPTs as 
(1) topological superconductor with both a real spin-singlet pairing ${(\hat c_{\up i}\hat c_{\down j}- \hat c_{\down i}\hat c_{\up j})}$ 
and a spin-order $\hat c_i^\dag \hat \si^z \hat c_j $,
or (2) topological insulator with $x$-$z$ plane coplanar spin order 
$\hat c_i^\dag   n_x \cdot  \hat \si^x \hat c_i +{\hat c_j^\dag  n_z \cdot \hat \si^z \hat c_j} $,
or (3) topological insulator with a spin-orbital coupling $\hat H= \sum_{{i,j,k},{i',j',k'}}(\im \hat c_i^\dag  n_x
\cdot \hat \si^x \hat c_{i'} +\im \hat c_j^\dag  n_y \cdot \hat \si^y \hat c_{j'}
+\im \hat c_k^\dag  n_z  \hat \si^z \hat c_{k'} ) + \dots$ with 
additional symmetry-preserving interaction terms. (See Sec.~\ref{sec:global-UV-H} and \cite{Wen1111.6341}.)

There are 2 different symmetry-protected vacua, forming a group structure  $\Omega^4_{\text{Pin}^{\tilde{c}-}}=\Z_2$ where $\text{Pin}^{\tilde{c}-}\equiv \mathrm{Pin}^{-}\ltimes_{\Z_2^F} U(1)$. 

\subsubsection{SPT vacua and topological terms}

The only non-trivial topological term (generating $\Z_2$ group) is of bosonic nature, $w_2^2(TM_4)$ \cite{1401.1142WS}. This is consistent with the fact the the Dirac operator can be defined (see Section \ref{sec:U1-Dirac-op}) for AII and AIII symmetry, but not AI (at least not in an obvious way).

\subsection{AII class: {$\frac{U(1)   \rtimes \Z^{{T}}_{4}}{\Z_2}$} and ${\mathrm{Pin}^{+}\ltimes_{\Z_2^F} U(1) }$}

The AII class corresponds to $U(1)$ and time reversal symmetries with $T^2=(-1)^F$.
We list one standard ways to realize the full symmetry group {$\frac{U(1)   \rtimes \Z^{{T}}_{4}}{\Z_2}$} on fermionic electrons:

\begin{enumerate}

\item UV lattice symmetry {$\frac{U(1)^{\text{c}}    \rtimes \Z^{{T}}_{4}}{\Z_2}$},
IR Euclidean ${\mathrm{Pin}^{+}\ltimes_{\Z_2^F} U(1) }$: Topological Insulator.

We can take the $U(1)^{\text{c}}$-fermion number charge symmetry operator as
$\e^{\imth {\theta}  \hat N_j}$, and the usual time reversal symmetry
$\hat T_{}$, so that $\hat T^2=(-1)^F$ 
and
$\hat T \e^{\imth \theta \hat N} \hat T^{-1} = \e^{-\imth \th \hat N}$.
The full onsite symmetry is {${G_{\text{Tot}}}={\frac{U(1)^{\text{c}}    \rtimes \Z^{{T}}_{4}}{\Z_2}}$},
again the denominator is the redundant factor ${\Z_2}={\Z_2^F}$.

By flipping $T^2=(-1)^F$ in Minkowski to $T^2= +1$ in Euclidean, we get the full symmetry 
${\mathrm{Pin}^{-}\ltimes_{\Z_2^F} U(1)}$
for the Cobordism theory. 

\end{enumerate}

Potentially we can realize AII class topological insulator with a spin-orbital coupling $\hat H= \sum_{{i,j,k},{i',j',k'}}(\im \hat c_i^\dag  n_x
\cdot  \si^x \hat c_{i'} +\im \hat c_j^\dag  n_y \cdot \hat \si^y \hat c_{j'}
+\im \hat c_k^\dag  n_z  \hat \si^z \hat c_{k'} ) + \dots$
 and additional symmetry-preserving interaction terms (See Sec.~\ref{sec:global-UV-H} and \cite{Wen1111.6341}).

There are 8 different symmetry-protected vacua, forming a group structure   $(\nu_{\text{AII}}, \alpha,\beta) \in$ {$\Omega^4_{\text{Pin}^{\tilde{c}+}}$} $=\Z_2 \times \Z_2 \times \Z_2$ where $\text{Pin}^{\tilde{c}+}\equiv \mathrm{Pin}^{+}\ltimes_{\Z_2^F} U(1) $.
       
\subsubsection{SPT vacua and topological terms}

The partition function of a general fSPT on a general 4-manifold reads \cite{Metlitski:2015yqa}:
\begin{equation}
 Z^{\nu,\alpha,\beta}=(-1)^{\nu N_0'+\alpha\,w_2^2(TM_4)+\beta\,w_1^4(TM_4)},\qquad (\nu,\alpha,\beta) \in \Z_2\times \Z_2\times \Z_2.
 \label{Z-AII}
\end{equation} 
where $N_0'$ is the mod 2 index of the Dirac operator defined using $T$ transform from Section \ref{sec:U1-Dirac-op}. The corresponding topological term arises after integrating out $\nu$ copies of massive Dirac fermions.

\subsection{AIII class: {$\frac{U(1)   \times \Z^{{T}}_{4}}{\Z_2}$} or {${U(1)} \times \Z^{{T'}}_{2}$}, and ${\mathrm{Pin}^{c} }={\mathrm{Pin}^{\pm}\times_{\Z_2^F} U(1)}$}

The AIII class corresponds to different $U(1)$ and time reversal symmetries with $T^2=(-1)^F$.
We list two standard ways to realize the full symmetry group {$\frac{U(1)   \times \Z^{{T}}_{4}}{\Z_2}$} on fermionic electrons:

\begin{enumerate}

\item UV lattice symmetry {$\frac{U(1)   \times \Z^{{T}}_{4}}{\Z_2}$},
IR Euclidean ${\mathrm{Pin}^{+}\times_{\Z_2^F} U(1)}$: Topological Superconductor.

We can take the $U(1)$-spin symmetry $\e^{\imth {\theta} (2\hat S_j^z)}$ as $U(1)^\text{spin}_{z}$, 
and the usual time reversal symmetry
$\hat T_{}$, so that $\hat T^2=(-1)^F$ thus a $\Z^{{T}}_{4}$, 
and
$\hat T \e^{\imth \theta (2\hat S_j^z)} \hat T^{-1} = \e^{\imth \theta (2\hat S_j^z)}$.
The full onsite symmetry is {$\frac{U(1)   \times \Z^{{T}}_{4}}{\Z_2}$},
again a mod-out redundant factor ${\Z_2}={\Z_2^F}$.

By flipping $T^2=(-1)^F$ in Minkowski to $T^2= +1$ in Euclidean, we get the full symmetry 
${\mathrm{Pin}^{+}\times_{\Z_2^F} U(1)}$
for the Cobordism theory. 

\item UV lattice symmetry {${U(1)} \times \Z^{{T'}}_{2}$},
IR Euclidean ${\mathrm{Pin}^{-}\times_{\Z_2^F} U(1)}$: Topological Superconductor.

Since $\hat T$ and $\e^{\imth \theta (2\hat S_j^z)}$ commute, we can define $\hat T'=\e^{\imth \frac{\pi}{2} (2\hat S_j^z)} \hat T$,
so that $(\hat T')^2=\e^{\imth {\pi} (2\hat S_j^z)} (\hat T)^2=(-1)^F(-1)^F=+1$.
The full onsite symmetry is {${U(1)} \times \Z^{{T'}}_{2}$}.

By flipping $T^2=+1$ in Minkowski to $T^2= (-1)^F$ in Euclidean, we get the full symmetry 
${\mathrm{Pin}^{-}\times_{\Z_2^F} U(1)}$
for the Cobordism theory. 

\end{enumerate}

Potentially we can realize AIII class topological superconductor with both a real spin-singlet pairing $(\hat c_{\up i}\hat c_{\down j}- \hat c_{\down i}\hat c_{\up j})$ and additional symmetry-preserving interaction terms (See Sec.~\ref{sec:global-UV-H} and \cite{Wen1111.6341}).

Note that ${\mathrm{Pin}^{c} }={\mathrm{Pin}^{\pm}\times_{\Z_2^F} U(1)}$. There are 16 different symmetry-protected vacua, forming a group structure  $(\nu_{\text{AIII}}, \alpha) \in$ {$\Omega^4_{\mathrm{Pin}^{c} }$}
$=\Z_8 \times \Z_2 $.

\subsubsection{SPT vacua and topological terms}

The partition function of a general fSPT on a general 4-manifold reads \cite{Metlitski:2015yqa}:
\begin{equation}
 Z^{\nu,\alpha,\beta}=e^{2\pi i\nu \eta_{\text{Pin}^c}}(-1)^{\alpha\,w_2^2(TM_4)},\qquad (\nu,\alpha) \in \Z_8\times \Z_2.
 \label{Z-AIII}
\end{equation} 
where $\eta_{\text{Pin}^c}$ is the $\eta$-invariant of the Dirac operator defined using $CT$ transform from Section \ref{sec:U1-Dirac-op}. The corresponding topological term arises after integrating out $\nu$ copies of massive Dirac fermions. Note that $8\eta_{\text{Pin}^c}=w_1^4(TM_4)$. 

Using the appropriate index theorem for Dirac operator,
on an oriented manifold both (\ref{Z-AII}) and (\ref{Z-AIII}) take values $\pm 1$ and can be written as 
\begin{equation}
 (-1)^{\nu(\sigma(M_4)-c_1^2(\det(\mathfrak{s})))/8+\alpha w_2^2(TM_4)} 
\end{equation} 
where $\det(\mathfrak{s})$ is the determinant line bundle of the Spin$^c$ structure $\mathfrak{s}$ (cf. footnote \ref{foot:Pinc}).

\subsection{A class: $U(1)$ and ${\mathrm{Spin}^{c} }$}

There are no non-trivial symmetry-protected vacua because {$\Omega^4_{{\mathrm{Spin}^c}}$} $=0$. 
          
\subsection{BDI class: {$\Z_2^T \times \Z_2^F$} and {${\mathrm{Pin}^{-}}$}}
 
There are no non-trivial symmetry-protected vacua because {$\Omega^4_{\mathrm{Pin}^{-}}$}=0.
 
\subsection{DIII class:    {$\Z_4^T $} and {${\mathrm{Pin}^{+}}$}}

The symmetry of DIII class fSPTs as $\Z_4^T  \supset \Z_2^F$ is already discussed in Sec.~\ref{sec:global-UV}.
Potentially we can realize DIII class topological superconductor 
with both a real spin-singlet pairing $(\hat c_{\up i}\hat c_{\down j}- \hat c_{\down i}\hat c_{\up j})$ 
and a spin-orbital coupling
$(\im \hat c_i^\dag  n_x
\cdot \hat \si^x \hat c_{i'} +\im \hat c_j^\dag  n_y \cdot \hat \si^y \hat c_{j'}
+\im \hat c_k^\dag  n_z  \hat \si^z \hat c_{k'} ) + \dots$ with 
 additional symmetry-preserving interaction terms (See Sec.~\ref{sec:global-UV-H} and \cite{Wen1111.6341}).)

There are 16 different symmetry-protected vacua, forming a group structure $(\nu_{\rm DIII})\in $ {$\Omega^4_{\mathrm{Pin}^{+}}$} $=\Z_{16}$

\subsubsection{SPT vacua and topological terms}

The partition function of a general fSPT on a general 4-manifold reads \cite{Kapustin:2014dxa,Witten:2015aba}
:
\begin{equation}
 Z^{\nu}=e^{\pi i\nu \eta_{\text{Pin}_+}}
\end{equation} 
where $\eta_{\text{Pin}_+}$ is the $\eta$-invariant of the Dirac operator without background gauge field that can be defined using $CT$ transform from Section \ref{sec:U1-Dirac-op}. Note that $8\eta_{\text{Pin}_+}=w_1^4(TM_4)$. 

On an oriented manifold it takes $\pm 1$ values and becomes
\begin{equation}
 (-1)^{\nu\sigma(M_4)/16}.
\end{equation}

\subsection{D class:    {$\Z_2^F $} and {${\mathrm{Spin}}$}}

There are no non-trivial nontrivial symmetry-protected vacua because {$\Omega^4_{{\mathrm{Spin}}}=0$}


\section{The Web of Symmetry Reduction and Embedding}

\label{sec:sym-reduction}

\subsection{Symmetry reduction and embedding through Hamiltonian approach}

\label{sec:sym-reduction-H}

By analyzing the global symmetry groups for 10 Cartan classes, 
we find the following symmetry embedding relations, presented in Table \ref{table:sym-web}.
An arrow directed from a group $G_1$ to a group $G_2$ means that $G_2$ symmetry is embedded inside $G_1$. Equivalently,
symmetry $G_1$ can be broken down to $G_2$.
We find our web relation also manifests their notations in terms of C, A and D, etc.\footnote{Again we remind the readers
that the Cartan symmetry class associate many distinct symmetry groups in a single class. 
So in our study here, we only pick up certain symmetry groups of Cartan class that can be related by embedding or symmetry-breaking.
Here we either consider the largest symmetry groups in Table \ref{10-sym-groups} (that we study their cobordism theory) 
or the symmetries realizable in fermionic electron condensed matter.
} 
 
\subsubsection{CI/CII/C $\to$ AI/AII/AIII/A $\to$ BDI/DIII/D}
Let us start by analyzing the total symmetries $G_{\text{Tot}}$ in terms of lattice Hamiltonian formalism (shown in the third column in Table \ref{10-sym-groups}).

\begin{table*}[!h] 
 \centering 
 {
\begin{center}
\begin{tikzpicture}\kern-5mm[>=stealth,->,shorten >=2pt,looseness=.5,auto]
      \matrix (M)[matrix of math nodes,row sep=1cm,column sep=8mm]{
          & 
          \begin{minipage}[c]{.85in} C: \ccblue{$SU(2)$},\\ 
          \ccred{$\Omega^4_{{\mathrm{Spin}^{} \times_{\Z_2^F} SU(2)}}$}\\
          $=0$ 
\end{minipage} &    &  
\begin{minipage}[c]{.58in} A: \ccblue{$U(1)$},\\ 
          \ccred{$\Omega^4_{{\mathrm{Spin}^c}}$}\\
          $=0$ 
\end{minipage} 
&    &  
\begin{minipage}[c]{.8in} D: \ccblue{$\Z_2^F$}, \\ 
          \ccred{$\Omega^4_{{\mathrm{Spin}}}$}\\
          $=0$ 
\end{minipage} 
\\
        \begin{minipage}[c]{1.in}
       CI: \\
       \ccblue{$\frac{SU(2)^{\text{spin}} \times \Z_4^T}{\Z_2}$}, \\
       $(\nu_{\text{CI}}, \alpha)$ 
       $\in$  \\
       \ccred{$\Omega^4_{\mathrm{Pin}^{+}\times_{\Z_2^F} SU(2)}$}\\
        $=\Z_4 \times \Z_2$
\end{minipage}  
        &   &  
   \begin{minipage}[c]{1.in}
   AI:\\
        \ccblue{$U(1)^{\text{c}}   \rtimes \Z^{{T}}_{2}$},\\
        $(\alpha)\in$\\
        \ccred{$\Omega^4_{\mathrm{Pin}^{-}\ltimes_{\Z_2^F} U(1)}$}\\
        $=\Z_2$
        \end{minipage}
        &   &  
           \begin{minipage}[c]{.8in}
        BDI:\\
\ccblue{$\Z_2^T \times \Z_2^F$},\\
 \ccred{$\Omega^4_{\mathrm{Pin}^{-}}$}=0
        \end{minipage}
        &\\
          &   & 
          \begin{minipage}[c]{1.in}
          AIII:\\
          \ccblue{$\frac{U(1)^{\text{spin}}_{z} \times \Z^{{T}}_{4}}{\Z_2}$},\\
$(\nu_{\text{AIII}}, \alpha) \in$\\
\ccred{$\Omega^4_{\mathrm{Pin}^{c} }$}\\
$=\Z_8 \times \Z_2 $
\end{minipage} 
          &   & 
          \begin{minipage}[c]{.8in}
          \rm{DIII}:
          \ccblue{$\Z_4^T $},\\
          $(\nu_{\rm DIII})\in $ \\
          \ccred{$\Omega^4_{\mathrm{Pin}^{+}}$}$\\
          =\Z_{16}$\\  
                  \end{minipage}
\\
       \begin{minipage}[c]{1.in}
       CII:\\
       \ccblue{${SU(2) \times \Z_2^{T}}$},\\
        \ccblue{$\frac{[U(1)^{\text{c}}   \times \Z^{{CT}}_{2} ]\rtimes \Z_4^C}{\Z_2}$}, \\
       $(\nu_{\text{CII}}, \alpha,\beta)$ 
       $\in$  \\
       \ccred{$\Omega^4_{\mathrm{Pin}^{-}\times_{\Z_2^F} SU(2)}$}\\
        $=\Z_2 \times \Z_2 \times \Z_2$
\end{minipage}  
       &   & 
              \begin{minipage}[c]{1.in}
       AII: \ccblue{$\frac{U(1)   \rtimes \Z^{{T}}_{4}}{\Z_2}$},\\
       $(\nu_{\text{AII}}, \alpha,\beta) \in$\\
       \ccred{$\Omega^4_{\mathrm{Pin}^{+}\ltimes_{\Z_2^F} U(1) }$}\\
       $=\Z_2 \times \Z_2 \times \Z_2$
       \end{minipage}  
       \\
       };
       \foreach \a/\b in {1-2/1-4, 1-4/1-6, 2-1/2-3, 2-3/2-5, 
                          3-3/3-5, 4-1/4-3, 2-1/1-2, 2-3/1-4, 
                          2-5/1-6, 3-3/2-5, 3-5/1-6, 4-1/1-2, 4-1/3-3, 
                          4-3/3-5, 2-1/3-3, 3-3/1-4, 4-3/1-4} {
          \draw[thick,->](M-\a)--(M-\b);
       }
    \end{tikzpicture}
    \end{center}
 }
\caption{We propose a web of symmetry group embedding of 10 particular global symmetries in Cartan class.
We list down their classifications and corresponding topological terms in terms of their indices studied in our Sec.~\ref{10SPT},
where all indices $\nu$, $\beta, \dots$ are meant to be integers.
They are related as follows:
$\nu_{\text{AII}}=2\nu_{\text{CII}}$ mod 2 and
$\nu_{\text{DIII}}=2\nu_{\text{AIII}}=4\nu_{\text{CI}}= 8 \beta$ mod 16.
}
 \label{table:sym-web}
\end{table*}

We would like to begin from the general C (CI, CII and C) classes and see what groups do they embed (or how they can be broken down).

Obviously, CI and CII both contain the full symmetry of C, A and D group. They are related by breaking time reversal. We denote in brief
$$\text{CI, CII $\overset{\text{Break}}{\Longrightarrow}$ C: break time reversal}$$
We have $SU(2) \supset U(1) \supset \Z_2^F$, thus
$$\text{C $\overset{\text{Break}}{\Longrightarrow}$ A $\overset{\text{Break}}{\Longrightarrow}$ D}.$$
We can consider taking a smaller total group of CI class, {$\frac{U(1)^\text{spin}_{z} \rtimes [\Z^\text{spin}_{4,y} \times \Z_2^{CT}]}{\Z_2}$} 
(instead of the larger total group {$\frac{SU(2)^{\text{spin}} \times \Z_4^T}{\Z_2}$}). 
This is related to $\frac{{ [U(1)^\text{spin}_{z} \rtimes \Z^\text{spin}_{4,y}] \times \Z_4^{T}}}{(\Z_2)^2}$ analyzed earlier in Sec.~\ref{sec:CI-sym} by
redefining $\hat T$ to $\hat C\hat T$ where $\hat C$ is generated by the $\pi$-spin rotation symmetry
along $y$, say $\e^{\im {\pi}   \hat S_j^y}$.  
Then breaking $\Z^\text{spin}_{4,y}$ but keeping $\Z_2^{CT}$, we obtain AI's $U(1)   \rtimes \Z^{{T}}_{2}$, so
$$\text{CI $\overset{\text{Break}}{\Longrightarrow}$ AI: break $\Z^\text{spin}_{4,y}$ but keep ${U(1)^\text{spin}_{z}}$ and $\Z_2^{CT}$}.$$
Take another smaller total group, $\frac{{ [U(1)^\text{spin}_{z} \rtimes \Z^\text{spin}_{4,y}] \times \Z_4^{T}}}{(\Z_2)^2}$,
then breaking $\Z^\text{spin}_{4,y}$, we obtain AIII's {$\frac{U(1)^{\text{spin}}_{z} \times \Z^{{T}}_{4}}{\Z_2}$}:
$$ \text{CI $\overset{\text{Break}}{\Longrightarrow}$ AIII: break $\Z^\text{spin}_{4,y}$, but keep $U(1)$ and $\Z_4^{T}$}.$$
For CII breaking, take a smaller {$G_{\text{Tot}}=\frac{[U(1)^{\text{c}}  \rtimes \Z_4^C ] }{\Z_2}  \times \Z^{{CT}}_{2} $}. Breaking $\Z^{{CT}}_{2}$,
we get AII's $\frac{[U(1)  \rtimes \Z_4 ] }{\Z_2}$ via 
$$\text{CII $\overset{\text{Break}}{\Longrightarrow}$ AII: break $\Z^{{CT}}_{2}$ but keep $U(1)^{\text{c}} $ and $\Z_4^C $}.$$
Alternatively breaking $\Z^{{C}}_{4}$ of {$G_{\text{Tot}}=\frac{[U(1)^{\text{c}}  \rtimes \Z_4^C ] }{\Z_2}  \times \Z^{{CT}}_{2} $},
we get AIII's {${U(1)} \times \Z^{{T'}}_{2}$} or its rewriting {$\frac{U(1)   \times \Z^{{T}}_{4}}{\Z_2}$} via
$$\text{CII $\overset{\text{Break}}{\Longrightarrow}$ AIII: break $\Z^{{C}}_{4}$ but keep $U(1) $ and $\Z^{{CT}}_{2}$}.$$

Now consider the general A (AI, AII, AIII and A) classes and see what groups do they embed (or how they can be broken down).
All A classes can be broken to A by removing time reversal.
The AI's {$U(1)   \rtimes \Z^{{T}}_{2}$} and AIII's {${U(1)^{\text{}}_{} \times \Z^{{T}}_{2}}$} embeds BDI's {$\Z_2^T \times \Z_2^F$} via:
$$\text{AI, AIII $\overset{\text{Break}}{\Longrightarrow}$ BDI: break $U(1)$ down to $\Z_2^F$}.$$
The AII's {$\frac{U(1)   \rtimes \Z^{{T}}_{4}}{\Z_2}$} and AIII's {$\frac{U(1)^{\text{}}_{} \times \Z^{{T}}_{4}}{\Z_2}$} embeds DIII's {$\Z_4^T$} via:
$$\text{AII, AIII $\overset{\text{Break}}{\Longrightarrow}$ DIII: break $U(1)$ but keep $\Z_2^F$}.$$
Lastly the general D (BDI, DIII and D) classes can be broken to D by removing time reversal.

\subsection{Relations of symmetries through field theories and  topological terms}
\label{sec:sym-emb-cob}
%


Consider in general the meaning of symmetry embedding $G_2\subset G_1$ from the point of view of cobordism theory. The embedding provides a natural map between the bordism groups of manifold with the corresponding structure:
\begin{equation}
	\Omega_d^{G_2} \longrightarrow \Omega_d^{G_1}
\end{equation}
realized by treating manifolds with structure $G_2 \subset G_1$ as special cases of manifolds with structure $G_1$. Note that in general this map is neither surjective nor injective. The dual map relates the Pontryagin dual groups classifying corresponding SPTs:
 \begin{equation}
 	\Omega^d_{G_1} \longrightarrow \Omega^d_{G_2}.
 \end{equation}

\subsubsection{CI $\rightarrow$ AIII}
This corresponds to embedding $\text{Pin}^c=\text{Pin}^+\times_{\Z_2} U(1)$ structure into $\text{Pin}^+\times_{\Z_2} SU(2)$ in an obvious way, by embedding $U(1)$ as a maximal torus of $SU(2)$. This was already discussed in Section \ref{sec:SU2-terms}.The topological invariants  reduce as follows:
\begin{equation}
\begin{array}{rcl}
 \eta_{SU(2)} & \rightarrow  & 2\eta_{\text{Pin}^c}, \\
w_2^2(TM_4) & \rightarrow & w_2^2(TM_4). \\
\end{array}
\end{equation} 
The corresponding map between SPTs is then given by
\begin{equation}
\begin{array}{rcl}
 \Z_4\times \Z_2 & \longrightarrow  & \Z_8\times \Z_2, \\
 (\nu,\alpha) & \longmapsto & (2\nu,\alpha). 
\end{array}
\end{equation}  

\subsubsection{CI $\rightarrow$ AI}
This corresponds to a slightly less obvious embedding of $\text{Pin}^{\tilde{c}-}=\text{Pin}^-\ltimes_{\Z_2} U(1)$ structure into $\text{Pin}^+\times_{\Z_2} SU(2)$ realized by embedding $U(1)$ as a maximal torus of $SU(2)$ and by identifying the orientation-reversal element (of order 4) of $\text{Pin}^{\tilde{c}-}$ with the orientation-reversal element (of order 2) of $\text{Pin}^+\times_{\Z_2} SU(2)$ times the charge conjugation $C_{SU(2)}\in SU(2)$ order 4 element which already appeared in Section \ref{sec:SU2-Dirac}. The topological invariants  reduce as follows:
\begin{equation}
\begin{array}{rcl}
 \eta_{SU(2)} & \rightarrow  & 0, \\
w_2^2(TM_4) & \rightarrow & w_2^2(TM_4). \\
\end{array}
\end{equation} 
The corresponding map between SPTs is then given by
\begin{equation}
\begin{array}{rcl}
 \Z_4\times \Z_2 & \longrightarrow  &  \Z_2, \\
 (\nu,\alpha) & \longmapsto & \alpha. 
\end{array}
\end{equation}  

\subsubsection{CII $\rightarrow$ AII}
This corresponds to embedding $\text{Pin}^{\tilde{c}+}=\text{Pin}^+\ltimes_{\Z_2} U(1)$ structure into $\text{Pin}^-\times_{\Z_2} SU(2)$. It can be realized by embedding $U(1)$ as a maximal torus of $SU(2)$ and by identifying the orientation-reversal element (of order 2) of $\text{Pin}^{\tilde{c}+}$ with the orientation-reversal element (of order 4) of $\text{Pin}^-\times_{\Z_2} SU(2)$ times the charge conjugation $C_{SU(2)}\in SU(2)$ order 4 element. The product of those two order 4 elements is indeed an order 2 element. The topological invariants  reduce as follows:
\begin{equation}
\begin{array}{rcl}
 N_0' & \rightarrow  & 0, \\
w_2^2(TM_4) & \rightarrow & w_2^2(TM_4), \\
w_1^4(TM_4) & \rightarrow & w_1^4(TM_4). \\
\end{array}
\end{equation} 
The corresponding map between SPTs is then given by
\begin{equation}
\begin{array}{rcl}
 \Z_2^3 & \longrightarrow  & \Z_2^3, \\
 (\nu,\alpha,\beta) & \longmapsto & (0,\alpha,\beta). 
\end{array}
\end{equation} 

\subsubsection{CII $\rightarrow$ AIII}
This corresponds to embedding $\text{Pin}^{{c}}=\text{Pin}^-\times_{\Z_2} U(1)$ structure into $\text{Pin}^-\times_{\Z_2} SU(2)$ in an obvious way by embedding $U(1)$ as a maximal torus of $SU(2)$. The topological invariants  reduce as follows:
\begin{equation}
\begin{array}{rcl}
 N_0' & \rightarrow  & 0, \\
w_2^2(TM_4) & \rightarrow & w_2^2(TM_4), \\
w_1^4(TM_4) & \rightarrow & w_1^4(TM_4)=8\eta_{\text{Pin}^c}. \\
\end{array}
\end{equation} 
The corresponding map between SPTs is then given by
\begin{equation}
\begin{array}{rcl}
 \Z_2^3 & \longrightarrow  & \Z_8\times \Z_2, \\
 (\nu,\alpha,\beta) & \longmapsto & (4\beta,\alpha). 
\end{array}
\end{equation} 



\subsection{Bosonic $\rightarrow$ Fermionic}

The full symmetry group $\text{Pin}^\pm(d)\times_{\Z_2^f} \tilde{G}$ (in the examples above $\tilde{G}=SU(2)$, $U(1)$ or $\Z_2$) is a central extension of the bosonic symmetry $O(d)\times (\tilde{G}/\Z_2^f)$ by fermionic parity:
\begin{equation}
	1\longrightarrow \Z_2^f \longrightarrow \text{Pin}^\pm(d)\times_{\Z_2^f} \tilde{G} \longrightarrow O(d)\times (\tilde{G}/\Z_2^f) \rightarrow 1\,.
\end{equation}
As in the case considered in section \ref{sec:sym-emb-cob} this provides a homomorphism between the cobordism groups classifying the corresponding fermionic/bosonic SPTs:
\begin{equation}
	\Omega^d_{O\times (\tilde{G}/\Z_2^f) }\equiv\Omega^d_{O}(B(\tilde{G}/\Z_2^f)) \longrightarrow \Omega^d_{\text{Pin}^\pm\times_{\Z_2^f} \tilde{G}}\,.
\end{equation}
In general, this map is neither surjective nor injective. Consider, for example the CI case in $d=4$ with $\text{Pin}^\pm\times_{\Z_2^f} SU(2)$ symmetry. The bosonic symmetry is $O(4)\times SO(3)$ symmetry. The $\nu=1,3$ classes are not in the image, so the map is indeed not surjective. The map is also not injective because, for example
\begin{equation}
	w_2(TM_4)^2+w_2(SO(3))^2 \longmapsto 0
\end{equation}
since $w_2(TM_4)=w_2(SO(3))$ on manifolds with $\text{Pin}^\pm\times_{\Z_2^f} SU(2)$ structure.

Physically speaking, we can also understand the above discussion as why some bosonic SPTs may become trivial in a fermionic system.
Because the fermionic system, which supports a bosonic SPTs of symmetry $G$, is indeed an extended symmetry $G_{\text{Tot}}$, under a
 fermion parity $\Z_2^F$ extension, as $G_{\text{Tot}}/\Z_2^F=G$.
In the different extended symmetry $G_{\text{Tot}}$ and a different (extended) Hilbert space, 
the original nontrivial bosonic SPTs in G may thus become deformable under LUT to a trivial insulator/vacuum, 
only after lifting the original state into the extended $G_{\text{Tot}}$ set-up.

\section{Time Reversal and $SU(N)$ Symmetry-Protected Topological Invariants}

\label{sec:SUN-Sym}

Following the setup in Sec.~\ref{sec:global}, now we would like to go beyond the 10 particular global symmetries within Cartan symmetry classes of Sec.~\ref{10SPT}. We would like to study global symmetries including $SU(N)$ flavor/color in QCD$_4$ or 
viewed as $SU(N)$ larger spin/orbital symmetries in cold atom systems (or more exotic orbitals in condensed matter).
Here we less rigorously use ``QCD$_4$'' in a \emph{more general} context, that either contains a $SU(N)$ global (so that later can be gauged) 
or a $SU(N)$ gauge symmetry.

Earlier in Sec.~\ref{sec:CI-sym}, we mentioned the {CI class} with a symmetry $G_{\text{Tot}}=${$\frac{SU(2) \times \Z_4^T}{\Z_2}$}
and cobordism group for ${\mathrm{Pin}^{+}\times_{\Z_2^F} SU(2)}$}. 
While in electronic condensed matter, it can be realized as a $SU(2)$-spin rotation and time reversal-invariant 
topological superconductor system, 
we can also regard it as a $N_f=2$-flavor $SU(2)$ QCD$_4$ 
without color gauge coupling. 
On the other hand, we can further dynamically gauge the $SU(2)$ to obtain a 
$N_c=2$-color $SU(2)$ QCD$_4$ 
without a flavor symmetry but only strong gauge coupling.

Here in Sec.~\ref{sec:SUN-Sym}, we would like to explore the SPTs associated to $SU(2) \times SU(2)$ color-flavor symmetry, 
$SU(3)$ symmetry, and $SU(4)$ symmetry with $\Z_2^T$ time-reversal.
Later in Sec.~\ref{sec:SUN-gauge}, we will explore the consequence of gauging $SU(N)$ 
for these SPT vacua.

We summarize the global symmetries and their notations at UV lattice/IR Minkowski/Euclidean signatures, and their
SPT invariants in Table \ref{table:SUN-groups}.

\begin{table*}[!h] 
 \centering
 \makebox[\textwidth][r]{
 \begin{tabular}{ |c| l | c|  c|}
\hline 
 \begin{minipage}[c]{2.2in} 
Particle Physics / QCD \\
(or Cold Atom) Realization 
\end{minipage}
& 
\begin{minipage}[c]{2.1in} Full Sym $G_{\text{Tot}}$: 
($G_{\text{Tot}}/ \Z_2^F = G $)\\
\ccblue{Minkowski} vs. \ccred{Euclidean}\\[-2mm]
\end{minipage} & 
\begin{minipage}[c]{1.678in}
Cobordism $\Omega^4$;  \\
Classification (3+1D)
\end{minipage} 
 \\
\hline
$SU(2)_{\text{color}} \times SU(2)_{\text{flavor}}$, ${T^2=(-1)^{F}}$&
\ccblue{$\frac{(SU(2))^2 \times \Z_4^{T}}{(\Z_2)^2}$}\,vs.\,
\ccred{$\frac{(SU(2))^2 }{\Z_2}\times \Z_2^{T}$}
& 
\begin{minipage}[c]{2.in} 
${(\mathrm{Pin}^{+}\times (SU(2))^2)/(\Z_2^F)^2}$\\
$= {\mathrm{Pin}^{+}\times_{\Z_2^F} SO(4)}$;\\
$(\nu,\alpha,\beta,\gamma) \in \Z_4\times \Z_2\times \Z_2\times \Z_2$
\end{minipage}
\\
\hline
$SU(2)_{\text{}}$, ${T^2=(-1)^{F}}$
&
\begin{minipage}[c]{2.2in} 
\ccblue{$\frac{SU(2) \times \Z_4^{T}}{\Z_2}$}\,vs.\,
\ccred{${SU(2)  \times \Z_2^{T}}$}
\end{minipage}
& 
\begin{minipage}[c]{2.in} 
${\mathrm{Pin}^{+}\times_{\Z_2^F} SU(2)}$;\\
$(\nu_{\text{CI}}, \alpha) \in \Z_4 \times \Z_2 $
\end{minipage}
\\
\hline
$SU(3)_{\text{}}$, ${T^2=(-1)^{F}}$&
\ccblue{${SU(3) \times \Z_4^{T}}$}\,vs.\,
\ccred{${SU(3) \times \Z_2^F \times \Z_2^{T}}$}
& 
\begin{minipage}[c]{2.in} 
${\mathrm{Pin}^{+}\times_{} SU(3)}$;\\
$(\nu,\alpha) \in \Z_{16}\times \Z_2$
\end{minipage}
\\
\hline
$SU(4)_{\text{}}$, ${T^2=(-1)^{F}}$
&
\begin{minipage}[c]{2.2in} 
\ccblue{$\frac{SU(4) \times \Z_4^{T}}{\Z_2}$}\,vs.\,
\ccred{${SU(4)  \times \Z_2^{T}}$}
\end{minipage}
& 
\begin{minipage}[c]{2.in} 
${\mathrm{Pin}^{+}\times_{\Z_2^F} SU(4)}$;\\
$(\alpha,\beta,\gamma) \in \Z_{2}\times \Z_2\times \Z_2$.
\end{minipage}
\\
\hline
$SU(2n+1)_{\text{}}$, ${T^2=(-1)^{F}}$
&
\begin{minipage}[c]{1.8in} 
\ccblue{${SU(2n+1) \times \Z_4^{T}}$}\,vs.\,\\
\ccred{${SU(2n+1) \times \Z_2^F \times \Z_2^{T}}$}
\end{minipage}
& 
\begin{minipage}[c]{2.in} 
${\mathrm{Pin}^{+}\times_{} SU(2n+1)}$;\\
$(\nu,\alpha) \in \Z_{16}\times \Z_2$
\end{minipage}
\\
\hline
\hline
 \end{tabular}
 } \hspace*{-22mm}
\caption{
{Time Reversal and $SU(N)$ Symmetry-Protected Topological Invariants}:
The first column shows the conventional (but less-precise or misused) notation for some $SU(N)$-symmetry group in 
UV lattice Hamiltonian (Minkowski signature with unitary time evolution).
The second column shows the total symmetry group $G_{\text{Tot}}$.
The last column shows the group for cobordism calculation, their cobordism classification,
and their indices for SPT invariants.
We remark that the more proper time reversal notations,
(1) for $\text{Pin}^+\times \dots$, would be a $T'=CT$-symmetry,
(2) for $\text{Pin}^+\ltimes \dots$, would be a $T$-symmetry;
see Sec.~\ref{sec:2-SU(N)} for discussions.
Here in Table \ref{table:SUN-groups} and in Sec.~\ref{sec:SUN-Sym}, we may simply \emph{rename} the $T'=CT$-symmetry as the $T$-symmetry.
}
 \label{table:SUN-groups}
\end{table*}

\subsection{$SU(2)_{\text{color}} \times SU(2)_{\text{flavor}}$:
{$G_{\text{Tot}}=\frac{(SU(2))^2 \times \Z_4^{T}}{(\Z_2)^2}$} and ${(\mathrm{Pin}^{+}\times (SU(2))^2)/(\Z_2^F)^2}$}

We can realize a $SU(2) \times SU(2)$-symmetry with time reversal in fermionic system as follows:
\begin{itemize}
\item UV lattice symmetry {$G_{\text{Tot}}=\frac{(SU(2))^2 \times \Z_4^{T}}{(\Z_2)^2}$},
IR Euclidean ${(\mathrm{Pin}^{+}\times (SU(2))^2)/(\Z_2^F)^2}$: 2-color 2-flavor QCD$_4$. 

Let us call the two $SU(2)$ as color and flavor onsite symmetries, 
$SU(2)_{\text{color}}\equiv SU(2_c)$ and $SU(2)_{\text{flavor}} \equiv SU(2_f)$.
Here when we denote a fermion $\hat c_j$ on site $j$, we actually implicitly mean
$\hat c_j \equiv \hat c_{\alpha_c, \beta_f, j}$ where
$\alpha_c$, $\beta_f$ are color($c$)/flavor($f$) indices. 
Fermion of fundamental representations
on site $j$ lives in a $2^{2\times 2} =16$-dimensional Hilbert space, subject to symmetry constraint.

 The $SU(2)$-color/flavor rotation onsite symmetry operator  
 acts on a site $j$ by two independent generators, analogous to two copies of \eqn{eq:SU(2)-rot}. 
Time reversal symmetry
has $\hat T^2=(-1)^F$ thus forms $\Z^{{T}}_{4}$ group,
whose normal subgroup ${\Z_2}={\Z_2^F}$ is a doubled redundant factor in the two centers of $(SU(2))^2$, so we mod out it twice.

The full onsite symmetry is {$G_{\text{Tot}}=\frac{(SU(2))^2 \times \Z_4^{T}}{(\Z_2)^2}$}.
By flipping $T^2=(-1)^F$ in Minkowski to $T^2=+1$ in Euclidean, we get the full symmetry 
${(\mathrm{Pin}^{+}\times (SU(2))^2)/(\Z_2^F)^2}={\mathrm{Pin}^{+}\times_{\Z_2^F} SO(4)}$
for the Cobordism theory.
\end{itemize}

There are 32 different symmetry-protected vacua, forming a group structure  
$(\nu,\alpha,\beta,\gamma) \in$ {$\Omega^4_{\mathrm{Pin}^{+}\times SU(2)^2/\Z_2^2}$} $=\Z_4\times \Z_2\times \Z_2\times \Z_2$
 for a classification, 
firstly shown in our Appendix \ref{sec:App} with further details and calculations.
We explore their field theories, topological terms and physics in the next subsection.

\subsubsection{SPT vacua and topological terms}
\label{sec:SU2SU2-terms}

To recap, consider possible fermionic SPTs protected by $(SU(2)^2 \times \Z_4^T )/\Z_2^2$ symmetry in Minkowski spacetime. In Euclidean spacetime the symmetry becomes  $( SU(2)^2 \times \Z_2^T)/\Z_2$. When one considers the classification of SPTs by cobordism, the corresponding structure group is $(\mathrm{Pin}^{+}\times SU(2)^2)/\Z_2^2$. As in CI/CII cases, consider the topological term that arises after integrating out massive fermions (normalized by the partition function of the same fermions but with the mass  of opposite sign). When putting the fermions on an unoriented space we will use $CT$ transformation, as in the CI case. Note that $CT$ transformation, unlike $T$, does not contain charge conjugation matrix and thus has a universal definition for Dirac fermions transforming in arbitrary representation of any symmetry group. It satisfies $(CT)^2=1$ which is indeed in agreement with $\mathrm{Pin}^+$ choice in the Euclidean structure group. 

Consider a possibly unoriented 4-manifold with $(\mathrm{Pin}^{+}\times SU(2)^2)/\Z_2^2$ structure. For each $SU(2)$ factor in the structure group, 
the corresponding forgetful map $\mathrm{Pin}^{+}\times SU(2)^2/\Z_2^2\rightarrow SU(2)/\Z_2\cong SO(3)$ defines an $SO(3)$ bundle. Denote the corresponding real rank 3 vector bundles as $V_1$ and $V_2$. They satisfy the following condition: $w_2(V_1)+w_2(V_2)+w_2(TM_4)=0$. When the manifold is $\mathrm{Pin}^+$ (that is $w_2(TM_4)=0$), 
the product $V_1\times V_2$ can be lifted an $SO(4)$ bundle $V_{SO(4)}$.

Because the fermionic parity is identified with $\Z_2$ centers of both $SU(2)$ symmetry groups, one needs to consider fermions in the tensor product of fundamental representations, that is $(\mathbf{2},\mathbf{2})$  representation of $SU(2)\times SU(2)$, where, as usual, $\mathbf{n}$ denotes the representation of dimension $n$. Equivalently,  $(\mathbf{2},\mathbf{2})$ is the vector representation of $SO(4)\cong SU(2) \times_{\Z_2} SU(2)$. Because this representation can be chosen to be real (4-dimensional), one can consider $\nu$ Majorana fermions, similarly to the case of DIII symmetry. The ratio of the determinants of the twisted Dirac operators (acting on the corresponding twisted Majorana spinor bundle) is given by 
\begin{equation}
 Z^{\nu}_{SU(2)^2/\Z_2}[a] = \left(\frac{\det(\slashed{D}_a-|m|)}{\det(\slashed{D}_a+|m|)}\right)^\nu
\stackrel{|m|\rightarrow \infty}{\longrightarrow} \exp(\pi \im \nu \eta_{SU(2)^2/\Z_2}[a])
\end{equation} 
where, for consistency, $\eta_{SU(2)^2/\Z_2}$ is still defined as the $\eta$-invariant of the Dirac operator acting on the twisted \textit{Dirac} spinor bundle (thus $1/2$ factor in the exponent compared to the CI case). The calculation of the bordism group (see Appendix \ref{sec:App}) tells us that $\eta_{SU(2)^2/\Z_2}\in \frac{1}{2}\Z$ so that effectively $\nu\in\Z_4$, and moreover\footnote{The symmetry $V_1\leftrightarrow V_2$ (which corresponds to the exchange of two $SU(2)$ groups) of the expression follows from the fact that $w_2(V_1)+w_2(V_2)+w_2(TM_4)=0$ and $w_1^2(TM_4)w_2(TM_4)=0$.}
\begin{equation}
 2\eta_{SU(2)^2/\Z_2}=w_1^4(TM_4)+w_1^2(TM_4)w_2(V_1)\mod 2.
 \label{twice-eta-SO4}
\end{equation}
When the manifold in Pin$^+$, 
this is in agreement with the general criterion of the presence of mixed TR-global anomaly \cite{Witten:2016cio}. 
Namely, when the manifold is Pin$^+$, a single (i.e. $\nu=1$) Majorana spinor is a section of a globally defined $V_{SO(4)}$ bundle tensored with a globally defined spinor bundle. The total Stiefel-Whitney class of the sum of two copies (i.e. $\nu=2$) of $V_{SO(4)}$ is given by $w(V_{SO(4)}\oplus V_{SO(4)})=1+w_2^2(V_{SO(4)})$. 
Therefore, $V_{SO(4)}^{\oplus 2}$ is stably trivial if and only if\footnote{Note that this is different from CI case, when the sum of two copies of $V_{SU(2)}$ (treated as a real 4-dimensional bundle) is always trivial. This is why in that case $2\eta_{SU(2)}=w_1^4(TM_4)\mod 2$ is independent of the choice of $V_{SU(2)}$ bundle.} $w_2(V_{SO(4)})=0$ (note that $w_2(V_{SO(4)})\equiv w_2(V_1)\equiv w_2(V_2)$ for Pin$^+$ manifolds). When this happens, the partition function for $\nu=2$ should coincide with the partition function of $2\dim V_{SO(4)}=8$ massive Majorana fermions in trivial representation (i.e. $\nu=8$ class of DIII), which is indeed $\exp(\pi \im w_1^4(TM_4))$. 
For the sum of four copies, we have $w(V_{SO(4)}^{\oplus 4})=1$, so that $\nu=4$ partition function should coincide with the partition function $4\dim V_{SO(4)}=16$ massive Majorana fermions in trivial representation, which is trivial. It follows that that indeed $\exp(4\pi \im \eta_{SU(2)^2/\Z_2})=1$.

Other possible topological terms correspond to purely bosonic SPTs. They can depend on combinations of Stiefel-Whitney classes of $TM_4$, $V_1$ and $V_2$ bundles. Taking into account the bosonic SPT with the action given by (\ref{twice-eta-SO4}), there are 3 more independent $\Z_2$ valued topological invariants. They can be chosen to be the following:
\begin{equation}
 w_1^4(TM_4)+w_2^2(V_1),
\end{equation}
\begin{equation}
 w_1^2(TM_4)w_2(V_2),
\end{equation}
\begin{equation}
 w_2^2(V_2).
\end{equation}
The fact that there are no more independent terms follows from the following relations (which can be shown using Wu's formula, for example): 
\begin{equation}
w_1^2(TM_4)w_2(TM_4)=w_1(TM_4)w_3(TM_4)=w_1^4(TM_4)+w_2^2(TM_4)+w_4(TM_4)=0,
\end{equation}
\begin{equation}
w_1(TM_4)w_3(V_1)=0,\qquad w_1(TM_4)w_3(V_2)=0,
\end{equation}
\begin{equation}
(w_1^2(TM_4)+w_2(TM_4)) w_2(V_1)=w_2^2(V_1),
\end{equation}
together with the condition
\begin{equation}
w_2(V_1)+w_2(V_2)+w_2(TM_4)=0.
\end{equation}

Therefore, the partition function of a general SPT on a general 4-manifold equipped with $(\mathrm{Pin}^{+}\times SU(2)^2)/\Z_2^2$ structure reads
\begin{multline}
 Z^{\nu,\alpha,\beta,\gamma}=e^{\pi i \nu \eta_{SU(2)^2/\Z_2}}\,(-1)^{\alpha\, (w_1^4(TM_4)+w_2^2(V_1))+\beta\, w_1^2(TM_4)w_2(V_2)+\gamma\, w_2^2(V_2)},\\ (\nu,\alpha,\beta,\gamma) \in \Z_4\times \Z_2\times \Z_2\times \Z_2.
\end{multline} 
On an oriented manifold one can use an index theorem for the twisted Dirac operator to obtain a more explicit expression:
\begin{multline}
 Z^{\nu,\alpha,\beta,\gamma}=(-1)^{\nu(\sigma(M_4)+p_1(V_1)+p_1(V_2))/4+\alpha\, w_2^2(V_1)+\gamma w_2^2(V_2)}\equiv\\
 (-1)^{\nu(\sigma(M_4)+p_1(V_1)+p_1(V_2))/4+\alpha\, p_1(V_1)+\gamma p_1(V_2)}
 ,\;\; (\nu,\alpha,\beta,\gamma) \in \Z_4\times \Z_2^3.
\end{multline} 
Note that the expression is well defined because $\sigma(M_4)+p_1(V_1)+p_1(V_2)$ is always a multiple of 4, which follows from the fact that 
\begin{multline}
 -\sigma(M_4) = p_1(TM_4)=\mathcal{P}_2(w_2(TM_4))= \mathcal{P}_2(w_2(V_1)+w_2(V_2)) =\\
 \mathcal{P}_2(w_2(V_1))+\mathcal{P}_2(w_2(V_2)) + 2w_2(V_1)w_2(V_2) = \\
 p_1(V_1)+p_1(V_2) + 2w_2(V_1)w_2(V_2)
 \mod 4
\end{multline}
together with
\begin{equation}
w_2(V_1)w_2(V_2)=w_2^2(V_1)+w_2(V_1)w_2(TM_4)=0.
\end{equation}

The following 4-manifolds equipped with $\mathrm{Pin}^{+}\times SU(2)^2/\Z_2^2$ structure can distinguish all $4\cdot 2^3=32$ different SPTs,
shown in Table \ref{table6}.
\begin{table}[h!]
\centering
    \begin{tabular}{|c | c|}
    \hline
$(M_4,\;V_1,\,V_2)$ & $Z^{\nu,\alpha,\beta,\gamma}$
 \\ \hline  \hline  
$(\mathbb{RP}^4, 3,3)$ & $e^{\pi \im\nu/2}\,(-1)^\alpha$
\\ \hline    
$(\mathbb{CP}^2, L_\C+1,3)$ & $(-1)^\alpha$
\\ \hline    
$(\mathbb{RP}^4, 2L_\R+1,2L_\R+1)$ & $(-1)^{\nu+\beta+\gamma}$
\\ \hline    
$(\mathbb{CP}^2,3, L_\C+1)$ & $(-1)^\gamma$
\\ \hline    
    \end{tabular}
    \caption{Some examples of non-trivial values of the partition function on 4-manifolds 
    $\mathbb{CP}^2$ and $\mathbb{RP}^4$ 
    that distinguish all SPT states 
    (of $SU(2)_{\text{color}} \times SU(2)_{\text{flavor}}$) with a particular global symmetry 
    {$G_{\text{Tot}}=\frac{(SU(2))^2 \times \Z_4^{T}}{(\Z_2)^2}$} for Hamiltonian, or based on
    Cobordism ${\mathrm{Pin}^{+}\times_{(\Z_2^F)^2} (SU(2))^2}$.}
                          \label{table6}
        \end{table}

\subsection{$SU(3)_{}$-symmetry: {$G_{\text{Tot}}={SU(3) \times \Z_4^{T}}$} and  ${\mathrm{Pin}^{+}\times_{} SU(3)}$}

Naively we may want to consider $SU(3)$ as a color/flavor onsite symmetry, whose symmetry operator is 
$
\e^{\imth \frac{\theta}{2} \sum_{a=1}^8 \hat n_a \cdot \hat c^\dag_j \hat \lambda^a \hat c_j}
$
(analogous to \eqn{eq:SU(2)-rot}, but replacing the $\si$ matrices to rank-3 Gell-Mann matrices). 
Such that 
a fermion of fundamental representation
on site $j$ lives in a $2^3=8$-dimensional Hilbert space on each site, subject to symmetry constraint.
In principle, we want $\hat T^2=(-1)^F$ thus $\Z^{{T}}_{4} \supset {\Z_2^F}$.
The $SU(3)$ contains neither $\Z^{{T}}_{4}$ nor ${\Z_2^F}$, so the total group is {$G_{\text{Tot}}={SU(3) \times \Z_4^{T}}$}.

However, $\hat T =\hat U_T K$ described in Sec.~\ref{sec:global} cannot be implemented for $SU(N)$ fundamental fermions for \emph{odd} $N$ consistently 
with $\hat T^2=(-1)^F$. At least $\hat T^2=(-1)^F$ does not manifest in the above lattice fermion construction. Nevertheless, 
there could be other lattice constructions at UV, or an effective field theory at intermediate energy scale to realize 
{$G_{\text{Tot}}={SU(3) \times \Z_4^{T}}$}. Here is one resolution:
\begin{itemize}
\item We can consider $SU(3)$ as a color/flavor onsite symmetry together with the $SU(2)$ spin-$\frac{1}{2}$ rotational symmetry for fermions.
There are ${3 \times 2}=6$ types of fermion creation/annihilation operators on each site.
Such that a fermion of fundamental representation
on site $j$ lives in a $2^{3 \times 2}=64$-dimensional Hilbert space on each site, subject to symmetry constraint.
The $SU(2)$ contains ${\Z_2^F}$ at its center. 
The total symmetry group is {$G_{\text{Tot}}=\frac{SU(3) \times SU(2) \times \Z_4^{T}}{\Z_2}$}.
We can implement on $\hat T^2=(-1)^F$ on 6 types of fermions per site $j$, then weakly breaking \emph{only} the $SU(2)$ symmetry. 
This can be a UV lattice realization. 
\end{itemize}

We like to switch gears now and directly consider the Euclidean path integral. By flipping $T^2=(-1)^F$ in Minkowski to $T^2=+1$ in Euclidean, we get the full symmetry 
${\mathrm{Pin}^{+}\times_{} SU(3)}$
for the Cobordism theory.

There are 32 different symmetry-protected vacua, forming a group structure  
$ (\nu,\alpha) \in$ {$\Omega^4_{\mathrm{Pin}^{+}\times SU(3)}$}$= \Z_{16} \times \Z_2$ for a classification, 
firstly shown in our Appendix \ref{sec:App} with further details and calculations. 
We explore their field theories, topological terms and physics in the next subsection.

\subsubsection{SPT vacua and topological terms}
\label{sec:SU3-terms}

In this case, there is no center shared between $SU(3)$ and Pin$^+(4)$, therefore $\mathrm{Pin}^{+}\times SU(3)$ structure just means the Pin$^+$ structure and $SU(3)$ bundle $V_{SU(3)}$ without any additional constraints. Therefore the corresponding bordism invariants are $8\eta_{\mathrm{Pin}^+}\in \Z_{16}$, which is the only invariant of $\Omega_{\mathrm{Pin}^{+}}$, and $w_4(V_{SU(3)})=c_2(V_{SU(3)})\mod 2$, which is the only non-trivial Stiefel-Whitney class of $V_{SU(3)}$ (treated as a real 6-dimensional bundle).

  Note that $\eta_{SU(3)}$, the $\eta$-invariant of the Dirac operator acting on Dirac spinors in fundamental representation of $SU(3)$, does not give a new independent invariant, because
  \begin{equation}
2\eta_{SU(3)} - 6\eta_{\text{Pin}^+} = w_4(V_{SU(3)}) \mod 2.
\label{eta-SU3-relation}
  \end{equation}
  This can be seen from the condition that $V_{SU(3)}$ is stably trivial (as a real vector bundle) if and only if $w_4(V_{SU(3)})=0$. The fact that the coefficient in front of $w_4(V_{SU(3)})$ in the hand side of (\ref{eta-SU3-relation}) is not zero can be checked on oriented manifolds, where $e^{2\pi i \eta_{SU(3)}}=(-1)^{c_2(V_{SU(3)})}$ from the index theorem. 
  
 On a general 4-manifold, the SPT partition function reads
  \begin{equation}
   Z^{\nu,\alpha}=e^{\pi i \nu \eta_{\text{Pin}^+}}\,(-1)^{\alpha w_4(V_{SU(3)})},\qquad (\nu,\alpha) \in \Z_{16}\times \Z_2.
  \end{equation} 
  On an oriented 4-manifold, it simplifies to 
   \begin{equation}
    Z^{\nu,\alpha}=(-1)^{\nu\frac{\sigma(M_4)}{16}}\,(-1)^{\alpha c_2(V_{SU(3)})},\qquad (\nu,\alpha) \in \Z_{16}\times \Z_2.
   \end{equation} 
  The following 4-manifolds can distinguish all different SPTs,
  shown in Table \ref{table7},
  where $H$ is the $SU(2)\subset SU(3)$ (complex rank 2) bundle with instanton number 1 (i.e. $c_2=1$) induced by Hopf fibration $S^7\rightarrow S^4$. 

\begin{table}[h!]
\centering
    \begin{tabular}{|c | c|}
    \hline
$(M_4,\;V_{SU(3)})$ & $Z^{\nu,\alpha}$
 \\ \hline  \hline  
$(\mathbb{RP}^4, 6)$ & $e^{\pi \im \nu/8}$
\\ \hline    
$(S^4, H+2)$ & $(-1)^\alpha$
\\ \hline    
    \end{tabular}
 \caption{Some examples of non-trivial values of the partition function on 4-manifolds 
    $S^4$ and $\mathbb{RP}^4$ 
    that distinguish all SPT states 
     with a particular global symmetry 
    {$G_{\text{Tot}}={SU(3) \times \Z_4^{T}}$} for Hamiltonian, or based on
    Cobordism  ${\mathrm{Pin}^{+}\times_{} SU(3)}$.}
\label{table7}
        \end{table} 

Note that the case of more general $SU(N)$ for odd $N$ is completely analogous. In particular the corresponding SPTs have the same classification $\Omega^4_{\text{Pin}^+\times SU(N)}\cong  \Z_{16}\times \Z_{2}$.

\subsection{$SU(4)_{\text{}}$-symmetry: {$G_{\text{Tot}}=\frac{SU(4) \times \Z_4^{T}}{\Z_2}$} and ${\mathrm{Pin}^{+}\times_{\Z_2^F} SU(4)}$}

We can realize a $SU(4)$-symmetry with time reversal in fermionic system as follows:

\begin{itemize}

\item UV lattice symmetry {$G_{\text{Tot}}=\frac{SU(4) \times \Z_4^{T}}{\Z_2}$},
IR Euclidean ${\mathrm{Pin}^{+}\times_{\Z_2^F} SU(4)}$: 4-flavor QCD$_4$. 

We consider $SU(4)$ as a color/flavor onsite symmetry, whose symmetry operator is 
$
\e^{\imth \frac{\theta}{2} \sum_{a=1}^{15} \hat n_a \cdot \hat c^\dag_j \hat \lambda^a \hat c_j}
$
(analogous to \eqn{eq:SU(2)-rot}, but replacing the Pauli's $\si$ matrices to rank-4 generalized Gell-Mann matrices),
such that 
a fermion of fundamental representation
on site $j$ lives in a $2^4=16$-dimensional Hilbert space on each site, subject to symmetry constraint.
In principle, we want $\hat T^2=(-1)^F$ and thus $\Z^{{T}}_{4} \supset {\Z_2^F}$.
The $\hat T =\hat U_T K$ described in Sec.~\ref{sec:global} can be implemented for $SU(N)$ fundamental fermions for \emph{even} $N$ consistently 
with $\hat T^2=(-1)^F$. We require that $\hat U_T \hat U_T^*=(-1)^F$ and
time reversal commutes with $SU(4)$, so 
$\hat T (\e^{\imth \frac{\theta}{2} \sum_{a=1}^{15} \hat n_a \cdot \hat c^\dag_j \hat \lambda^a \hat c_j})
 \hat T^{-1}  =( \e^{\imth \frac{\theta}{2} \sum_{a=1}^{15} \hat n_a \cdot \hat c^\dag_j \hat \lambda^a \hat c_j})$.
The $SU(4)$ has a $\Z_4$ center inside that contains ${\Z_2^F}$  but not $\Z^{{T}}_{4}$, so 
the full onsite symmetry total group is {$G_{\text{Tot}}=\frac{SU(4) \times \Z_4^{T}}{\Z_2}$}.

By flipping $T^2=(-1)^F$ in Minkowski to $T^2=+1$ in Euclidean, we get the full symmetry 
${\mathrm{Pin}^{+}\times_{\Z_2^F} SU(4)}$
for the cobordism theory.

\end{itemize}

There are 8 different symmetry-protected vacua, forming a group structure  
$(\alpha,\beta,\gamma) \in$
{$\Omega^4_{\mathrm{Pin}^{+}\times_{\Z_2} SU(4)}$} $= \Z_2 \times \Z_2 \times \Z_2$ for a classification, 
firstly shown in our Appendix \ref{sec:App} with further details and calculations. 
We explore their field theories, topological terms and physics in the next subsection.

\subsubsection{SPT vacua and topological terms}
\label{sec:SU4-terms}

The situation is very similar to the cases with $\mathrm{Pin}^{+}\times_{\Z_2} SU(2)/\Z_2$ (CI class) and $(\mathrm{Pin}^{+}\times SU(2)^2)/\Z_2^2$ structure groups. Here one can consider  the forgetful map $\mathrm{Pin}^{+}\times SU(4)/\Z_2\rightarrow SU(4)/\Z_2\cong SO(6)$. This defines an $SO(6)$ bundle $V_{SO(6)}$ (real 6-dimensional). It satisfies the following condition: $w_2(V_{SO(6)})+w_2(TM_4)=0$. When the manifold is $\mathrm{Pin}^+$ (that is $w_2(TM_4)=0$), 
the bundle can be lifted to an $SU(4)$ bundle $V_{SU(4)}$ (complex 4-dimensional).

As before, consider $\nu$ massive Dirac fermion in the fundamental representation of $SU(4)$. The ratio of the partition functions for positive/negative masses is again given in terms of the corresponding $\eta$-invariant:
\begin{equation}
 Z^{\nu}_{SU(4)}[a] = \left(\frac{\det(\slashed{D}_a-|m|)}{\det(\slashed{D}_a+|m|)}\right)^\nu
\stackrel{|m|\rightarrow \infty}{\longrightarrow} \exp(2\pi \im \nu \eta_{SU(4)}[a]).
\end{equation} 

From the calculation of the bordism group (see Appendix \ref{sec:App}), 
it follows that there are only $\Z_2$ valued invariants. In particular, $\nu=\nu_\text{max}\equiv 2$ SPT should be trivial. When 4-manifold is Pin$^+$, and there is a $V_{SU(4)}$ bundle, this is consistent with the fact that $V_{SU(4)}^{\oplus 2}$ is stably trivial and of real dimension $16$. This implies that indeed $ \exp(4\pi \im \eta_{SU(4)})= \exp(16\pi \im \eta_{\text{Pin}^+})=1$. Moreover, as before, one expects that if $\nu$ is taken to be $\nu_\text{max}/2=1$, the corresponding SPT is purely bosonic. This is indeed what the bordism group calculation tells us, since all the invariants are expressed via Stiefel-Whitney classes of $TM_4$ and $V_{SO(6)}$. It is easy to see that there are 3 independent degree 4 combinations of them. Therefore the partition function of a general SPT on a general 4-manifold reads as follows:
  \begin{equation}
   Z^{\alpha,\beta,\gamma}=(-1)^{\alpha w_1^4(TM_4)+\beta w_4(V_1)+\gamma w_2^2(V_1)},\qquad (\alpha,\beta,\gamma) \in \Z_{2}\times \Z_2\times \Z_2.
  \end{equation} 
  
  The following 4-manifolds can distinguish all different SPTs, shown in Table \ref{table8}.
\begin{table}[!h] 
\centering
      \begin{tabular}{|c | c|}
      \hline
  $(M_4,\;V_{SO(6)})$ & $Z^{\nu,\alpha}$
   \\ \hline  \hline  
  $(\mathbb{RP}^4, 6)$ & $(-1)^\alpha$
  \\ \hline    
  $(S^4, H+2)$ & $(-1)^\beta$
  \\ \hline   
    $(\mathbb{CP}^2, L_\C+4)$ & $(-1)^\gamma$
    \\ \hline  
    \end{tabular}
 \caption{Some examples of non-trivial values of the partition function on 4-manifolds 
    $S^4$, $\mathbb{CP}^2$ and $\mathbb{RP}^4$ 
    that distinguish all SPT states 
     with a particular global symmetry 
    {$G_{\text{Tot}}=\frac{SU(4) \times \Z_4^{T}}{\Z_2}$} 
    for Hamiltonian, or based on
    Cobordism  ${\mathrm{Pin}^{+}\times_{\Z_2^F} SU(4)}$.}
    \label{table8}
        \end{table} 
As before, $H$ is the $SU(2)\subset SO(4)\subset SO(6)$ (complex dimension 2) bundle with instanton number 1 induced by Hopf fibration $S^7\rightarrow S^4$.


\section{The Web of  Symmetry Reduction and Embedding for $SU(N)$ with Time-Reversal Symmetry}

\label{sec:sym-reduction-SU(N)}

\subsection{$(\text{Pin}^+\times SU(2)^2)/\Z_2^2$ symmetry}

By turning off one of the $SU(2)$ background fields (that is by considering trivial $V_2=3$) we get a an SPT with $(\text{Pin}^+\times SU(2))/\Z_2$ symmetry, that is of CI class. Namely, the topological invariants in Section \ref{sec:SU2SU2-terms} reduce as follows:
\begin{equation}
\begin{array}{rcl}
 \eta_{SU(2)^2/\Z_2} & \rightarrow  & 2\eta_{SU(2)} \\
w_1^4(TM_4)+w_2^2(V_1) & \rightarrow & w_1^4(TM_4)+w_2^2(TM_4)=4\eta_{SU(2)}+w_2^2(TM_4) \\
w_1^2(TM_4)w_2(V_2) & \rightarrow & 0 \\
w_2^2(V_2) & \rightarrow & 0 \\
\end{array}.
\end{equation} 
The corresponding map between SPTs is then given by
\begin{equation}
\begin{array}{rcl}
 \Z_4\times \Z_2^3 & \longrightarrow  & \Z_4\times \Z_2, \\
 (\nu,\alpha,\beta,\gamma) & \longmapsto & (\nu+2\alpha,\alpha). 
\end{array}
\end{equation} 

\subsection{$\text{Pin}^+\times SU(3)$ symmetry}

By turning off one of the $SU(3)$ background fields (that is by considering trivial $V_{SU(3)}=6$) we get a an SPT with $\text{Pin}^+$ symmetry, that is of DIII class. The topological invariants in Section \ref{sec:SU3-terms} reduce as follows:
\begin{equation}
\begin{array}{rcl}
 \eta_{\text{Pin}^+} & \rightarrow  & \eta_{\text{Pin}^+}, \\
w_4(V_{SU(3)}) & \rightarrow & w_1^4(TM_4)+w_2^2(TM_4)=4\eta_{SU(2)}+w_2^2(TM_4). \end{array}
\end{equation} 

 The corresponding map between SPTs is then given by
\begin{equation}
\begin{array}{rcl}
 \Z_{16}\times \Z_2 & \longrightarrow  & \Z_{16}, \\
 (\nu,\alpha) & \longmapsto & \nu. 
\end{array}
\end{equation} 

\subsection{$\text{Pin}^+\times_{\Z_2} SU(4)$ symmetry}
Consider embedding $SU(2)$ into $SU(4)$ in a block diagonal way. That is so that the corresponding unitary matrices are related as follows 
\begin{equation}
U_{SU(4)}=\left(
\begin{array}{cc}
 U_{SU(2)} & 0 \\
 0 & U_{SU(2)} 
\end{array}
\right).  
\end{equation} 
This defines reduction of $\text{Pin}^+\times_{\Z_2} SU(4)$ symmetry to $\text{Pin}^+\times_{\Z_2} SU(2)$, that is of CI class. The fundamental representation of $SU(4)$ then decomposes as $\mathbf{4}\rightarrow 2\cdot \mathbf{2}$. The vector representation of $SO(6)\cong SU(4)/\Z_2$ decomposes as $\mathbf{6}\rightarrow \mathbf{3}+3\cdot \mathbf{1}$, which means that the corresponding real vector bundles are reduced as $V_{SO(6)}\rightarrow V_{SO(3)}+3$. The topological invariants in Section \ref{sec:SU4-terms} then reduce as follows:
\begin{equation}
\begin{array}{rcl}
 w_1^4(TM_4) & \rightarrow  & w_1^4(TM_4)=4\eta_{SU(2)} \\
w_4(V_{SO(6)})& \rightarrow & 0 \\
w_2^2(V_{SO(6)}) & \rightarrow & w_2^2(V_{SO(6)})=w_2^2(TM_4) \\
\end{array}.
\end{equation} 
and the corresponding map between SPTs is  given by
\begin{equation}
\begin{array}{rcl}
  \Z_2^3 & \longrightarrow  & \Z_{4}\times \Z_2, \\
 (\alpha,\beta,\gamma) & \longmapsto & (2\alpha,\gamma). 
\end{array}
\end{equation} 


\section{Time Reversal and $SU(N)$ Yang-Mills Theory}

\label{sec:SUN-gauge}


\subsection{Gauging $SU(2)$: From CI and CII SPTs to $SU(2)$ gauge theories}

\label{sec:SUN-gauge-SU(2)}

Consider gauging of $SU(2)$ symmetry. This involves summation over all (allowed) classes of $V_{SO(3)}$ bundles ($V_{SO(3)}$ is the adjoint bundle associated to the $SO(3)$ principle bundle) and (path) integration over connection. To get a (physically) well defined integral we need to add a Yang-Mills (YM) term $\sim 1/(4g^2)\int_{M_4}\text{Tr}\,F_a\star F_a$ to the action in order to suppress contribution from large values of connection 1-form of $a$. Since the fermionic symmetry $\Z_2^F$ is identified with the center of $SU(2)$, the gauging of $SU(2)$ involves bosonization, so that the IR theory effectively becomes bosonic. This is in agreement with the fact that it can be put on a manifold $M_4$ which has $\mathrm{Pin}^{\pm}\times_{\Z_2} SU(2)$ structure, but no $\mathrm{Pin}^{\pm}$ structure.

The fSPT part in such coupled YM-fSPT system can be understood as something that appears after first integrating out some gapped theory 
{that has unique vacuum (SPTs)}
coupled to $SU(2)$ gauge field. 
{
On oriented spin manifolds or on a flat spacetime, the path integral becomes just\footnote{We normalize it so that 
integrand of the path integral is $e^{-S_E}$.} 
\begin{equation}
\int [{\cal D} {a}] \exp\big(-S[a] \big)=
\int [{\cal D} {a}] \exp\big(- \int\limits_{M_4} (\frac{1}{4g^2}\text{Tr}\,F_a\wedge \star F_a)
+  \int\limits_{M_4} (\frac{\rm i \theta}{8 \pi^2}  \text{Tr}\,F_a\wedge F_a) \big)
 \end{equation} 
both in CI and CII case. We stress that, in contrast to the previous \eqn{eq:SU(2)-theta-pi-probe} for background field probing SPTs, here we do have
\emph{dynamical gauge field} $a$, that is we are summing over $a$ with the path integral measure $\int [{\cal D} {a}]$.
}

The cases $\nu =0,1 \mod 2$ correspond to the values of $\theta=0,\pi$ in the theta-term, both known to preserve time-reversal symmetry on the classical level. On the quantum level, these two cases are very different. 

\subsubsection{$\nu=0 \mod 2$}

Consider first the simpler case of $\nu =0 \mod 2$, i.e. $\theta=0$. Then, on a flat space the we have a pure $SU(2)$ YM theory which is believed to be gapped in the IR with a single vacuum preserving time-reversal symmetry. Above we saw that on a general $M_4$, when  $\nu =0 \mod 2$, the actions of CI and CII fSPTs become the same, and, moreover, they coincide with the action 
\begin{equation}
  \pi \im (\alpha\,w_2^2+\beta\,w_1^4)
\label{bSPT}
\end{equation} 
of bosonic SPTs labelled by $(\alpha,\beta)\in {(\Z_2)}^2 \cong \Omega_{O}^4$, where $\beta=\nu/2$ in the CI case. Therefore coupled YM-fSPT systems in the IR become corresponding bosonic SPTs.

\subsubsection{$\nu=1 \mod 2$}

The case of $\nu =0 \mod 2$ is much more subtle. On a flat space we have a YM action with $\theta=\pi$ term. This situation has been analyzed in detail in \cite{Gaiotto:2017yup,Gaiotto:2017tne}. In particular, the authors argued that in this case the theory has non-trivial 't Hooft anomaly for time-reversal symmetry.  Therefore there are two natural possibilities of what can happen in IR. 

The first possibility is that the theory is still gapped, the time-reversal symmetry is spontaneously broken, and is there multiple (two, in the simplest scenario) vacua not invariant under time reversal symmetry. We discuss this first scenario in Sec.~\ref{sec:TR-break}. 
One more possibility is that the theory is actually a TQFT also symmetry-protected, thus a symmetry-enriched TQFT, as the second scenario.
Another possibility is that the theory is actually gapless, as the third scenario in Sec.~\ref{sec:SU2-pi-gapless}. 

\subsubsection{Spontaneous time reversal symmetry breaking}
\label{sec:TR-break}

The case of the spontaneously broken symmetry is the one considered in detail in \cite{Gaiotto:2017yup,Gaiotto:2017tne}, where, in particular, the 2+1D (3d) theories living on the domain walls were identified. When the time-reversal theory is spontaneously broken all SPT classes should collapse to one, because there is no non-trivial 3+1D SPT with no (including time-reversal) symmetry. This corresponds to the fact that there is no torsion in the oriented bordism group $\Omega_4^{SO}(\text{pt})$. In other words, the theory can be only put on an oriented manifold, where $w_1^4(TM_4)=0$ (so that no dependence on choice of $\nu$ and $\beta$ in CI and CII case respectively) and $w_2^2(TM_4)=p_1(TM_4) \mod 2$, so that $\alpha$ can be continuously deformed to zero. 

{
In summary, in this scenario, 
the dynamical gauging of $SU(2)$ results in {spontaneous time reversal symmetry breaking} for {$\nu=1 \mod 2$} case,
which suggests that there are only two trivial vacua ($T$-breaking vacua) and there can be no bosonic SPTs attached to them. }

\subsubsection{Deconfined gapless and time-reversal symmetric CFTs}

\label{sec:SU2-pi-gapless}

When the theory is gapless, the time-reversal symmetry can be still preserved. Then, naively, the IR theory can be universally described as a direct product of a certain fixed CFT with a bosonic SPT labelled by $(\alpha',\beta')$ as in (\ref{bSPT}), so that $(\alpha',\beta')=(\alpha,(\nu-1)/2)$ in the CI case and $(\alpha',\beta')=(\alpha,\beta)$ in the CII case. 

We propose a scenario that there are actually two distinct gapless deconfined time-reversal invariant $SU(2)$-gauge theories, 
protected by two different topological terms (\eqn{eq:SU(2)-CI-top-term} and \eqn{eq:SU(2)-CII-top-term}) respectively. The two gapless deconfined states should be two different time-reversal invariant CFTs.
 
How do we support our proposal of two different time-reversal symmetric CFTs for these strong-coupled $SU(2)$ gauge theories?
Here are our arguments and justifications:
\begin{enumerate}
\item The partition functions for two theories ${Z_{\text{YM+fSPT-CI}}}$ and ${Z_{\text{YM+fSPT-CII}}}$
are defined differently, effectively as follows
\bea \label{eq:SU2-CFT1}
&&\int [{\cal D} {a}] \exp\big(-S_{\text{YM+fSPT-CI}} \big)=
\int [{\cal D} {a}] \exp\big(-\int\limits_{M_4} (\frac{1}{4g^2}\text{Tr}\,F_a\wedge \star F_a) \big) 
\exp(2\pi \im \nu \eta_{SU(2)}[a]), \;\;\;\;\;\; \;\;\;\\
&& \label{eq:SU2-CFT2}
\int [{\cal D} {a}] \exp\big(-S_{\text{YM+fSPT-CII}} \big)=
\int [{\cal D} {a}] \exp\big(-\int\limits_{M_4} (\frac{1}{4g^2}\text{Tr}\,F_a\wedge \star F_a) \big) 
(-1)^{\nu N_0'[a]}, \;\;\;\;\;\; \;\;\;
\eea
but different only on non-orientable manifolds. We expect that the ratio of two partition functions
is not equal to 1 in certain non-orientable manifolds $M_4$ generically:
\bea
\frac{Z_{\text{YM+fSPT-CI}}}{Z_{\text{YM+fSPT-CII}}}
=\frac{\int [{\cal D} {a}] \exp\big(-S_{\text{YM+fSPT-CI}} \big)}{\int [{\cal D} {a}] \exp\big(-S_{\text{YM+fSPT-CII}} \big)} \neq 1.
\label{SU2-CFT-ratio}
\eea

Note that even though it maybe hard to have unambiguous (even at the physical level of rigor due to the ambiguous overall factor as well as renormalization counter terms) definition of individual path integrals (\ref{eq:SU2-CFT1}) and (\ref{eq:SU2-CFT2}), 
their ratio is much more controlled since one can expect the the possible ambiguities cancel.

\item {
There is no evidence for the field theory duality (in any sense of duality) for these non-supersymmetric $SU(2)$ gauge theories on non-orientable manifolds (that is between path integrals  between \eqn{eq:SU2-CFT1} and \eqn{eq:SU2-CFT2}).}

\item Numerical results have not ruled out the scenarios that $SU(2)$ gauge theories with topological terms can be gapless.
If they are indeed gapless, providing all the Lorentz/Euclidean rotational symmetries endorsed to \eqn{eq:SU2-CFT1} and \eqn{eq:SU2-CFT2},
they should be CFTs. 

\item Finally, but importantly, we can propose ideal numerical tests, starting from lattice Hamiltonian models of CI and CII SPTs (TSC/TI described in 
Sec.\ref{sec:CI-sym} and Sec.\ref{sec:CII-sym}), 
where they have \emph{onsite} $SU(2)$ and time reversal symmetry. Crucially \emph{onsite} $SU(2)$ global symmetry can be dynamically gauged by 
inputting dynamical gauge variables on the links between the sites.
There are two general methods to consider the numerical simulations: 
One is by the emergent gauge field construction through the ``\emph{soft gauging}'' method 
(e.g. the continuous group formalism analogous to Sec.~4.7.1 of \cite{1705.06728WWW}, 
both for the spatial Hamiltonian or spacetime lattice path integral), 
another is by the spacetime lattice path integral method (e.g. through quantum Monte Carlo simulation).
\end{enumerate}


\subsection{Gauging $SU(2)_{\text{color}}$ of $SU(2)_{\text{color}} \times SU(2)_{\text{flavor}}$}

\label{sec:SUN-gauge-SU(2)2}

Now we consider gauging one of $SU(2)$ (color) out of fSPTs of $SU(2)_{\text{color}} \times SU(2)_{\text{flavor}}$.

\subsubsection*{$\nu$ is even}

When $\nu$ is even, the theory has $\theta=0$ term for both $SU(2)$ factors in a flat spacetime. Consider gauging the first $SU(2)$ factor by coupling the SPT to the corresponding $SU(2)$ Yang-Mills action. Similarly to the CI case, the theory is expected to flow to a trivial gapped theory tensored with a bosonic SPT protected by time-reversal and $SO(3)$ global symmetry, where the global $SO(3)$ is the second $SU(2)$ factor divided by $\Z_2$ center. Such SPTs are known to be classified by $\Z_2^4$ with the corresponding topological terms being $w_1^4(TM_4),w_2^2(TM_4),w_1^2(TM_4)w_2(V_2),w_2^2(V_2)$. The corresponding coefficients of the action of the bosonic SPTs in the IR (i.e. at strong gauge coupling) are given by the following map
\begin{equation}
\begin{array}{rcl}
\Z_4\times \Z_2\times \Z_2\times \Z_2 &\rightarrow &\Z_2^4, \\
(\nu,\alpha,\beta,\gamma)&\mapsto &(\nu/2+\alpha,\alpha,\nu/2+\beta,\alpha+\gamma).
\end{array}
\end{equation}

\subsubsection*{$\nu$ is odd}

When $\nu$ is odd, the theory has $\theta=\pi$ term for both $SU(2)$ factors in the flat space. Consider again gauging the first $SU(2)$ factor by coupling the SPT to the $SU(2)$ Yang-Mills action. Now the theory is expected to flow to either a gapped theory with a spontaneously broken time-reversal symmetry, a gapped TQFT, or a gapless theory (presumably a CFT, discussed in 
Sec.~\ref{sec:SU2-pi-gapless}). 

In the first case, all different choices of the SPTs in the UV (i.e. at weak gauge coupling) become equivalent in the IR because all the topological terms can be deformed to each other when restricted to oriented manifolds (i.e. breaking time-reversal symmetry). 

In the last case, we get a fixed (that is independent on choice of parameters $\nu,\alpha,\beta,\gamma$) CFT 
tensored with a bosonic SPT protected by time-reversal and $SO(3)$ global symmetry. The corresponding coefficients of the action of the bosonic SPT in the IR are given by the following map
\begin{equation}
\begin{array}{rcl}
\Z_4\times \Z_2\times \Z_2\times \Z_2 &\rightarrow &\Z_2^4, \\
(\nu,\alpha,\beta,\gamma)&\mapsto &((\nu-1)/2+\alpha,\alpha,(\nu-1)/2+\beta,\alpha+\gamma).
\end{array}
\end{equation}
The fixed part depends on the background $SO(3)$ gauge field. If the background field is turned off it is the same as the (hypothetical) CFT for CI class at $\nu=1$ and $\alpha=0$ (i.e. the one  described by $Z_\text{YM+fSPT-CI}$ in eqn. (\ref{eq:SU2-CFT1}). 

\subsection{Gauging $SU(3)$: 3-color Yang-Mills theory + topological terms
}

Consider gauging $SU(3)$ group. 
On a flat spacetime, the theory has theta-term with $\theta=\pi\alpha$. Note that unlike in previous cases, the gauge group does not contain $\Z_2^F$ fermionic parity and therefore gauging does \emph{not} bosonize the theory. When $\alpha=0$ in the IR we expect to get a trivial gapped theory tensored with an SPT of DIII class specified by the same $\nu \in \Z_{16}$. 
When $\alpha=1$, if the theory in the IR turns out to be a non-trivial CFT with unbroken time-reversal symmetry, it is a fixed CFT tensored by the same SPT of DIII type.
But certainly, the less-interesting but a highly possible scenario is that time-reversal symmetry is broken spontaneously.

\subsection{Gauging $SU(4)$: 4-color Yang-Mills theory + topological terms
}

Consider gauging $SU(4)$ group. On a flat space the theory has theta-term with $\theta=\pi\beta$. Here the gauge group does contain $\Z_2^F$ fermionic parity and therefore gauging bosonizes the theory. When $\alpha=0$ in the IR we expect to get a trivial gapped theory tensored with a bosonic SPT protected by time-reversal symmetry. Such SPTs have $(\Z_2)^2$ classification and the corresponding topological terms are 
$w_1^4(TM_4),w_2^2(TM_4)$. The coefficients are given by the following map
\begin{equation}
\begin{array}{rcl}
\Z_2^3 &\rightarrow &\Z_2^2, \\
(\alpha,\beta,\gamma)&\mapsto &(\alpha,\gamma).
\end{array}
\end{equation}
 When $\beta=1$, if the theory in the IR turns out to be a non-trivial CFT with unbroken time-reversal symmetry, it is a fixed CFT tensored by the same bosonic SPT.
 But again a highly possible scenario is that time-reversal symmetry is broken spontaneously.




\section{Conclusion}

\label{sec:Conclusion}

\subsection{Discussion in a Gauging Framework}

\label{sec:conclude-gauge}

In the Introduction, we introduce 5 possible outcome quantum states, labeled from (1) to (5),
then we leave some questions as an \emph{overture}, one of them is 
``How do these states appear in our study?''
We would like to first reminder ourselves and readers, Sec.~\ref{sec:SUN-gauge}'s gauging SPTs to $SU(N)$ Yang-Mills, 
in the framework discussed in the introduction, Sec.~\ref{sec:intro}.

First, in the case of $SU(2)$ gauge theory invariant under time-reversal in Sec.~\ref{sec:SUN-gauge-SU(2)} (gauging $SU(2)$, which includes fermion parity $\Z_2^F \subset SU(2)$, thus also doing bosonization), the case of
$\nu = 0 \mod 2$ corresponds to the case (1) as [SPTs] $\otimes$ [a trivial gapped vacuum] (i.e.~SPTs $\otimes$ a pure-Yang-Mills gauge theory at $\theta=0$ preserving time reversal).

For {$\nu=1 \mod 2$}, we discuss three scenarios:\\
\noindent
$\diamond$ First scenario, {spontaneous time reversal breaking} in Sec.\ref{sec:TR-break}, stands for the (2) case, where we have 
[SSB Landau-Ginzburg order] $\otimes$ [a trivial gapped vacuum].\\
\noindent
$\diamond$ Second scenario, symmetry-enriched TQFT, stands for the (3) case SETs.\\
\noindent
$\diamond$ Third scenario, {deconfined gapless and time-reversal symmetric CFTs} in Sec.\ref{sec:SU2-pi-gapless}, 
stands for the (4) or (5) case. For (4), there could be two distinct time-reversal symmetric CFT fixed points that one can distinguish on non-orientable manifolds. 

Now we would like to address ``What are the justifications and the sharp distinctions of states as (4) SP-gapless/CFT and (5) SET-gapless/CFT?'' question.

The outcome of (4) is, in some sense, related to %
gapless theories at quantum critical points with enhanced global symmetries (see References in \cite{sachdev2011quantum}, and other recent work 
\cite{1702.07035-Benini-Hsin-Seiberg, 1703.02426-Senthil}). 
We stress that outcome states of (5) seem to be overlooked. 

In principle, even though gapless states naively have infinitely many degenerate ground states,
there are finite volume/size effects (say the length scale of the system is $L$), that distinguish the states in the spectrum by energy gap scaling as the order 
$\Delta E \sim O(L^{-\# })$, related to the scaling dimensions ($\Delta$) of primary operators.
However, the topological degeneracy has the smaller energy gap scaling as the order 
$\delta E \sim O(e^{-\# L})$ of an exponentially decaying tiny gap. 
Such a data of small gaps $\Delta E$ and $\delta E$ could be encoded in the path integral calculations $Z(M_4, \dots)$
on some topology, such 
as $M_3 \times S_1$ or more generic $M_4$. We could potentially read this data. One could compare a generic $M_3$ with a simpler 3-sphere $S^3$,
and the contributions for the (5) case, SET-gapless/CFT have \emph{topology-dependent} terms appearing at $\delta E$. 
In contrast, the (4) case's SP-gapless/CFT, the $\delta E$ term either does not appear, or the absolute value (related to the order of $\delta E$)
is insensitive to the spacetime \emph{topology}.

Similarly, the $SU(2_c)$ gauge theory with $SU(2_f)$ time-reversal in Sec.~\ref{sec:SUN-gauge-SU(2)2} can be phrased in this framework.
The $\nu = 0 \mod 2$ again corresponds to the case (1) as [$SO(3)\times \Z_2^T$-bosonic SPTs] $\otimes$ [a trivial gapped vacuum ($\theta=0$)].

For {$\nu=1 \mod 2$}, we again discuss three scenarios.
In the first scenario, {spontaneous time reversal breaking},
we have  [SSB Landau-Ginzburg order] $\otimes$ [a trivial gapped vacuum] as in the case (2).    
In the second scenario, 
we have a TQFT of [SET] $\otimes$ [$SO(3)\times \Z_2^T$-bosonic SPTs] as in the case (3). 
Again this SET can be enriched by $SU(2_f)$ not just $\Z_2^T$.
Third scenario,  
{[deconfined gapless time-reversal symmetric CFTs]}
 $\otimes$ [$SO(3)\times \Z_2^T$-bosonic SPTs] as in the case (4)/(5).
This particular CFT is symmetry-enriched by $\Z_2^T$ and $SU(2_f)$.
This state can be reduced to, by removing $SU(2_f)$-symmetry, 
but is in general \emph{different} from, the simpler naive 
[gauged CI's CFT in Sec.\ref{sec:SU2-pi-gapless}]  $\otimes$ [$SO(3)\times \Z_2^T$-bosonic SPTs].

At this moment, the analytical calculation of the path integral $Z(M_4, \dots)$ for non-Abelian gauge theories 
seems to be challenging. 
However, certainly, other justifications of the states of (4) and (5) can be from computer numerical simulations (with finite volume/size effect).
For (5), it should be that local OPEs cannot distinguish two CFTs, but the extended OPEs can distinguish them. 
Another physical motivated approach is that one could look into the \emph{vortices} and \emph{their sub-gap} structures to see the data.
We leave these issues for the future exploration.


\subsection{Comments and Relations to other recent work}

We conclude with some final remarks:

\begin{enumerate}


\item 
In our work, we have introduced the $SU(N)$ generalization of fermionic SPTs of TI/TSC.
Along this development, we discuss the fate of gauging the $SU(N)$ symmetry dynamically, resulting in 
non-Abelian $SU(N)$ Yang-Mills-like or QCD-like theories, which we term them as $SU(N)$-generalization of quantum ``spin'' liquids in condensed matter.
(We leave detail explorations of the relation of our work on time-reversal symmetric non-Abelian $SU(N)$ Yang-Mills to 
time-reversal symmetric Abelian $U(1)$ Maxwell theories (related to \cite{Metlitski:2015yqa, 1505.03520WS, 1710.00743ZWSenthil}), for future study.)

We study the possible scenarios of $SU(N)$ Yang-Mills gauge theories with topological terms under time-reversal symmetry,
obtained from dynamically gauging $SU(N)$ SPTs.
Our systems can be regarded as QFTs coupling to topological terms (e.g. TQFTs), see \cite{Kapustin:2014gua} and References therein.
We emphasize that, in the lattice Hamiltonian formalism at UV, 
the $SU(N)$ SPTs (specified by some topological terms in field theory) are \emph{gapped}, \emph{unitary}, 
and have onsite $SU(N)$ and onsite time-reversal symmetries. Thus
the local \emph{onsite} symmetry (e.g. $SU(N)$) guarantees that it can be gauged, 
but the gauging would drastically change the dynamics of the ground states and energy spectra (see discussion in the main text).
The consequence could be of all kinds, including gapped or gapless, topological or not, symmetry-preserving or symmetry-breaking, etc (see Sec.~\ref{sec:conclude-gauge}).

In contrast, the effective boundary theories of SPTs have \emph{non-onsite} symmetry thus have 't Hooft anomalies. 
This results in obstruction for dynamically gauging the boundary theories of SPTs alone.

\item In our work, 
the fermionic operators at UV and the effective fermionic fields at IR are not the same operators.
Although the effective descriptions at UV (e.g. lattice and  cutoff scales) and IR (e.g. field theory) may be different, the assumption is that
if the RG flow preserves the symmetry, then we can use the same symmetry group description $G_{\text{Tot}}$ and study the symmetry protected ground state.
 
\item Topological terms (SPT invariants) define the bulk theory. 
So we show the bulk topological terms (thus more than given only the group structure classification)
that characterize the boundary anomalies of SPTs and other physical observables.

\item We provide some remarks for the related work here.
Ref.~\cite{1401.1142WS} provides various helpful physical signatures and insightful intuitions behind for fTI/fTSC.
(However it does not provide precise bulk background field theory or topological terms for SPTs, few of their prediction slightly mismatches with the cobordism classifications.) It will be interesting to compare 
other physical observables based on our topological terms set-up.
Ref.~\cite{1505.06341TMorimoto} analyzes the stability of boundary gapless states of SPTs under quartic interactions and obtain the
classifications similar to the cobordism result.
There are also other strong coupling gauge theory applications studied recently in \cite{1604.06184Yonekura,1607.01873Yonekura}, using condensed-matter ideas. 
The anomalies of gauge theories studied in \cite{Gaiotto:2017yup, Gaiotto:2017tne} have been applied in 
Ref.~\cite{1704.05852Yamazaki, 1705.01949Tanizaki, 1706.06104Yonekura, 1711.04360Yamazaki, 1711.10487Tanizaki}
to constrain the phases of QCD-like theories.

\item Other than Ref.\cite{1604.06527FH}, there are other cobordism/cohomology calculations recently motivated by related physics studied in
Ref.\cite{1612.02860BrumfielMorgan,1708.04264Campbell}.

\end{enumerate}

\newpage

\section{Acknowledgements}

We thank Michael Hopkins,  Zohar Komargodski, Kantaro Ohmori, Nathan Seiberg, Zheyan Wan, Zhenghan Wang, Xiao-Gang Wen, Edward Witten, Shing-Tung Yau 
and Kazuya Yonekura for conversations.
MG thanks the support from U.S.-Israel Binational Science Foundation grant \cred{2014108}. 
PP gratefully acknowledges the support from Marvin L. Goldberger Fellowship and the DOE Grant DE-SC0009988. 
JW gratefully acknowledges the Corning Glass Works Foundation Fellowship and NSF Grant PHY-1314311 and PHY-1606531.
JW thanks National Taiwan University, Chinese University of Hong Kong, and ISSP and Kavli IPMU of The University of Tokyo, 
for their hospitality during the work progress in the summer. 
%
This work is also supported by the NSF Grant PHY-1306313, PHY-0937443, DMS-1308244, DMS-0804454, DMS-1159412 and Center for Mathematical Sciences and Applications at Harvard University.

\newpage

\begin{center}
{\bf\LARGE{Appendix}}
\end{center}

\appendix

\section{The plan of the article and the convention of notation}

\label{sec:plan-convention}

The plan of our article is organized in terms of the Table of Contents, from page 1-4.

The convention for our notations and terminology is as follows.

\begin{itemize}
 \item The term ``spin'' can have different meanings in our article, including the $SU(2)$-spin rotation, or the spin group, or the spin manifold.
\item  For condensed matter realization, fTI/fTSC stand for fermionic topological insulator/superconductor, bTI/bTSC stand for their bosonic counterparts.
\item The fSPTs/bSPTs stand for fermionic/bosonic Symmetry Protected Topological states (SPTs) respectively. 
\item Non-orientable and unorientable manifolds both mean that the manifolds \emph{cannot} be oriented.
An unoriented manifold means that an orientation has not been chosen (that is, even though it might be possible to orient the manifold, the transition functions in the atlas do not necessarily preserve orientation).
\item {$(-1)^{F}$ is the generator of $\Z_2^F$ fermionic parity, where $F$ is the fermion number (or $\hat{N}$)}.
\item Mathematically $\Z_2^F$, $\Z_2$, $\Z/2$, $\Z/2\Z$, $\{\pm 1\}$ all mean the same, the cyclic group of order 2. Notation $\Z_2^F$ is used when we want to emphasize its physical meaning as fermionic parity symmetry. Notation $\{\pm 1\}$ is sometimes used to emphasize that is considered as  a multiplicative group.
\item Rank $r$ real (complex) vector bundle $V$ is a bundle with fibers being real (complex) vector spaces of real (complex) dimension $r$.
\item $w_i(V)$ is the $i$-th Stiefel-Whitney class of a  real vector bundle $V$ (which maybe be also complex rank $r$ but considered as real rank $2r$).
\item $p_i(V)$ is the $i$-th Pontryagin class of a real vector bundle $V$. 
\item $c_i(V)$ is the $i$-th Chern class of a complex vector bundle $V$.
\item $M_d$ (or simply $M$) is a $d$-dimensional (possibly non-orientable) manifold.
\item $TM_d$ (or simply $TM$) is the tangent bundle over $M_d$ (or $M$).
\item For a top degree element of cohomology we often suppress explicit integration over the manifold (i.e. pairing with the fundamental class $[M_d]$), for example: $w_2^2(TM_4)\equiv \int_{M_4}w_2^2(TM_4)$.
\item $\mathcal{P}_2$ is the Pontryagin square operation $H^{2i}(M,\Z_{2^k})\rightarrow H^{4i}(M,\Z_{2^{k+1}})$. 
\item $\sigma(M)$ is the signature of manifold $M$.
\item $L_\R$ is the tautological (real) line bundle over real projective space $\RP^d$.
\item $L_\C$ is the tautological (complex) line bundle over complex projective space $\CP^d$.
\item The $n$ (e.g. $1,2,3$)  represents trivial real vector bundle of dimension $n$ over the base space.
\item $V_1-V_2+V_3$ is an example of an operation on the vector bundles $V_1,V_2,V_3$ in the Grothendieck group.\footnote{Let us remind that one way to define it is as isomorphism classes of ``virtual'' vector bundles, that is pairs of bundles $(V,V')$ modulo relation $(V\oplus W,V'\oplus W)\sim (V,V')$. The operation $+$ then descends from $\oplus$ so that $-(V,V')=(V',V)$. Then $V_i\equiv (V_i,0)$.}
\item $\Omega_d^H$ the bordism group of $d$-manifolds with structure $H$ (e.g. $H=\text{Spin}$).
\item $\Omega^d_H\equiv \text{Hom}(\text{Tor}\,\Omega_d^H,U(1))$ the cobordism group classifying deformation classes of SPTs.
\item $A \times_{\Z_2} B\equiv (A\times B)/\Z_2$ where the quotient is w.r.t. the diagonal center $\Z_2$ subgroup.
\item \cred{Integration of characteristic classes of top degree over the base manifold (that is paired with the fundamental class in homology) is implicitly assumed throughout the paper.}
\end{itemize}

\section{Computation of Spin/Pin$^{+/-}$ Cobordism: {Bordism groups $\pi_4MTH$}}

\label{sec:App}
This Appendix aims to fill calculations of bordism groups in more detail.\footnote{
Here we adopt the notations widely used in mathematics community. 
We write $\Z_n$ (or $\Z/n$ or $\Z/(n\Z)$) for the finite group of order $n$.
We write ${\{\pm 1\}}$ for a $\Z_2$ finite group.} We suggest a nice recent review (which requires a minimal background) \cite{2018arXiv180107530B} as a reference  for notions and techniques used in this appendix. Below we will only briefly review some basic notions before we proceed with calculations.

In \cite{Kapustin1403.1467,Kapustin:2014dxa}, it was proposed that the $d$-dimensional SPTs (modulo continous deformations) with symmetry $H_d$ (which includes space-time symmetry, that is $H_d$ is an extension of $O(d)$ or $SO(d)$) are classified by the Pontryagin dual of the torsion part of the cobordism group of manifolds with the corresponding tangential $H$-structure (the corresponding lift of the $O(d)$ or $SO(d)$ orthonormal frame bundle of the tangent bundle):
\begin{equation}
	\Omega_H^d\equiv \text{Hom}(\Omega^H_d,U(1))
\end{equation}
The Pontryagin-Thom isomorphism provides a relation between the bordism groups of manifolds with (stable) tangential structure and homotopy groups of the Madsen-Tillman spectrum $MTH$ (which is a close cousin of the more usual Thom spectrum $MH$) associated to tangential structure $H$:
\begin{equation}
	\Omega^H_d = \pi_d MTH\equiv \text{colim}_{k\rightarrow \infty} \pi_{d+k} MTH_k.
	\label{bordism-MTH}
\end{equation} 
Roughly speaking, a spectrum $X$ is collection of (pointed) topological spaces $X_n$ together with inclusions $\Sigma X_n\rightarrow X_{n+1}$. Here $\Sigma$ denotes a suspension $\Sigma X_n =S^1 \wedge X_n \equiv (S^1\times X_n)/(X_n\vee S^1)$ where $\wedge$ and $\vee$ are smash product and wedge sum (one point union) of pointed topological spaces respectively. The colimit above can be understood as a limiting group in the sequence $\pi_{d} MTH_0\rightarrow \pi_{d+1} MTH_1\rightarrow \pi_{d+2} MTH_2\rightarrow \ldots$.

On the other hand, in the work of Freed-Hopkins\cite{1604.06527FH}, there is a 1:1 correspondence 
  \begin{equation*}
       \left\{ \vcenter{\hbox{deformation classes of reflection positive
     }\hbox{invertible $n$-dimensional extended topological}\hbox{field
     theories with 
     symmetry group~$H_n$}} \right\} \cong [MTH,\Sigma^{n+1}I\Z]_{\textrm{tors}}.  
  \end{equation*}
In particular, $[MTH,\Sigma^{n+1}I\Z]_{\textrm{tors}}$ stands for the torsion part of homotopy classes of maps from spectrum $MTH$ to the $(n+1)$-th suspension of spectrum $I\Z$. The Anderson dual $I\Z$ is a spectrum that is the fibration of $I\C\to I\C^{\times}$ where $I\C$ ($ I\C^{\times}$) is the Brown-Comenetz dual spectrum defined by 
$$[X, I\C] =\textrm{Hom}(\pi_0X, \C),$$
$$[X, I\C^{\times}] =\textrm{Hom}(\pi_0X, \C^{\times}).$$ 
There is an exact sequence 
 \begin{equation*}
     0 \longrightarrow \textrm{Ext}^1(\pi _n MTH,\Z) \longrightarrow
     [MTH,\Sigma^{n+1}I\Z]\longrightarrow \textrm{Hom}(\pi _{n+1}MTH,\Z)\longrightarrow 0. 
  \end{equation*}

The torsion part $[MTH,\Sigma^{n+1}I\Z]_{\textrm{tors}}$ is 
$$\textrm{Ext}^1((\pi _n MTH)_{\textrm{tors}},\Z)=\textrm{Hom}((\pi _n MTH)_{\textrm{tors}},U(1)).$$
This provides relation to the bordism groups in (\ref{bordism-MTH}).

In this section we compute homotopy groups $\pi_4MTH$ for groups $H=\pin^+\times_{\{\pm1\}} SU(2)$, $H=\pin^-\times_{\{\pm1\}} SU(2)$, $H=\pin^+\times_{\{\pm1\}} SO(4)$, $H=\pin^+ \times SU(3)$, and $H=\pin^+ \times_{\{\pm1\}} SU(4)$. In the following note, $BG$ stands for the classifying space associated to a group $G$.

As mentioned above, we can think of $\pi_4MTH$ as bordism group of 4-manifolds with $H$-principal structure on stable tangent bundles. In particular, $MTH$ is the colimit of $\Sigma^{n}MTH_n$, where $\Sigma^{n}MTH_n=\text{Thom}(BH_n; \R^n-V_n)$, where $V_n$ is the induced vector bundle (of dimension n) by the map $BH_n\to BO_n$. In the cases we are interested in, $BH_n\to BO_n$ is the projection 
$$H_n\xrightarrow{pr_1} \pin_n^{\pm}/{\{\pm1\}}=O(n).$$
In another way, we can think of $MTH=\text{Thom}(BH, -V)$, where $V$ is the induced virtual bundle (of dimension 0) by the map $BH\to BO$. In the case we are interested in, $BH\to BO$ is the projection 
$$H \xrightarrow{pr_1} \pin^{\pm}/{\{\pm1\}}=O.$$

Note:  ``$T$'' in $MTH$ denotes that the $H$ structures are on tangent bundles instead of normal bundles. In the following sections, $w_i$ denotes the $i$th Stiefel-Whitney class. $H^*(-)$ stands for mod 2 cohomology $H^*(-;\Z_2)$.

\subsection{$H=\pin^+\times_{\{\pm1\}} SU(2)$}
\subsubsection{Understanding $BH$}

\indent

Recall that $\pin^+$ is an extensions of $O$ by $\Z_2$. In particular, the classifying space $B\pin^+$ is classified by the following fibration
\begin{center}
\begin{tikzcd}
B\pin^+\ar[d]\\
BO\ar[r,"w_2"] &K(\Z_2,2)
\end{tikzcd}
\end{center}
where $K(\Z_2,2)$ is the Eilenberg-MacLane space.

Note that $SU(2)=\spin(3)=S^3$ so it has the following fibration
\begin{center}
\begin{tikzcd}
BSU(2)\ar[d]\\
BSO(3)\ar[r,"w_2"] &K(\Z_2,2)
\end{tikzcd}
\end{center}
We have a commutative diagram that each square is a homotopy pullback square
\begin{center}
\begin{tikzcd}
BH \ar[r] \ar[d] & BSO(3) \ar[d,"w'_2"] \\
BO  \ar[r, "w_2"']  & K(\Z_2,2)
\end{tikzcd}
\end{center}
There is a homotopy equivalent $f: BO\times BSO(3) \xrightarrow{\sim} BO\times BSO(3)$ by $(V,W) \to (V-W+3,W)$. Note that $f^*(w_2)=w_2(V-W) = w_2(V) +w_1(V)w_1(W)+w_2(W) = w_2+w'_2$ since $W$ is oriented. Then we have the following homotopy pullback

\begin{center}
 \begin{tikzcd}
BH \ar[r,"\sim"] \ar[d] & B\pin^+\times BSO(3) \ar[d]\\
BO\times BSO(3) \ar[r,"f"] \ar[d,"{pr_1,V}"] \ar[rr,bend right=10,"w_2+w'_2"' ] & BO\times BSO(3) \ar[r,"{w_2+0}"] \ar[dl,"V+W-3"]& K(\Z_2,2) \\
BO
\end{tikzcd}
\end{center}

This implies that 
\begin{eqnarray} \label{eq:1}
BH \sim B\pin^+\times BSO(3).
\end{eqnarray}

We note that there is a pullback diagram\\
\begin{center}
\begin{tikzcd}{\label{fig:diagram1}}
B\spin\times BO(1) \ar[r,"\sim","V+L-1"'] \ar[d] & B\pin^+  \ar[d]\\
BSO\times BO(1) \ar[r,"V+ L-1"] \ar[rr,bend right=10,"w_2"' ]& BO \ar[r,"{w_2}"] & K(\Z_2,2) \\
\end{tikzcd}
\end{center}

and we have $BO(3)\sim BSO(3)\times BO(1)$ by $V\mapsto (V\otimes \text{Det}\,V, \text{Det}\,V)$.  Thus, we have the following homotopy pullback square
\begin{center}
\begin{tikzcd}{\label{fig:diagram1}}
BH \ar[r,"\sim"] \ar[d] &[4.5em] B\pin^+\times BSO(3) \ar[r] \ar[d]&[7em] B\spin\times BO_3 \ar[d]\\
BO\times BSO(3) \ar[r,"{(V,W)\mapsto (V+ W-3,W)}"']\ar[d, "{(V,W)\mapsto V}"] &[6em] BO\times BSO(3) \ar[r,"{(V,W)\mapsto(V-\text{Det}V+1,W\otimes \text{Det}V)}"'] & BSO\times BO(3) \ar[r,"{w_2}"]\ar[dll,bend left=10,"{(E,F)\mapsto E+\text{Det}F-F\otimes \text{Det}F+2}"] & K(\Z_2,2) \\
BO
\end{tikzcd}
\end{center}

\subsubsection{Understanding $B\hat{H}$}
\indent

Write $P=K(\Z_2,1)\times K(\Z_2,2)$ with the group structre 
$$(x_1,x_2)*(y_1,y_2)=(x_1+y_1, x_2+y_2+x_1y_1),$$
in which $x_i$, $y_i\in H^i(-)$.
With this choice the map $BO\xrightarrow{(w_1,w_2)} P$ is a group homomorphism.

Define $B\hat{H}\to BO$ by the homotopy pullback square
\begin{center}
\begin{tikzcd}[row sep=large, column sep=large]
 B\hat{H} \ar[r] \ar[d] & BO(3) \ar[d,"{(w_1,w_2)}"] \\
 BO  \ar[r,"{(w_1,w_2+w_1^2)}"'] & P 
\end{tikzcd}
\end{center}

Then we have a homotopy square involving $B\hat{H}\to BO$ like below 
\begin{center}
\begin{tikzcd}[row sep=large, column sep=large]
 B\hat{H} \ar[r] \ar[d] & B\spin\times BO(3) \ar[d]  \\
BO\times BO_3  \ar[r,"{id-(V_3-s)}"']\ar[d,"{(V,W)\mapsto V}"'] & BO\times BO(3) \ar[r,"{(w_1,w_2)}"']\ar[dl,"{(W,V_3)\mapsto-W-(V_3-3)}"] &P\\
BO
\end{tikzcd}
\end{center}

Thus $B\hat{H}\to BO$ can be identified with the map 
$$B\spin\times BO(3) \to BO,$$ 
$$(W,V_3)\mapsto -W-(V_3-3).$$ 

This leads to the following equivalence
\begin{eqnarray} 
MT\hat{H} \sim M\spin\wedge \Sigma^{-3} MO(3).
\end{eqnarray}

\subsubsection{Identification of $B\hat{H}\to BO$ with $BH\to BO$}
\indent

 The homotopy fiber of $B\hat{H}\to BO$ being the same as the homotopy fiber of $BO(3)\to P$ is $B\spin_3$.
We can identify $B\hat{H}$ with $BH$.
We know that $BH$ sit in a homotopy pullback
\begin{center}
 \begin{tikzcd}
BH\ar[r] \ar[d] & BSO_3 \ar[d,"w'_2"]\\
BO \ar[r,"w_2"]  & K(\Z_2,2)
\end{tikzcd}
\end{center}

and $B\hat{H}$ sit in the pullback 
\begin{center}
\begin{tikzcd}[row sep=large, column sep=large]
 B\hat{H} \ar[r] \ar[d,"{(w_1,w_2)}"'] & BO_3 \ar[d,"{(w_1,w_2)}"] \\
 BO_n  \ar[r,"{(w_1,w_2+w_1^2)}"'] & P 
\end{tikzcd}
\end{center}

To identify them, expand the first homotopy pullback to the diagram
\begin{center}
\begin{tikzcd}[row sep=large, column sep=large]
 B\hat{H} \ar[r] \ar[d] & BO(3)\ar[d,"{(w_1,w_2)}"] \ar[r,"V_3\otimes \text{Det}V_3"]& BSO(3) \ar[d,"w'_2"]\\
 BO  \ar[r,"{(w_1,w_2+w_1^2)}"'] \ar[rr,bend right=20,"w_2"' ] & P \ar[r,"{w_1^2+w_2}"'] & K(\Z_2,2)
\end{tikzcd}
\end{center}

Thus, we can identify $B\hat{H}\to BO$ with $BH\to BO$. With these identification, we have 
\begin{eqnarray} 
MTH\sim  M\spin\wedge \Sigma^{-3} MO(3).
\end{eqnarray}
These are useful for computing homotopy groups of $MTH$.

\rmk From the following diagram,
\begin{center}
 \begin{tikzcd}
BH\ar[r] \ar[d] & BSO(3) \ar[d,"w'_2"]\\
BO \ar[r,"w_2"]  & K(\Z_2,2)
\end{tikzcd}
\end{center}
we can think of the $n$th homotopy group $\pi_n MTH$ as the bordism group of $n$-manifolds with a $SO(3)$-bundle $V_{SO(3)}$ such that the 2nd Stiefel-Whitney classes of tangent bundle $TM$ and of $V_{SO(3)}$ agrees, $w_2(TM) = w_2(V_{SO(3)})$. If we use the other model $B\hat{H}\simeq B\spin\times BO(3)\to BO$ by $(W, V_3) \mapsto -W - (V_3-3)$, then $V_3$ can be identified by $V_{SO(3)}\otimes (TM-n)$.

\subsubsection{Computation}
\indent

Since the computation involves no odd torsion, we can use the Adams spectral sequence 
$$E_2^{s,t}=\text{Ext}_{\mathcal{A}}^{s,t}(H^*(MTH), \Z_2) \Rightarrow \pi_{t-s} MTH_2^\wedge. $$

Since the mod 2 cohomology of Thom spectrum $M\spin$ is 
$$H^*(M\spin) = \mathcal{A}\otimes_{\mathcal{A}(1)}\{\Z_2\oplus M\},$$
where $M$ is a graded $\mathcal{A}(1)$-module with the degree $i$ homogeneous part $M_i=0$ for $i<8$. Here $\mathcal{A}$ stands for Steenrod algebra and $\mathcal{A}(1)$ stands for $\mathbb{F}_2$-algebra generated by $Sq^1$ and $Sq^2$. $\mathcal{A}(1)$ is a subalgebra of  $\mathcal{A}$. Thus, for $t-s<8$, we can identify the $E_2$-page with 
$$\text{Ext}_{\mathcal{A}(1)}^{s,t}(H^{*+3}(MO(3)), \Z_2).$$

 We need to identify the $\mathcal{A}(1)$-module structure of $H^{*+3}(MO(3))$. This can be done by Thom isomorphism and Wu formula of Stiefel-Whitney classes. By Thom isomorphism, we have 
 $$H^{*+3}(MO(3)) = \Z/2[w_1,w_2,w_3]U,$$ 
 where $U$ stands for Thom class of the universal 3-bundle $E_3$ over $BO(3)$ and $w_i$ is the $i$th Stiefel-Whitney class of $E_3$ over $BO(3)$. 
 
The $\mathcal{A}(1)$-module structure of $H^{*+3}(MO(3))$ below degree 8 and $E_2$ page are as Fig.~\ref{MTH1}:

  \begin{figure}[!h]
  \centering
    \hspace*{-.8cm}
\includegraphics[scale=1.4]{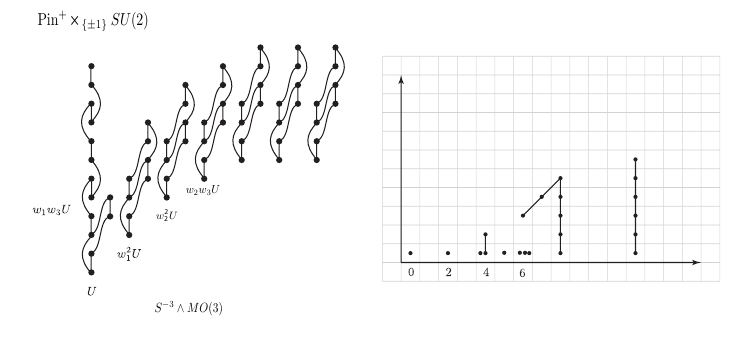}
\caption{$H=\pin^+\times_{\{\pm1\}} SU(2)$} \label{MTH1}
  \end{figure}

From the above spectral sequence, we have  
\begin{thm}
The bordism groups of $MTH$ are 
 
  \begin{center}
     \begin{tabular}{ c@{\hspace{2em}} c @{\hspace{2em}} } 
    \toprule
      $i$&$\;\;\pi_i MTH$\\
        \midrule \\[-8pt]
      $4$&$\Z/4\oplus\Z/2$\\
     $5$&$\Z/2$\\
      \bottomrule
     \end{tabular} 
  \end{center}
\end{thm}

\subsubsection{Manifold generators}

\claim $(\mathbb{CP}^2, L_{\C}+1)$ and $(\mathbb{RP}^4, 3)$ generate $\pi_4MTH$, where $L_{\C}$ is tautological complex line bundle over $\mathbb{CP}^2$.

First, check that $(\mathbb{CP}^2, L_{\C}+1)$ and $(\mathbb{RP}^4, 3)$ are elements in $\pi_4MTH$.
\begin{center}
 \begin{tikzcd}
\mathbb{CP}^2\ar[r,"L_{\C}+1"] \ar[d,"T\mathbb{CP}^2"'] & BSO(3) \ar[d,"w'_2"]\\
BO \ar[r,"w_2"]  & K(\Z_2,2)
\end{tikzcd}
\end{center}
with $w_2(T\mathbb{CP}^2) = w_2(L_\C+1)$.
\begin{center}
 \begin{tikzcd}
\mathbb{RP}^4\ar[r,"3"] \ar[d,"T\mathbb{RP}^4"'] & BSO(3) \ar[d,"w'_2"]\\
BO \ar[r,"w_2"]  & K(\Z_2,2)
\end{tikzcd}
\end{center}
with $w_2(T\mathbb{RP}^4) = w_2(3)$.

From the above spectral sequence, we have a map 
$$\pi_4 MTH =\Z_4\oplus \Z_2 \to \Z_2\oplus \Z_2,$$
$$(M, V_{SO(3)}) \mapsto (\int_M w_1w_3(V_{SO(3)}\otimes (TM-4)), \int_M w_2^2(V_{SO(3)}\otimes (TM-4)) ).$$

In particular, $$w_1(V_{SO(3)}\otimes (TM-4))= w_1(TM),$$
$$w_2(V_{SO(3)}\otimes (TM-4))=w_1^2(TM) + w_2(TM),$$
$$w_3(V_{SO(3)}\otimes (TM-4)) = w_1^3(TM)+ w_1(TM)w_2(TM) + w_3(V_{SO(3)}).$$

$(\RP^4, 3)$ is sent to (1,1) and $(\CP^2, L+1)$ is sent to $(0,1)$. 
\cred{So they generate the bordism group $ \pi_4 MTH$.}
If the invariants are chosen to be $w_1^4(TM)$ and $w_2^2(TM)$, it gives the same results.

\subsection{$H=\pin^-\times_{\{\pm1\}} SU(2)$}
\subsubsection{Understanding $BH$}
\indent
Recall that $\pin^-$ is an extensions of $O$ by $\Z_2$ with the following fibration
\begin{center}
\begin{tikzcd}
B\pin^-\ar[d]\\
BO\ar[r,"w_1^2+w_2"] &K(\Z_2,2).
\end{tikzcd}
\end{center}

Thus, the case of $H=\pin^-\times_{\{\pm1\}} SU(2)$ is analogous to case of $H=\pin^+\times_{\{\pm1\}} SU(2)$ by just exchanging $w_2$ and $w_1^2+w_2$.

We have a commutative diagram that each square is a homotopy pullback square

\begin{center}
 \begin{tikzcd}
BH \ar[r,"\sim"] \ar[d] & B\pin^-\times BSO(3)  \ar[d]\\
BO\times BSO(3) \ar[r,"f","\sim"'] \ar[d,"{pr_1=V}"] \ar[rr,bend right=10,"w_1^2+w_2+w'_2"' ] & BO\times BSO(3) \ar[r,"{(w_1^2+w_2+0)}"] \ar[dl,"V+W-3"]& K(\Z_2,2) \\
BO
\end{tikzcd}
\end{center}

This implies that 
\begin{eqnarray} \label{eq:2}
BH \sim B\pin^-\times BSO(3).
\end{eqnarray}

We have the following homotopy pullback square
\begin{center}
\begin{tikzcd}{\label{fig:diagram1}}
BH \ar[r,"\sim"] \ar[d] &[4.5em] BPin^-\times BSO_3 \ar[r] \ar[d]&[7em] BSpin\times BO_3 \ar[d]\\
BO\times BSO(3) \ar[r,"{(V,W)\mapsto (V- W+3,W)}"']\ar[d, "{(V,W)\mapsto V}"] &[6em] BO\times BSO(3) \ar[r,"{(V,W)\mapsto(V+\text{Det}V-1,W\otimes \text{Det}V)}"'] & BSO\times BO(3) \ar[r,"{(w_1,w_2)}"]\ar[dll,bend left=10,"{(E,F)\mapsto E-\text{Det}F+F\otimes \text{Det}F-2}"] & K(\Z_2,2) \\
BO
\end{tikzcd}
\end{center}

\subsubsection{Understanding $B\hat{H}$}
\indent

Write $P=K(\Z_2,1)\times K(\Z_2,2)$ with the group structre 
$$(x_1,x_2)*(y_1,y_2)=(x_1+y_1, x_2+y_2+x_1y_1)$$
in which $x_i$, $y_i\in H^i(-)$
With this choice the map $BO\xrightarrow{(w_1,w_2)} P$ is a group homomorphism.

Then define $B\hat{H}\to BO$ to be the composition of $B\hat{H}\to BO$ of case $H=\pin^+\times_{\{\pm1\}} SU(2)$ with $BO \xrightarrow{-id} BO$, so we have the following homotopy pullback square 
\begin{center}
\begin{tikzcd}[row sep=large, column sep=large]
 B\hat{H} \ar[r] \ar[d] & BO(3) \ar[d,"{(w_1,w_2)}"] \\
 BO  \ar[r,"{(w_1,w_2)}"'] & P 
\end{tikzcd}
\end{center}
Then we have a homotopy square involving $B\hat{H}\to BO$ like below 
\begin{center}
\begin{tikzcd}[row sep=large, column sep=large]
 B\hat{H} \ar[r] \ar[d] & B\spin\times BO(3) \ar[d,"{(-id,id)}"]  \\
BO\times BO_3  \ar[r,"{id-(V_3-s)}"'] \ar[d,"{(V,W)\mapsto V}"']& BO\times BO(3) \ar[r,"{(w_1,w_2)}"']\ar[dl,"{(W,V_3)\mapsto W+(V_3-3)}"] &P\\
BO
\end{tikzcd}
\end{center}

Thus $B\hat{H}\to BO$ can be identified with the map $$B\spin\times BO(3) \to BO,$$ 
$$(W,V_3)\mapsto -W+(V_3-3).$$ 

This leads to the following equivalence
\begin{eqnarray} 
MT\hat{H} \sim M\spin\wedge \Sigma^{3} MTO(3).
\end{eqnarray}

\subsubsection{Identification of $B\hat{H}\to BO$ with $BH\to BO$}
\indent

 The homotopy fiber of $B\hat{H}\to BO$ being the same as the homotopy fiber of $BO(3)\to P$ is $B\spin_3$.
We can identify $B\hat{H}$ with $BH$.
We know that $BH$ sit in a homotopy pullback
\begin{center}
 \begin{tikzcd}
BH\ar[r] \ar[d] & BSO(3) \ar[d,"w'_2"]\\
BO \ar[r,"w_1^2+w_2"]  & K(\Z_2,2)
\end{tikzcd}
\end{center}
and $B\hat{H}$ sit in the pullback 
\begin{center}
\begin{tikzcd}[row sep=large, column sep=large]
 B\hat{H} \ar[r] \ar[d] & BO(3) \ar[d,"{(w_1,w_2)}"] \\
 BO  \ar[r,"{(w_1,w_2)}"'] & P 
\end{tikzcd}
\end{center}

To identify them, expand the first homotopy pullback to the diagram

\begin{center}
\begin{tikzcd}[row sep=large, column sep=large]
 B\hat{H}\ar[r] \ar[d] & BO(3) \ar[d,"{(w_1,w_2)}"] \ar[r,"V_3\otimes \text{Det}V_3"]& BSO(3) \ar[d,"w'_2"]\\
 BO  \ar[r,"{(w_1,w_2)}"'] \ar[rr,bend right=20,"w_1^2+w_2"' ] & P \ar[r,"{w_1^2+w_2}"'] & K(\Z_2,2)
\end{tikzcd}
\end{center} 

Thus, we can identify $B\hat{H}\to BO$ with $BH\to BO$. With these identification, we have 
\begin{eqnarray} 
MTH^-\sim  M\text{Spin}\wedge \Sigma^{3} MTO(3).
\end{eqnarray}

\rmk From the following diagram,
\begin{center}
 \begin{tikzcd}
BH\ar[r] \ar[d] & BSO(3) \ar[d,"w'_2"]\\
BO \ar[r]  & K(\Z_2,2)
\end{tikzcd}
\end{center}
we can think of $\pi_n MTH$ as the bordism group of $n$-manifolds with a $SO(3)$-bundle $V_{SO(3)}$ such that $w_1^2+w_2(TM) = w_2(V_{SO(3)})$. If we use the other model $B\hat{H}\simeq B\spin\times BO(3)\to BO$ by $(W, V_3) \mapsto -W+ (V_3-3)$, then $V_3$ can be identified by $V_{SO(3)}\otimes (TM-n)$.

\subsubsection{Computation}
\indent

Similarly, we can use the Adams spectral sequence 
$$E_2^{s,t}=\text{Ext}_{\mathcal{A}}^{s,t}(H^*(MTH), \Z_2) \Rightarrow \pi_{t-s} MTH_2^\wedge .$$

Since $H^*(MSpin) = \mathcal{A}\otimes_{\mathcal{A}(1)}\{\Z_2\oplus M\}$, where $M$, $\mathcal{A}$ and $\mathcal{A}(1)$ as in case $H=\pin^+\times_{\{\pm1\}} SU(2)$. Thus, for $t-s<8$, we can identify the $E_2$-page with 
$$\text{Ext}_{\mathcal{A}(1)}^{s,t}(H^{*-3}(MTO(3)), \Z_2) .$$
 We need to identify the $\mathcal{A}(1)$-module structure of $H^{*-3}(MTO(3))$. This can be done by Thom isomorphism and Wu formula of Stiefel-Whitney classes. By Thom isomorphism, we have  
  $$H^{*}(MTO(3)) = \Z_2[w_1,w_2,w_3]U,$$
 where $U$ stands for Thom class of $-E_3$ over $BO(3)$ and $w_i$ is the $i$th Stiefel-Whitney class of $E_3$ over $BO(3)$. 

The $\mathcal{A}(1)$-module structure of $H^{*-3}(MTO(3))$ below degree 8 and $E_2$ page are as Fig.~\ref{MTH2}:
  \begin{figure}[!h]
  \centering
    \hspace*{-.8cm}
\includegraphics[scale=.74]{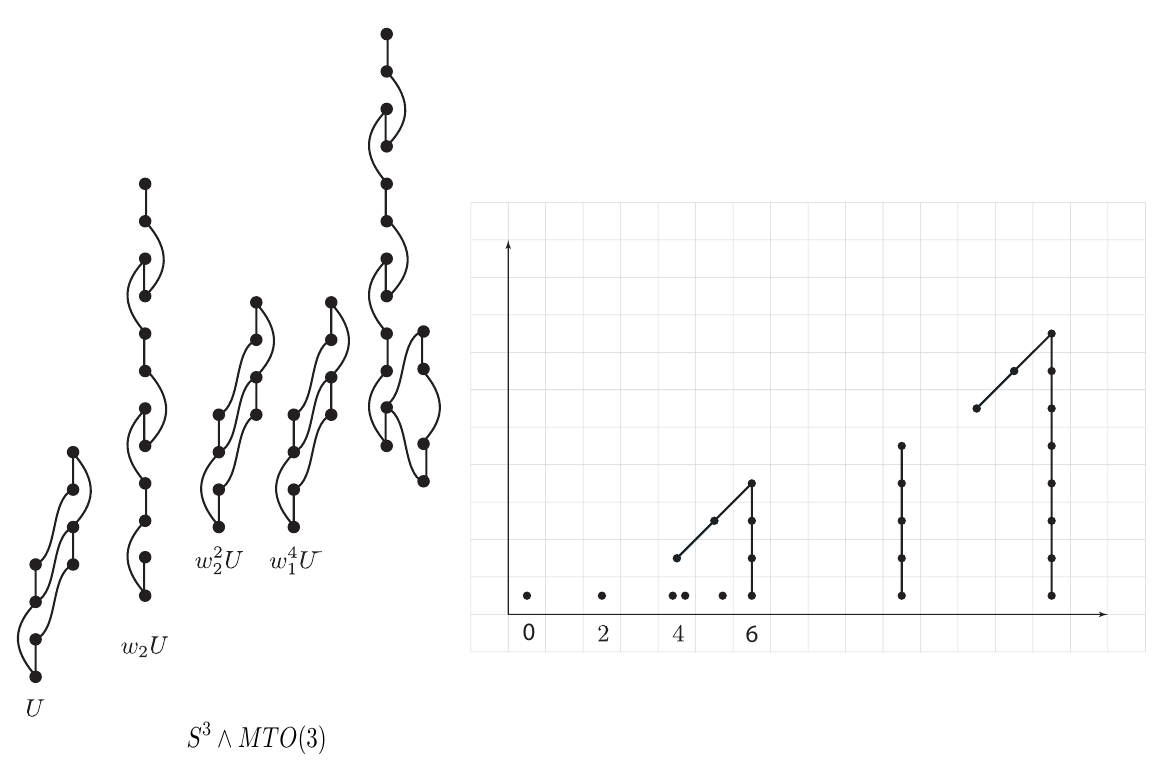} 
\caption{$H=\pin^-\times_{\{\pm1\}} SU(2)$} \label{MTH2}
  \end{figure}
\begin{thm}
The bordism groups of $MTH$ are
 
  \begin{center}
     \begin{tabular}{ c@{\hspace{2em}} c @{\hspace{2em}} } 
    \toprule
      $i$&$\;\;\pi_i MTH$\\
        \midrule \\[-8pt]
      $4$&$\Z_2\oplus\Z_2\oplus\Z_2$\\
     $5$&$\Z_2\oplus\Z_2$\\
      \bottomrule
     \end{tabular} 
  \end{center}
\end{thm}

\subsubsection{Manifold generators}
\indent

\claim The generators of $\pi_4MTH=\Z_2\oplus\Z_2\oplus\Z_2$ are $(S^4, H)$, $(\mathbb{CP}^2, L_{\C}+1)$, and $(\mathbb{RP}^4, 2L_{\R}+1)$, where  $H$ is the induced oriented 3-dimensional vector bundle from Hopf bundle $S^7\to S^4$, $L_{\C}$ is the tautological complex line bundle over $\mathbb{CP}^2$ and $L_{\R}$ is the tautological real line bundle over $\mathbb{RP}^4$.

First, check that $(S^4, H)$, $(\mathbb{CP}^2, L_{\C}+1)$, and $(\mathbb{RP}^4, 2L_{\R}+1)$ are elements in $\pi_4MTH$.
\begin{center}
 \begin{tikzcd}
\mathbb{CP}^2\ar[r,"L_{\C}+1"] \ar[d,"T\mathbb{CP}^2"'] & BSO(3) \ar[d,"w'_2"]\\
BO \ar[r,"w_1^2+w_2"]  & K(\Z_2,2)
\end{tikzcd}
\end{center}
with $w_1^2+w_2(T\mathbb{CP}^2) = w_2(L_{\C}+1)$
\begin{center}
 \begin{tikzcd}
\mathbb{RP}^4\ar[r,"2L_{\R}+1"] \ar[d,"T\mathbb{RP}^4"'] & BSO(3) \ar[d,"w'_2"]\\
BO \ar[r,"w_1^2+w_2"]  & K(\Z_2,2)
\end{tikzcd}
\end{center}
with $w_1^2+w_2(T\mathbb{RP}^4) = w_2(2L_{\R}+1)$.
\begin{center}
 \begin{tikzcd}
S^4\ar[r,"H"] \ar[d,"TS^4"'] & BSO(3) \ar[d,"w'_2"]\\
BO \ar[r,"w_1^2+w_2"]  & K(\Z_2,2)
\end{tikzcd}
\end{center}
with $w_1^2+w_2(TS^4) = w_2(H)$.

From the spectral sequence in the previous section, we have a map 
$$\pi_4 MTH \to \Z_2\oplus \Z_2\oplus \Z_2$$
$$(M, V_{SO(3)}) \mapsto (\text{mod 2 index of Dirac operator}, \int_M w_1^4(TM), \int_M w_2^2(TM) )$$
We can see $(S^4, H)$, $(\mathbb{CP}^2, L+1)$, and $(\mathbb{RP}^4, 2L_{\R}+1)$ are the generators.

\rmk   There is a map $MTH\to MTO$ if we forget the $H$-structure on stable tangent bundles. We know the latter one is isomorphic to $\Z_2\oplus\Z_2$ generated by $\mathbb{CP}^2$ and $\mathbb{RP}^4$. The kernel of this map is generated by $(S^4, H)$ where $H$ is the induced $SO(3)$ bundle from Hopf bundle $S^7\to S^4$.

\subsection{$H=\pin^+\times_{\{\pm1\}} SO(4)$}
\indent
\subsubsection{Understanding $BH$ and $B\hat{H}$}
There is a homotopy pullback square:
\begin{center}
\begin{tikzcd}
BH^+ \ar[r] \ar[d] & BSO(3)\times  BSO(3)   \ar[d,"w'_2+w''_2"] \\
BO  \ar[r, "{w_2}"'] & K(\Z_2,2)
\end{tikzcd}
\end{center}

Similar to computation of $MT\pin^+\times SU(2)$, if we define a new space $B\hat{H}$ to sit in the following homotopy pullback 
\begin{center}
\begin{tikzcd}
B\hat{H} \ar[r] \ar[d] & BO(3)\times  BSO(3)   \ar[d,"{(w'_1,w'_2)+(w''_1,w''_2)}"] \\
BO  \ar[r, "{w_1^2+w_2}"'] & P
\end{tikzcd}
\end{center}

\rmk From the following diagram,
\begin{center}
 \begin{tikzcd}
BH\ar[r] \ar[d] & BSO(3)\times BSO(3) \ar[d,"w'_2+w''_2"]\\
BO \ar[r,"{w_2}"']  & K(\Z_2,2)
\end{tikzcd}
\end{center}
we can think of $\pi_n MTH$ as the bordism group of $n$-manifolds with two oriented 3-dimensional vector bundle $V_1$ and $V_2$ such that then second Stiefel-Whitney class $w_2(TM)$ of tangent bundle $TM$ agrees with the sum of second Stiefel-Whitney classes $ w_2(V_1)+w_2(V_2)$ of $V_1$ and $V_2$. If we use the other model $B\hat{H}\simeq B\text{Spin}\times BO(3)\times BSO(3)\to BO$ by $(W, V_3, V_2) \mapsto -W- (V_3+V_2-6)$, then $V_3$ can be identified by $V_{1}\otimes (TM-n)$.

\subsubsection{Computation}
$B\hat{H}\to BO$ is identified with $BH\to BO$. We can see that the spectrum $MTH$ is homotopy equivalent to the spectrum $M\spin\wedge MO(3)\wedge MSO(3)$. Use the Adams spectral sequence 
$$\textrm{Ext}_{\mathcal{A}(1)}^{s,t}(H^{*+3}(MO(3))\otimes H^{*+3}(MSO(3)), \Z_2)\Rightarrow \pi_{t-s} MTH.$$

The $\mathcal{A}(1)$-structure and $E_2$ page are as Fig.~\ref{MTH3}:

  \begin{figure}[!hb]
  \centering
    \hspace*{-1.39cm}
\includegraphics[scale=.53]{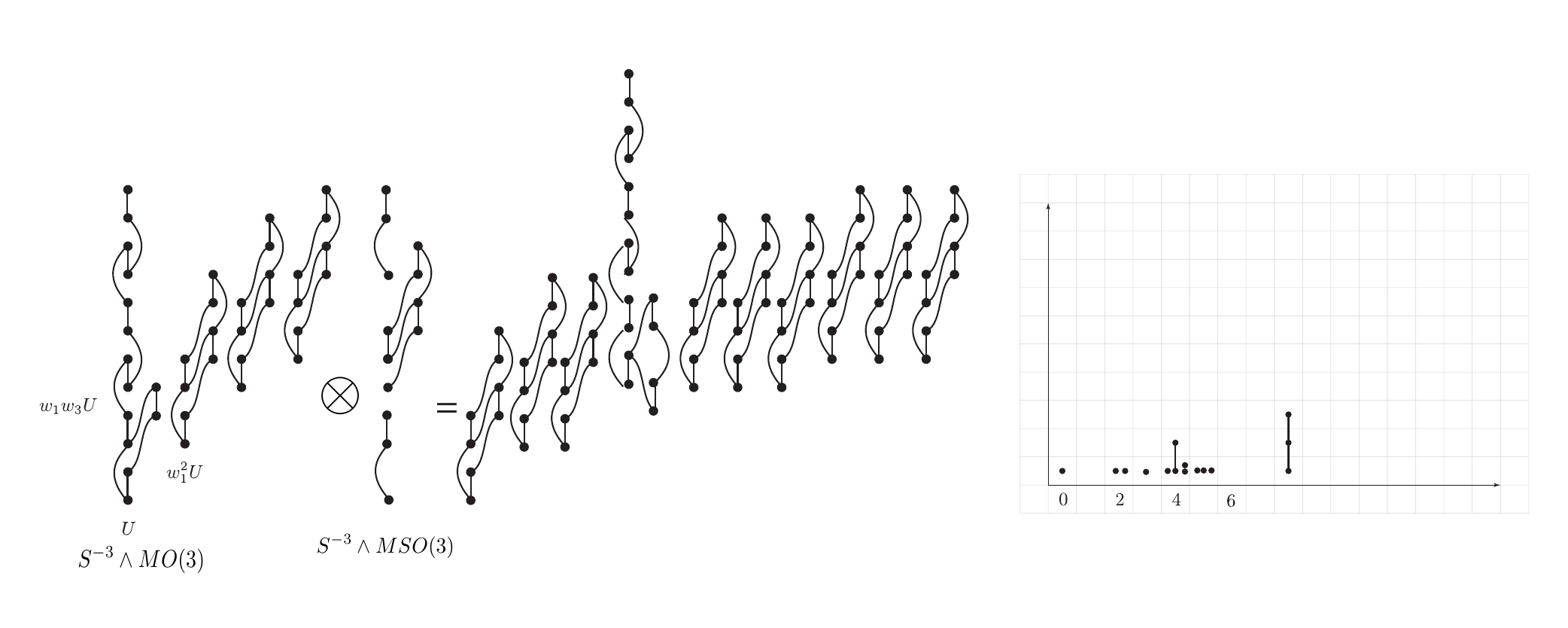} 
\caption{$H=\pin^+\times_{\{\pm1\}} SO(4)$} \label{MTH3}
  \end{figure}

 From the picture, we can read 
 \begin{thm}
The bordism groups of $MTH$ are
   \begin{center}
     \begin{tabular}{ c@{\hspace{2em}} c @{\hspace{2em}} } 
    \toprule
      $i$&$\;\;\pi_i MTH$\\
        \midrule \\[-8pt]
     $0$&$\Z_2$\\  
     $1$&$0$\\ 
     $2$&$\Z_2\oplus\Z_2$\\   
     $3$&$\Z_2$\\   
     $4$&$\Z_4\oplus\Z_2\oplus\Z_2\oplus\Z_2$\\
     $5$&$\Z_2\oplus\Z_2\oplus\Z_2$\\
      \bottomrule
     \end{tabular} 
  \end{center}
  \end{thm}

\subsubsection{Manifold generators}
\indent

\claim The generators of $\pi_4MTH=\Z_4\oplus\Z_2\oplus\Z_2\oplus\Z_2$ are $(\mathbb{RP}^4, 3,3)$, $(\mathbb{CP}^2, L_{\C}+1,3)$, $(\mathbb{RP}^4,2L_{\R}+1,2L_{\R}+1)$, $(\mathbb{CP}^2, 3, L_{\C}+1)$. $L_{\R}$ ($L_{\C}$) is the tautological (complex) line bundle over $\mathbb{RP}^4$ ($\mathbb{CP}^2$).

First check that they are elements in $\pi_4MTH$
\begin{center}
 \begin{tikzcd}
\mathbb{RP}^4\ar[r,"{(3, 3)}"] \ar[d,"TM"'] &[3.5em] BSO(3)\times BSO(3) \ar[d,"w'_2+w''_2"]\\
BO \ar[r,"{w_2}"']  & K(\Z_2,2)
\end{tikzcd}
\end{center}
\begin{center}
 \begin{tikzcd}
\mathbb{CP}^2\ar[r,"{(L_{\C}+1, 3)}"] \ar[d,"TM"'] &[3.5em] BSO(3)\times BSO(3) \ar[d,"w'_2+w''_2"]\\
BO \ar[r,"{w_2}"']  & K(\Z_2,2)
\end{tikzcd}
\end{center}
\begin{center}
 \begin{tikzcd}
\mathbb{RP}^4\ar[r,"{(2L_{\R}+1,2L_{\R}+1)}"] \ar[d,"TM"'] &[3.5em] BSO(3)\times BSO(3) \ar[d,"w'_2+w''_2"]\\
BO \ar[r,"{w_2}"']  & K(\Z_2,2)
\end{tikzcd}
\end{center}
\begin{center}
 \begin{tikzcd}
\mathbb{CP}^2\ar[r,"{(3, L_{\C}+1)}"] \ar[d,"TM"'] &[3.5em] BSO(3)\times BSO(3) \ar[d,"w'_2+w''_2"]\\
BO \ar[r,"{w_2}"']  & K(\Z_2,2)
\end{tikzcd}
\end{center}

The corresponding invariants mapping to $\Z_2\oplus\Z_2\oplus\Z_2\oplus\Z_2$ are 
$w_1^4(TM) + w_1^2(TM)w2(V_1)$, $w_1^4(TM)+w_2^2(V_1)$, $w_1^2(TM)w_2(V_2)$ and $w_2^2(V_2)$ respectively. 
\cred{We can check that the corresponding invariants generate the target $\Z_2\oplus\Z_2\oplus\Z_2\oplus\Z_2$, so they also generate the bordism group $\pi_4MTH$.}

\subsection{$H=\pin^+ \times SU(3)$}
\subsubsection{Computation}

The spectrum $MTH$ is homotopy equivalent to $MT\pin^+\wedge BSU(3)_{+}$. Use the Adams spectral sequence 
$$\textrm{Ext}_{\mathcal{A}(1)}^{s,t}(H^{*-1}(MTO(1))\otimes H^{*}(BSU(3)_{+}), \Z_2)\Rightarrow \pi_{t-s} MTH.$$

The $\mathcal{A}(1)$-structure and $E_2$ page are as Fig.~\ref{MTH4}:

  \begin{figure}[!h]
  \centering
    \hspace*{-.8cm}
\includegraphics[scale=.6]{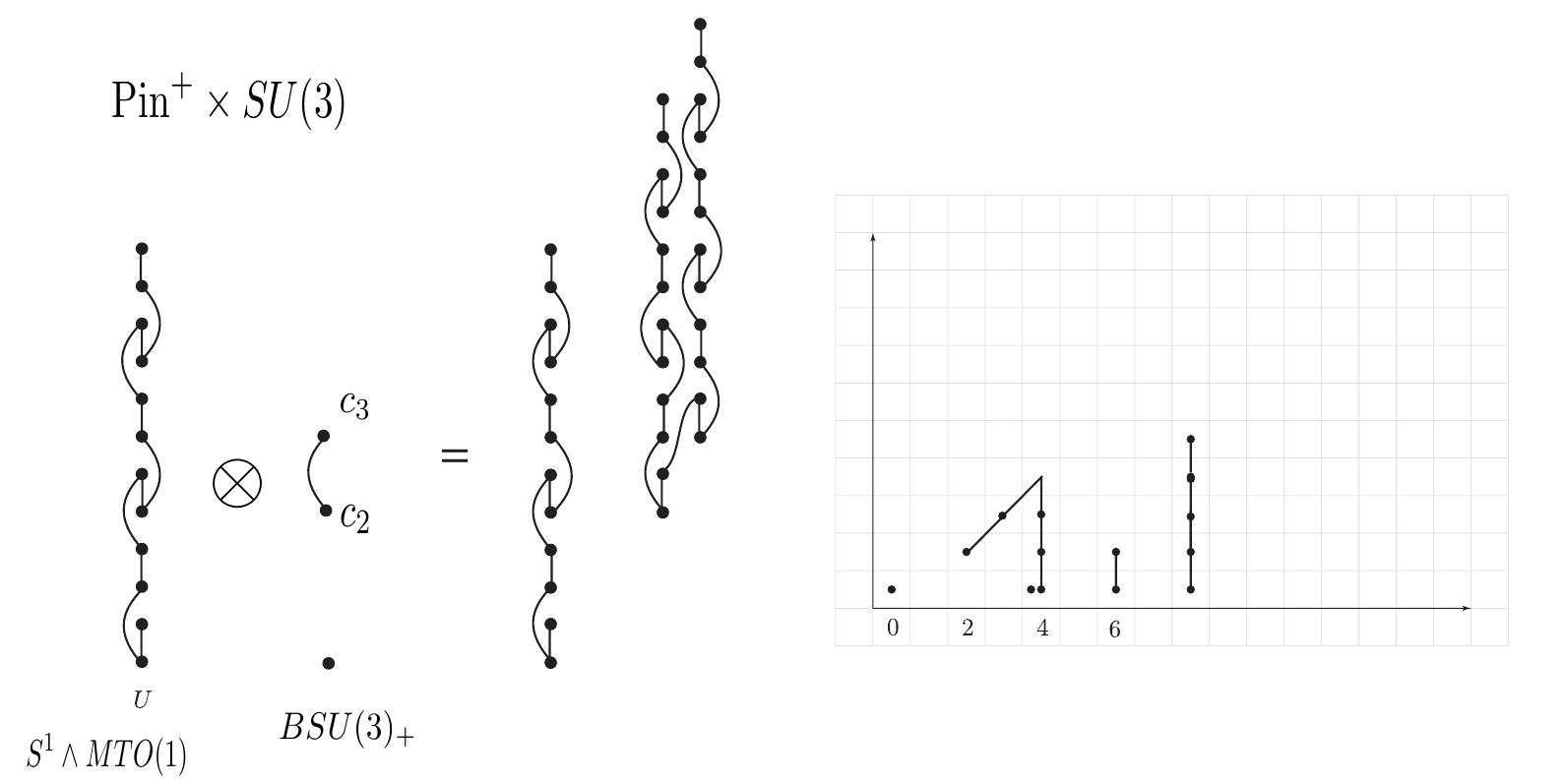} 
\caption{$H=\pin^+\times SU(3)$} \label{MTH4}
  \end{figure}

 From the picture, we can read 
 \begin{thm}
The bordism groups of $MTH$ are
   \begin{center}
     \begin{tabular}{ c@{\hspace{2em}} c @{\hspace{2em}} } 
    \toprule
      $i$&$\;\;\pi_i MTH$\\
        \midrule \\[-8pt]
     $0$&$\Z_2$\\  
     $1$&$0$\\ 
     $2$&$\Z_2$\\   
     $3$&$\Z_2$\\   
     $4$&$\Z_{16}\oplus\Z_2$\\
     $5$&$0$\\
      \bottomrule
     \end{tabular} 
  \end{center}
  \end{thm}

\subsubsection{Manifold generators}
\indent

\claim The generators of $\pi_4MTH=\Z_{16}\oplus\Z_2$ are $(\mathbb{RP}^4, \mathbb{RP}^4\times SU(3))$  and $(S^4,H)$ where $H$ is the Hopf fibration $S^7\to S^4$ considered as a $SU(2)$ bundle by $SU(2)\to SU(3)$.

If we think of $MTH$ as a $\pin^+$ 4-manifold with a $SU(3)$- bundles $W$, the corresponding invariants are eta invariant and 
$c_2$ (mod 2) of $W$. 
They generate the bordism groups of $MTH$.

\subsection{$H=\pin^+ \times_{\{\pm1\}} SU(4)$}
\subsubsection{Understanding $BH$}

There is a homotopy pullback square:
\begin{center}
\begin{tikzcd}
BH \ar[r] \ar[d] & BSO(6)   \ar[d,"w'_2"] \\
BO  \ar[r, "{w_2}"'] & K(\Z_2,2)
\end{tikzcd}
\end{center}

We can identity $BH\xrightarrow{projection} BO$ with $B\pin^+\times BSO(6)\xrightarrow{V-V_6+6} BO.$

The spectrum $MTH$ is homotopy equivalent to $\text{Thom}(B\pin^+\times BSO(6), -(V-V_6+6))$, which is $\Sigma^{-6}MT\pin^+\wedge MSO(6)$.

\rmk From the following diagram,
\begin{center}
 \begin{tikzcd}
BH\ar[r] \ar[d] & BSO(3)\times BSO(6) \ar[d,"w'_2"]\\
BO \ar[r,"{w_2}"']  & K(\Z_2,2)
\end{tikzcd}
\end{center}
we can think of $\pi_n MTH$ as the bordism group of $n$-manifolds with two oriented 6-dimensional vector bundle $V_1$ such that then second Stiefel-Whitney class $w_2(TM)$ of tangent bundle $TM$ agrees with the second Stiefel-Whitney classes $ w_2(V_1)$ of $V_1$.

\subsubsection{Computation}
Use the Adams spectral sequence 
$$\textrm{Ext}_{\mathcal{A}(1)}^{s,t}(H^{*-1}(MTO(1))\otimes H^{*+6}(BSO(6)), \Z_2)\Rightarrow \pi_{t-s} MTH.$$

The $\mathcal{A}(1)$-structure and $E_2$ page are as Fig.~\ref{MTH5}:

  \begin{figure}[!h]
  \centering
  \hspace*{-.9cm}
\includegraphics[scale=.52]{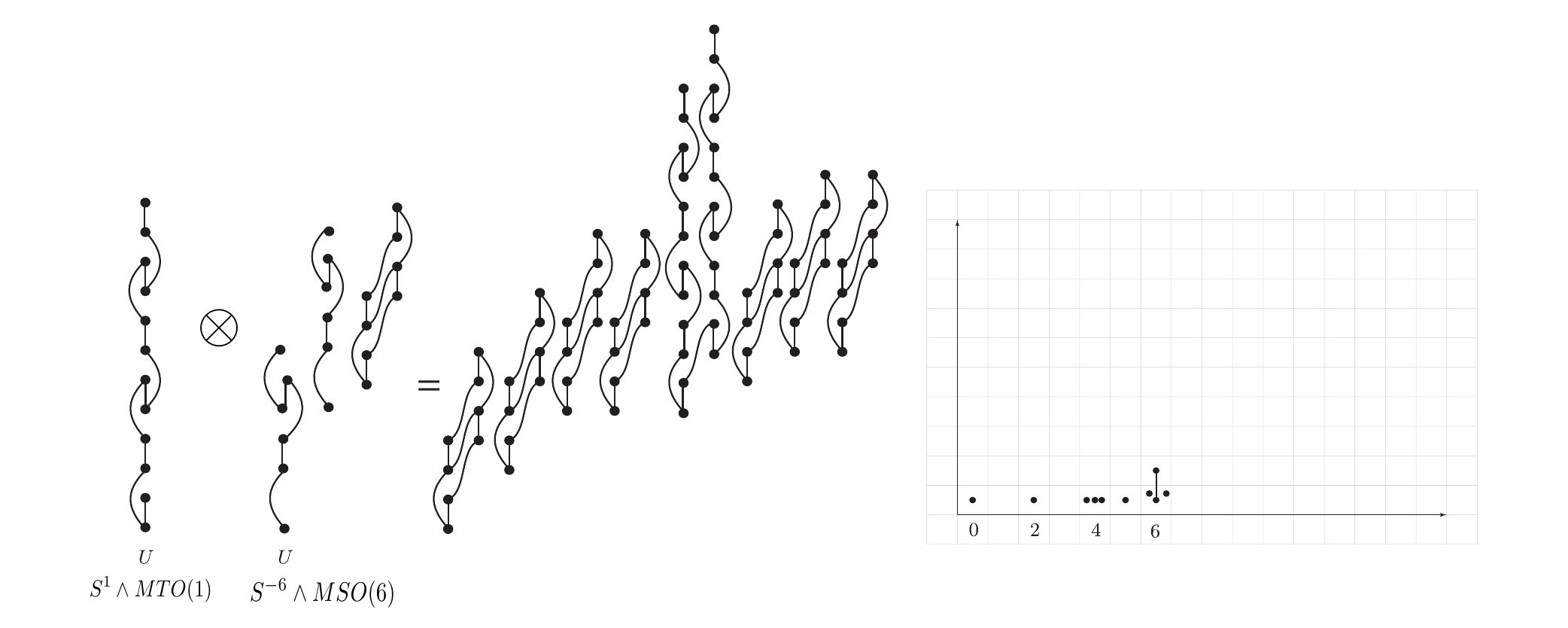} 
\caption{$H=\pin^+\times_{\{\pm1\}} SU(4)$} \label{MTH5}
  \end{figure}

 From above picture, we can read 
  \begin{thm}
The bordism groups of $MTH$ are
   \begin{center}
     \begin{tabular}{ c@{\hspace{2em}} c @{\hspace{2em}} } 
    \toprule
      $i$&$\;\;\pi_i MTH$\\
        \midrule \\[-8pt]
     $0$&$\Z_2$\\  
     $1$&$0$\\ 
     $2$&$\Z_2$\\   
     $3$&$0$\\   
     $4$&$\Z_2\oplus\Z_2\oplus\Z_2$\\
     $5$&$\Z_2$\\
      \bottomrule
     \end{tabular} 
  \end{center}
  \end{thm}

\subsubsection{Manifold generators}
\indent

\claim The generators of $\pi_4MTH=\Z_2\oplus\Z_2\oplus\Z_2$ are  $(\mathbb{RP}^4, 6)$, $(S^4, H+ 2)$, $(\mathbb{CP}^2, L_{\C}+4)$, where $H$ is the induced complex 2-dimensional vector bundle from Hopf fibration over $S^4$. 

First check that they are elements in $\pi_4MTH$
\begin{center}
 \begin{tikzcd}
\mathbb{RP}^4\ar[r,"{6}"] \ar[d,"TM"'] &[2.5em] BSO(6) \ar[d,"w'_2"]\\
BO \ar[r,"{w_2}"']  & K(\Z_2,2)
\end{tikzcd}
\end{center}

\begin{center}
 \begin{tikzcd}
S^4\ar[r,"{H+2}"] \ar[d,"TM"'] &[2.5em] BSO(6) \ar[d,"w'_2"]\\
BO \ar[r,"{w_2}"']  & K(\Z_2,2)
\end{tikzcd}
\end{center}

\begin{center}
 \begin{tikzcd}
\mathbb{CP}^2\ar[r,"{L_{\C}+4}"] \ar[d,"TM"'] &[2.5em] BSO(6) \ar[d,"w'_2"]\\
BO \ar[r,"{w_2}"']  & K(\Z_,2)
\end{tikzcd}.
\end{center}

  If we think of $MTH$ as a  4-manifold with an oriented 6-vector bundles $V_6$, the corresponding invariants are $w_1^4(TM)$, $w_4(V_6)$, $w_2^2(V_6)$. They generate the bordism groups of $MTH$.


\bibliographystyle{arXiv-new} 
\bibliography{JCW-ref,all,mybib,REf_new} 


\end{document}